\documentclass[twocolumn]{aastex62}
\newcommand{\tblspc}{-0.1cm}
\usepackage[normalem]{ulem}
\graphicspath{{./}{figures/}}

\received{...}
\revised{...}
\accepted{...}
\shorttitle{[OI] kinematics and evolution}
\shortauthors{Banzatti et al.}

\begin{document}

\title{Kinematic links and the co-evolution of MHD winds, jets, and inner disks from a high-resolution optical [OI] survey}

\correspondingauthor{Andrea Banzatti}
\email{banzatti@lpl.arizona.edu}

\author{Andrea Banzatti}
\affil{Department of Planetary Sciences, University of Arizona, 1629 East University Boulevard, Tucson, AZ 85721, USA}
\affil{Earths in Other Solar Systems Team, NASA Nexus for Exoplanet System Science}

\author{Ilaria Pascucci}
\affil{Department of Planetary Sciences, University of Arizona, 1629 East University Boulevard, Tucson, AZ 85721, USA}
\affil{Earths in Other Solar Systems Team, NASA Nexus for Exoplanet System Science}
\affil{Max Planck Institute for Astronomy (MPIA), K\"onigsthul 17, 69117, Heidelberg, Germany}

\author{Suzan Edwards}
\affiliation{Five College Astronomy Department, Smith College, Northampton, MA 01063, USA}

\author{Min Fang}
\affil{Department of Astronomy, University of Arizona, 933 North Cherry Avenue, Tucson, AZ 85721, USA}
\affil{Earths in Other Solar Systems Team, NASA Nexus for Exoplanet System Science}

\author{Uma Gorti}
\affiliation{SETI Institute/NASA Ames Research Center, Mail Stop 245-3, Moffett Field, CA 94035-1000, USA} 

\author{Mario Flock}
\affil{Max Planck Institute for Astronomy (MPIA), K\"onigsthul 17, 69117, Heidelberg, Germany}

\begin{abstract}
We present a survey of optical [OI] emission at 6300 \AA\ toward 65 T~Tauri stars at the spectral resolution of $\sim$7\,km/s. Past work identified a highly blueshifted velocity component (HVC) tracing microjets, and a less blueshifted low-velocity-component (LVC) attributed to winds. We focus here on the LVC kinematics to investigate links between winds, jets, accretion, and disk dispersal. We track the behavior of four types of LVC components: a broad and narrow component (``BC" and ``NC") in LVCs that are decomposed into two Gaussians, which typically have an HVC, and the single-Gaussian LVC profiles separated into those that have an HVC (``SCJ") and those that do not (``SC"). LVC centroid velocities and line widths correlate with HVC equivalent width and accretion luminosity, suggesting that LVC/winds and HVC/jets are kinematically linked and connected to accretion. The deprojected HVC velocity correlates with accretion luminosity, showing that faster jets come with higher accretion. BC and NC kinematics correlate and their blueshifts are maximum at $\sim 35^{\circ}$, suggesting a conical wind geometry with this semi-opening angle. Only SCs include $n_{13-31}$ up to $\sim3$ and their properties correlate with this infrared index, showing that [OI] emission recedes to larger radii as the inner dust is depleted, tracing less dense/hot gas and a decrease in wind velocity. All together, these findings support a scenario where optically thick, accreting inner disks launch radially-extended MHD disk winds that feed jets, and where inner disk winds recede to larger radii and jets disappear in concert with dust depletion.
\end{abstract}


\section{INTRODUCTION} \label{sec:intro}
Several processes are thought to drive or contribute to the evolution and dispersal of protoplanetary disks: accretion of disk material onto the star, jets and winds, dust evolution and planet formation \citep[e.g. reviews by][]{alex14,frank14,testi14,turner14}. The fundamental observation that theories attempt to explain is that gas-rich optically thick dust disks dissipate in a few Myr and that they seem to do so from inside out \citep[e.g.][]{fed10,ribas14,bp15}. 

Over the past 30 years, redistribution of angular momentum outward via magnetorotational instability(MRI)-induced turbulence \citep[e.g.][]{balb91} has been the prime mechanism to explain the early evolution of protoplanetary disks. While MRI-induced turbulence was believed to drive disk accretion, the presence of radially confined (out to a few au) MHD disk winds that further facilitate accretion were also proposed. These disk winds, after re-collimating, could also explain large scale jets that are often observed \citep[e.g.][]{pudr07}. In this classic picture, the faster inside-out dispersal phase was attained after the main accretion stage by thermal (or ``photoevaporative'') winds driven by high-energy photons from the central star \citep[e.g.][]{clarke01}. Depending on the heating agent (EUV, X-ray or FUV photons), these winds could be launched as close as $\sim$1\,au for gas at 10,000K, or as far out as tens of au for gas at a few 100--1,000 K \citep[e.g.][for a recent review]{gorti16}.

The above scenario for disk evolution has recently come into question. Theoretical simulations that take into account non-ideal MHD effects have found that disks are not turbulent over a large range of disk radii; instead, it is the development of radially extended, $\sim 1-30$\,au, MHD winds that may drive accretion, hence disk evolution, by extracting angular momentum from the disk \citep[e.g.][]{gressel15,bet17}. These winds also remove mass and could potentially disperse disks, but their role at various stages of evolution is presently unclear. The availability of disk wind tracers that probe different physical conditions and hence regions, in conjunction with the theoretical results, now enables a more systematic study of disk winds (photoevaporative and MHD) to investigate their relative contributions to disk evolution and dispersal \citep[e.g., review by][]{EP17}.

\subsection{[OI] optical emission as a tracer of jets and winds} \label{sec: OIhistory}
Optical forbidden emission from oxygen is a long-established tracer of outflows from TTauri stars. The emission is often blueshifted, implying origin in outflowing gas that is approaching the observer, where the receding part of the outflow is obscured by dust in a circumstellar disk \citep{jank83,appenz84,edwards87}. The prominent [OI] 6300~\AA\ line shows a dual profile structure consisting of two distinct velocity components, a ``high-velocity" component (HVC) with line peaks and wings as blueshifted as a few hundred km/s, and a ``low-velocity" component (LVC) with blueshifts less than $-50$\,km/s \citep{kwan88,ham94,hartig95,hirt97}. These components have order-of-magnitude differences in [OI] 5577/6300 line ratios, which through the different critical densities of the two lines indicate denser gas in the LVC than in the HVC \citep{edwards89,hartig95}; moreover, a different spatial extent shows the HVC to be at larger offsets from the star than LVC \citep{hirt97}. These properties suggested early on an HVC origin in high-velocity (micro-)jets collimated by magnetic fields \citep{kwan88}, as later confirmed by spatially-resolved observations \citep{dougados00}.  The LVC was instead proposed to be physically distinct from the HVC, and possibly tracing disk winds \citep{kwan95,hartig95,rigl13,natta14}.

Recently, several studies have focused on the LVC to better understand its origin and potential connection to winds that disperse inner disks. High resolution spectroscopy ($\sim 4$--7\,km/s) showed that the LVC sometimes has broad line wings and a narrow peak, which can be modeled as the combination of two Gaussian components \citep[a ``broad component'' BC and a ``narrow component" NC;][]{rigl13,simon16}. \citet{simon16} analyzed [OI] emission in a sample of 30 TTauri stars in Taurus and found that $\approx 40\%$ showed the BC+NC composite profile structure. The 5577/6300 line ratios were found to be higher in BC than in NC, supporting their origin in physically distinct gas at different density/temperature and spanning different disk radii (from the different FWHM). The launching radii of $\lesssim 0.5$\,au inferred for BC excluded a photoevaporative wind and instead pointed to an origin in MHD disk winds \citep{simon16}. The origin of the NC, launched at larger disk radii of $\gtrsim 1$~au as inferred from the smaller line widths, is to date still debated.

Correlations between forbidden emission and H$\alpha$ luminosities suggested early on that the outflow processes traced by [OI] lines are powered by accretion of disk material onto the star \citep{edwards89,cabrit90}. These luminosity correlations have been also found for the  LVC-BC and LVC-NC separately \citep{simon16,fang18}. Based on these luminosity correlations, \citet{nisini18} recently suggested that the excitation of [OI] lines is provided by accretion for all velocity components. 
Moreover, the HVC was found only in objects with a large infrared excess while the LVC in any object with infrared excess, even toward disks with a dust cavity (transition disks) where [OI] emission typically shows only a single-peaked and narrow LVC component \citep{hartig95,pasc11,simon16,mcginnis18}. This suggested that HVC/jets disappear with the decrease of disk accretion onto the star and the depletion of warm dust, while LVC/winds persist to later phases of disk evolution.

\subsection{This work} \label{sec: thiswork}
We started this work as a continuation of the analysis presented in \citet{simon16} with the main goal of clarifying the kinematic link between different components and their evolution as disks disperse. The main contributions of this work include: 1) a larger sample of 65 objects (Table \ref{tab: sample}), doubling the sample included in \citet{simon16}, 2) an improved correction for photospheric absorption by using a finer grid of photospheric spectral templates from weak line TTauri stars (Section \ref{sec: OI_correction}), 3) the individual analysis of four empirical LVC components identified in the data, tracking the behavior also of single-Gaussian lines (here called SC/SCJ, Section \ref{sec: OI_class}), 4) the systematic analysis of correlations for all [OI] components (both HVC and LVC) in the multi-dimensional space of their properties, and the investigation of relations between LVC properties (especially the kinematics) and HVC, accretion, and some key properties of their disks (specifically the inclination to the line of sight and the $n_{13-31}$ infrared index; Section \ref{sec: results}). 

Our main findings are: 1) the LVC is kinematically linked to HVC and accretion, 2) there is evidence for an origin of both BC and NC in the same MHD-driven wind, and 3) the wind kinematics co-evolve with the dust content in inner disks (Section \ref{sec: discuss}). In a parallel work, \citet{fang18} describes the combined analysis of [OI] (at 5577 and 6300 \AA) and [SII] (at 4068 \AA) for part of this sample, finding that [OI] emission is thermally excited in the LVC, and that the mass outflow rate is larger in the BC than in the NC.

\begin{deluxetable*}{r l c c c c c c c c c c c c}
\tabletypesize{\footnotesize}
\tablewidth{0pt}
\tablecaption{\label{tab: sample} Sample properties.}
\tablehead{\colhead{ID} & \colhead{Name} & \colhead{Instr.} & \colhead{SpT} & \colhead{Phot.} & \colhead{M$_\star$} & \colhead{$r_{6300}$} & \colhead{$v_{\rm rad}$} & \colhead{LVC} & \colhead{HVC} & \colhead{log$L_{acc}$} & \colhead{$n_{13-31}$} & \colhead{Incl.} & \colhead{Incl.}
\\ 
 &  &  &  & templ. & (M$_\odot$) &  & (km/s) & comp. &  &  &  &  (deg) &  ref. }
\tablecolumns{14}
\startdata
1 & AATau & HI & M0.6 & K8 & 0.60 & 0.1 & 15.4 $\pm$ 0.6 & BC+NC & yes & -2.40 & -0.36 & 71. $\pm$ 1. & \textit{(i20)} \\[\tblspc]
2 & AS205N & HI & K5 & K8 & 1.50 & 2.1 & -5.4 $\pm$ 0.5 & BC+NC & yes & -0.16 & -0.19 & 20. $\pm$ 5. & \textit{(i3)} \\[\tblspc]
3 & AS209 & MI & K5 & K3 & 1.40 & 0.1 & -9.1 $\pm$ 0.5 & SCJ & yes & -1.12 & -0.28 & 35. $\pm$ 1. & \textit{(i13)} \\[\tblspc]
4 & AS353A & HI & M3 & K8 & 0.32 & 3.3 & -9.0 $\pm$ 0.9 & BC+NC & yes & -0.13 & -0.65 & 18. $\pm$ 10. & \textit{(i19)} \\[\tblspc]
5 & BPTau & HI & M0.5 & K8 & 0.60 & 0.6 & 14.5 $\pm$ 0.5 & BC+NC & yes & -1.12 & -0.36 & 39. $\pm$ 3. & \textit{(i5)} \\[\tblspc]
6 & CITau & HI & K5.5 & K8 & 0.90 & 0.5 & 19.9 $\pm$ 0.8 & -- & -- & -- & -0.17 & 44. $\pm$ 2. & \textit{(i1)} \\[\tblspc]
7 & CoKuTau4 & HI & M1.1 & K9 & 0.54 & 0.0 & 18.2 $\pm$ 1.7 & SC & -- & -3.67 & 1.96 & -- & \textit{--} \\[\tblspc]
8 & CWTau & HI & K3 & K3 & 1.01 & 1.0 & 15.0 $\pm$ 0.5 & BC+NC & yes & -0.64 & -0.75 & 65. $\pm$ 2. & \textit{(i6)} \\[\tblspc]
9 & CXTau & HI & M2.5 & M3 & 0.39 & 0.0 & 19.3 $\pm$ 1.3 & -- & -- & -2.56 & -0.15 & 60. $\pm$ 5. & \textit{(i2)} \\[\tblspc]
10 & CYTau & HI & M2.3 & M2 & 0.41 & 1.0 & 15.1 $\pm$ 2.5 & SC & -- & -1.33 & -1.19 & 27. $\pm$ 3. & \textit{(i1)} \\[\tblspc]
11 & DFTau & HI & M2.7 & M2 & 0.60 & 4.2 & 16.4 $\pm$ 5.0 & BC+NC & yes & -0.98 & -1.09 & 60. $\pm$ 13. & \textit{(i10)} \\[\tblspc]
12 & DGTau & HI & K7 & K7 & 0.80 & 0.7 & 14.6 $\pm$ 0.5 & BC+NC & yes & -1.03 & 0.33 & 32. $\pm$ 2. & \textit{(i5)} \\[\tblspc]
13 & DHTau & HI & M2.3 & M2 & 0.41 & 0.5 & 16.1 $\pm$ 0.5 & BC+NC & -- & -2.02 & 0.26 & -- & \textit{--} \\[\tblspc]
14 & DKTau & HI & K8.5 & K8 & 0.68 & 0.3 & 16.6 $\pm$ 0.5 & BC+NC & yes & -1.50 & -0.68 & 26. $\pm$ 10. & \textit{(i1)} \\[\tblspc]
15 & DLTau & HI & K5.5 & K8 & 0.92 & 1.9 & 13.8 $\pm$ 0.6 & SCJ & yes & -0.85 & -0.74 & 38. $\pm$ 2. & \textit{(i5)} \\[\tblspc]
16 & DMTau & HI & M3 & M3 & 0.35 & 0.1 & 20.1 $\pm$ 0.5 & SC & -- & -2.02 & 1.29 & 34. $\pm$ 2. & \textit{(i1)} \\[\tblspc]
17 & DNTau & HI & M0.3 & K9 & 0.55 & 0.1 & 18.8 $\pm$ 0.5 & -- & -- & -1.93 & -0.13 & 28. $\pm$ 5. & \textit{(i1)} \\[\tblspc]
18 & DoAr24ES & MI & K0 & K3 & 0.70 & 0.1 & -7.6 $\pm$ 1.3 & -- & -- & -- & -0.49 & 20. $\pm$ 5. & \textit{(i3)} \\[\tblspc]
19 & DoAr44 & HI & K2 & K3 & 1.40 & 0.2 & -4.4 $\pm$ 0.5 & SC & -- & -0.73 & 0.80 & 55. $\pm$ 15. & \textit{(i12)} \\[\tblspc]
20 & DOTau & HI & M0.3 & K9 & 0.70 & 1.7 & 17.1 $\pm$ 5.0 & SCJ & yes & -1.00 & -0.15 & 37. $\pm$ 5. & \textit{(i1)} \\[\tblspc]
21 & DPTau & HI & M0.8 & M1 & 0.57 & 1.2 & 16.8 $\pm$ 0.5 & BC+NC & yes & -1.69 & -0.31 & -- & \textit{--} \\[\tblspc]
22 & DRTau & HI & K6 & K8 & 0.90 & 11.0 & 23.0 $\pm$ 5.0 & BC+NC & yes & -0.85 & -0.34 & 9. $\pm$ 5. & \textit{(i3)} \\[\tblspc]
23 & DSTau & HI & M0.4 & K8 & 0.65 & 0.5 & 16.9 $\pm$ 0.6 & SC & -- & -1.82 & -0.96 & 70. $\pm$ 3. & \textit{(i6)} \\[\tblspc]
24 & EXLup & MI & M0 & K8 & 0.50 & 0.2 & -1.0 $\pm$ 0.5 & SCJ & yes & -1.30 & -0.14 & 38. $\pm$ 4. & \textit{(i8)} \\[\tblspc]
25 & EXLup08 & HI & M0 & K8 & 0.50 & 10.6 & -1.0 $\pm$ 5.0 & SCJ & yes & 0.30 & -0.56 & 38. $\pm$ 4. & \textit{(i8)} \\[\tblspc]
26 & FMTau & HI & M4.5 & M3 & 0.15 & 4.4 & 14.4 $\pm$ 5.0 & BC+NC & -- & -2.07 & -0.09 & 55. $\pm$ 2. & \textit{(i2)} \\[\tblspc]
27 & FNTau & HI & M3.5 & M3 & 0.30 & 0.5 & 15.4 $\pm$ 0.8 & SCJ & yes & -1.53 & -0.02 & 20. $\pm$ 10. & \textit{(i17)} \\[\tblspc]
28 & FPTau & HI & M2.6 & M3 & 0.39 & 0.4 & 18.2 $\pm$ 2.7 & SC & -- & -2.27 & -0.10 & 66. $\pm$ 4. & \textit{(i2)} \\[\tblspc]
29 & FZTau & HI & M0.5 & M2 & 0.63 & 1.7 & 15.8 $\pm$ 3.5 & BC+NC & yes & -0.68 & -0.98 & 38. $\pm$ 15. & \textit{(i10)} \\[\tblspc]
30 & GHTau & HI & M2.3 & M2 & 0.36 & 0.0 & 18.2 $\pm$ 0.5 & BC+NC & -- & -2.18 & -0.33 & -- & \textit{--} \\[\tblspc]
31 & GITau & HI & M0.4 & K8 & 0.58 & 0.3 & 17.1 $\pm$ 0.5 & BC+NC & yes & -1.69 & -0.79 & -- & \textit{--} \\[\tblspc]
32 & GKTau & HI & K6.5 & K8 & 0.69 & 0.0 & 17.0 $\pm$ 0.5 & BC+NC & yes & -1.71 & -0.37 & -- & \textit{--} \\[\tblspc]
33 & GMAur & HI & K6 & K7 & 0.90 & 0.3 & 16.5 $\pm$ 0.5 & SC & -- & -0.95 & 1.75 & 55. $\pm$ 1. & \textit{(i1)} \\[\tblspc]
34 & GOTau & HI & M2.3 & M3 & 0.42 & 0.1 & 17.1 $\pm$ 1.3 & SC & -- & -2.00 & 0.03 & 53. $\pm$ 2. & \textit{(i1)} \\[\tblspc]
35 & GQLup & MI & K5 & K8 & 0.89 & 0.3 & -2.9 $\pm$ 0.5 & SC & -- & -0.36 & -0.18 & 60. $\pm$ 0. & \textit{(i15)} \\[\tblspc]
36 & GWLup & HI & M2.3 & M2 & 0.41 & 0.3 & -1.5 $\pm$ 0.5 & SC & -- & -1.87 & -0.22 & 40. $\pm$ 1. & \textit{(i4)} \\[\tblspc]
37 & HMLup & HI & M3 & M3 & 0.36 & 1.0 & -1.6 $\pm$ 1.8 & SCJ & yes & -1.61 & -0.31 & 53. $\pm$ 19. & \textit{(i4)} \\[\tblspc]
38 & HNTau & HI & K3 & K3 & 0.70 & 0.7 & 20.8 $\pm$ 3.6 & SCJ & yes & -0.93 & -0.62 & 75. $\pm$ 4. & \textit{(i2)} \\[\tblspc]
39 & HQTau & HI & K2 & K2 & 1.53 & 0.0 & 18.5 $\pm$ 9.7 & SC & -- & -1.60 & -0.50 & -- & \textit{--} \\[\tblspc]
40 & IPTau & HI & M0.6 & K9 & 0.59 & 0.1 & 17.2 $\pm$ 0.8 & -- & yes & -2.29 & 0.14 & 41. $\pm$ 10. & \textit{(i1)} \\[\tblspc]
41 & ITTau & HI & K6 & K3 & 0.76 & 0.0 & 16.0 $\pm$ 5.0 & BC+NC & -- & -1.67 & -0.87 & 66. $\pm$ 12. & \textit{(i2)} \\[\tblspc]
42 & LkCa15 & HI & K5 & K7 & 0.87 & 0.2 & 18.5 $\pm$ 0.5 & SC & -- & -1.70 & 0.53 & 51. $\pm$ 1. & \textit{(i1)} \\[\tblspc]
43 & LkHa330 & HI & F7 & G0 & 2.20 & 0.5 & 19.5 $\pm$ 0.5 & SC & -- & -0.46 & 1.88 & 12. $\pm$ 2. & \textit{(i3)} \\[\tblspc]
44 & RNO90 & MI & G8 & K2 & 1.50 & 0.8 & -10.1 $\pm$ 0.6 & BC+NC & -- & 0.06 & -0.47 & 37. $\pm$ 4. & \textit{(i3)} \\[\tblspc]
45 & RULup & HI & K7 & K8 & 0.70 & 9.1 & 0.0 $\pm$ 5.0 & BC+NC & yes & -0.01 & -0.53 & 35. $\pm$ 5. & \textit{(i3)} \\[\tblspc]
46 & RXJ1615 & MI & K5 & K7 & 1.10 & 0.1 & -2.9 $\pm$ 0.5 & SC & -- & -- & 1.17 & 44. $\pm$ 2. & \textit{(i1)} \\[\tblspc]
47 & RXJ1842 & HI & K3 & K5 & 0.99 & 0.1 & -0.9 $\pm$ 0.5 & BC+NC & yes & -1.18 & 0.72 & 54. $\pm$ 20. & \textit{(i9)} \\[\tblspc]
48 & RXJ1852 & MI & K4 & K3 & 0.93 & 0.0 & -1.1 $\pm$ 0.5 & SC & -- & -1.69 & 2.63 & 16. $\pm$ 20. & \textit{(i9)} \\[\tblspc]
49 & RYLup & HI & K2 & K2 & 1.40 & 0.5 & 0.8 $\pm$ 2.3 & SC & -- & -1.40 & 0.87 & 68. $\pm$ 7. & \textit{(i7)} \\[\tblspc]
50 & SCrAN & MI & K3 & K8 & 1.50 & 2.5 & 2.5 $\pm$ 1.0 & SCJ & yes & -0.66 & -0.09 & 10. $\pm$ 5. & \textit{(i3)} \\[\tblspc]
51 & Sz73 & MI & M0 & M0 & 0.76 & 0.3 & -12.2 $\pm$ 5.0 & SCJ & yes & -1.22 & -0.06 & 48. $\pm$ 3. & \textit{(i4)} \\[\tblspc]
52 & Sz76 & HI & M3.2 & M3 & 0.32 & 0.3 & -1.6 $\pm$ 1.1 & SC & -- & -2.23 & 0.00 & -- & \textit{--} \\[\tblspc]
53 & Sz98 & HI & M0.4 & M1 & 0.58 & 0.1 & -0.3 $\pm$ 0.7 & SCJ & yes & -1.53 & -0.56 & 47. $\pm$ 1. & \textit{(i14)} \\[\tblspc]
54 & Sz102 & HI & K2 & K2 & -- & 9.0 & 12.0 $\pm$ 2.0 & -- & yes & -1.10 & 0.64 & 73. $\pm$ 9. & \textit{(i18)} \\[\tblspc]
55 & Sz111 & HI & M1.2 & M3 & 0.51 & 0.0 & -0.3 $\pm$ 0.5 & SC & -- & -1.74 & -- & 53. $\pm$ 5. & \textit{(i7)} \\[\tblspc]
56 & TWA3A & HI & M4 & M3 & 0.21 & 0.7 & 12.3 $\pm$ 0.6 & SC & -- & -3.48 & -- & -- & \textit{--} \\[\tblspc]
57 & TWHya & HI & M0.5 & K8 & 0.69 & 0.5 & 10.1 $\pm$ 0.5 & SC & -- & -1.10 & 0.96 & 7. $\pm$ 1. & \textit{(i11)} \\[\tblspc]
58 & UXTauA & HI & K0 & K2 & 1.51 & 0.1 & 18.4 $\pm$ 0.5 & SC & -- & -1.52 & 1.82 & 39. $\pm$ 2. & \textit{(i1)} \\[\tblspc]
59 & V409Tau & HI & M0.6 & M2 & 0.53 & 0.7 & 17.6 $\pm$ 0.5 & -- & yes & -1.64 & -- & -- & \textit{--} \\[\tblspc]
60 & V773Tau & HI & K4 & K3 & 0.98 & 0.0 & 16.4 $\pm$ 5.0 & BC+NC & -- & -1.88 & -0.85 & -- & \textit{--} \\[\tblspc]
61 & V836Tau & HI & M0.8 & K8 & 0.58 & 0.1 & 20.6 $\pm$ 0.6 & SC & -- & -2.51 & -0.07 & 51. $\pm$ 10. & \textit{(i1)} \\[\tblspc]
62 & V853Oph & HI & M2.5 & M3 & 0.33 & 1.6 & -5.8 $\pm$ 1.1 & BC+NC & yes & -2.02 & -0.17 & 54. $\pm$ 5. & \textit{(i1)} \\[\tblspc]
63 & VVCrAS & MI & K1 & K8 & 0.53 & 5.3 & -5.7 $\pm$ 5.0 & BC+NC & yes & 0.21 & 0.04 & 50. $\pm$ 20. & \textit{(i16)} \\[\tblspc]
64 & VYTau & HI & M1.5 & M2 & 0.47 & 2.2 & 13.3 $\pm$ 5.0 & SC & -- & -2.93 & -0.13 & -- & \textit{--} \\[\tblspc]
65 & WaOph6 & MI & K6 & K7 & 0.90 & 0.1 & -7.6 $\pm$ 10.0 & SCJ & yes & -1.26 & -0.40 & 41. $\pm$ 3. & \textit{(i1)} \\
\enddata

\tablecomments{
Columns in the table --
1st: identifier used in figures in this paper; 3rd: instrument used (HI = Keck-HIRES, MI = Magellan-Mike); 4th: stellar spectral type from the literature; 5th: photospheric template spectral type (this work); 6th: stellar mass from the literature; 7th: veiling measured at 6300 \AA\  (this work); 8th: stellar radial velocity \citep[heliocentric, this work or][]{fang18}; 9th: empirical class of LVC component (this work); 10th: presence of HVC component (this work); 11th: accretion luminosity from the literature; 12th: $n_{13-31}$ infrared index (this work); 13th and14th: disk inclination from the literature. References for literature values are listed in Appendix \ref{App: sample_refs}.
}
\end{deluxetable*}

\section{OBSERVATIONS} \label{sec: data}
\subsection{Sample}
The sample analyzed in this work, presented in Table \ref{tab: sample}, is comprised of previously published Keck-HIRES spectra \citep{pasc15,simon16}, archival spectra from Keck-HIRES \citep{fang18}, and a new observational campaign from Magellan-MIKE, which is published here for the first time. All spectra have a similar spectral resolution of $\sim 7$\,km/s near the [OI] emission. 
The sample includes 65 spectra for 64 TTauri stars; one object, EXLup, has been observed twice, once during a strong accretion outburst in 2008 and once in 2012. Half of the sample is from the Taurus star-forming region, the other half from Lupus, Ophiucus, Corona Australis, and TWHya. 
[OI] emission is detected at 6300 \AA\ in 61 objects, and at 5577 \AA\ in 38 objects. 
The sample does not aim to be statistically complete, but instead is one representation of the range of emission properties from young stars of masses mostly within 0.3--0.8 M$_{\odot}$ and spectral types of late-K and early-M. We also include in Table \ref{tab: sample} three parameters that we will use in the analysis of [OI] kinematic components: the accretion luminosity, the disk inclination, and the mid infrared spectral index $n_{13-31}$, as described in Section \ref{sec: OI_params}.

\subsection{Keck-HIRES spectra}
We have included in this work 31 spectra from \citet{simon16}\footnote{From their original sample of 33, we have excluded V710Tau, where [OI] is not detected, and DKTau, where a better spectrum is available from \citet{fang18}.}, plus 21 spectra from \citet{fang18}. The spectra of AS353A and LkHa330 were added and reduced as explained in \citet{simon16}. The spectra were taken with the Keck-HIRES spectrograph \citep{hires} using slits of $0.9'' \times 7''$ or $1.1'' \times 7''$. The data reduction is described in \citet{fang18} and \citet{pasc15}, which also determined the spectral resolution to be $\sim6.6$\,km/s in this dataset. These works also analyzed and discussed the achieved precision in radial velocity for these datasets. We have re-corrected all these spectra for telluric and photospheric absorption, using the procedure described in Section \ref{sec: OI_correction}. 

\subsection{Magellan-MIKE spectra}
The 11 MIKE spectra in this sample were taken over three nights in July 2012 with the MIKE echelle spectrograph \citep{MIKE} mounted on the Magellan II 6.5m telescope at Las Campanas Observatory in Chile. These spectra cover the region at 4800--9200 \AA\ with 34 echelle orders, but in this work we focus on the [OI] emission lines at 5577 and 6300 \AA. The slit was $0.7'' \times 5''$, and we determine that the spectral resolution was $7.4 \pm 0.2$\,km/s near the 6300 \AA\ line, consistent with the nominal instrumental resolution of $\sim 7.1$\,km/s \citep{MIKE}, by measuring the width of nearby telluric lines in the spectrum of the telluric standard star observed at the lowest airmass and highest S/N. The spectra reduction has been done using the CarnegiePython MIKE pipeline\footnote{http://code.obs.carnegiescience.edu/mike}, which applies flat fields, removes scattered light, and subtracts sky background and the telluric emission lines. Order by order the pipeline extracts the stellar spectrum and applies a wavelength calibration based on Th-Ar lamp exposures taken before and after each science spectrum.

\begin{deluxetable}{l l l}
\tabletypesize{\small}
\tablewidth{0pt}
\tablecaption{\label{tab: phot_stds} Photospheric standards.}
\tablehead{\colhead{Star} & SpT & Type}
\tablecolumns{3}
\startdata
J155812-232836 & G2  & WTTS \\ 
J155548-251224 & G3  & WTTS \\ 
LAH597 & G9  & WTTS \\ 
J162307-230059 & K2  & WTTS \\ 
J155847-175759 & K3  & WTTS \\ 
1SWASPJ1407 & K5  & WTTS \\ 
HBC427 & K6  & WTTS \\ 
TWA9A & K7  & WTTS \\ 
V819Tau & K8  & WTTS \\ 
2M121530-394842 & K9  & WTTS \\ 
J160013-241810 & M0  & WTTS \\ 
J161018-250232 & M1  & WTTS \\ 
v1321Tau & M2  & WTTS \\ 
TWA8A & M3  & WTTS \\ 
\enddata
\end{deluxetable}

\begin{figure*}[!ht]
\includegraphics[width=1\textwidth]{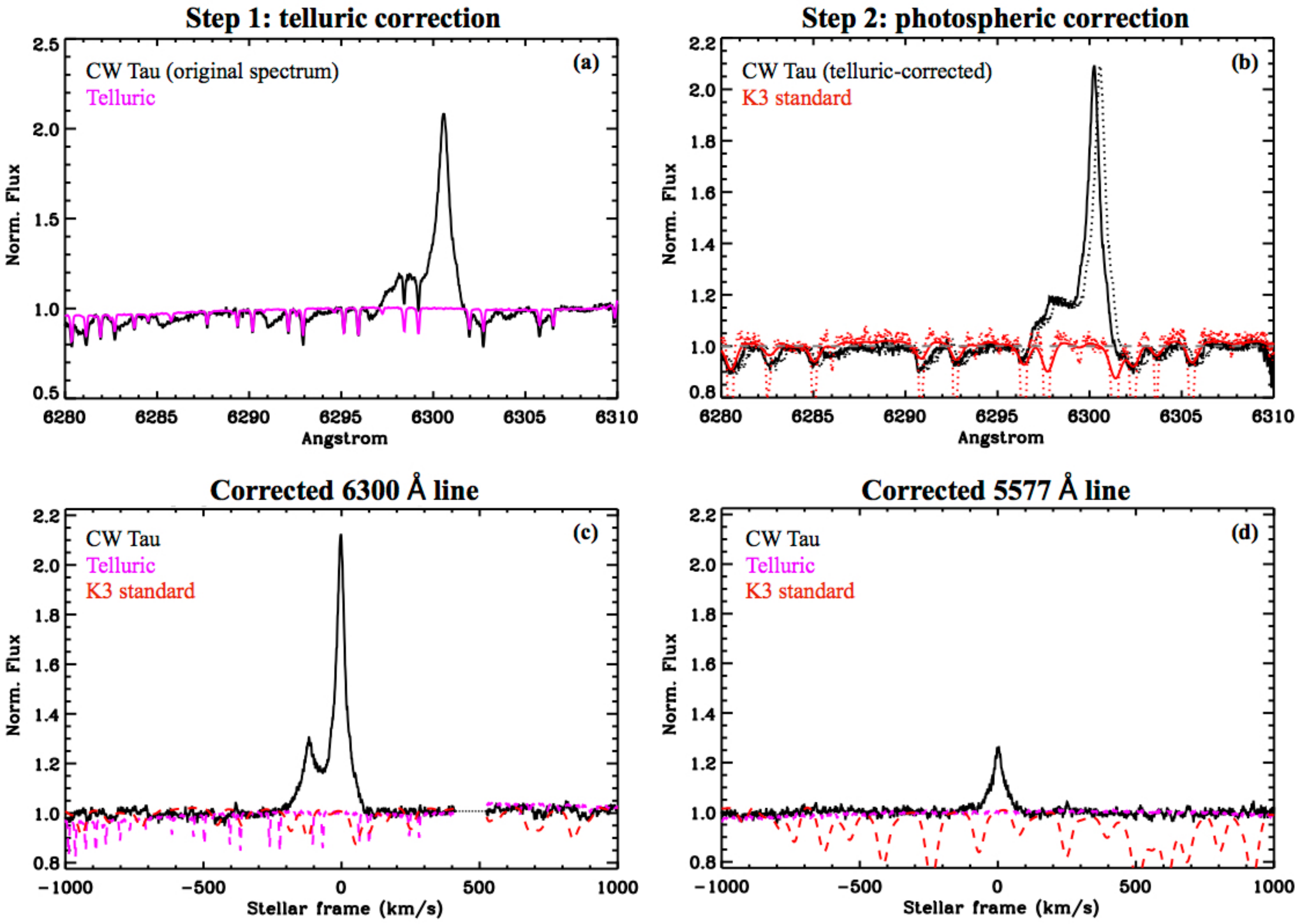} 
\caption{Correction of [OI] spectra, showing CW Tau as an example of the procedure described in the text (Section \ref{sec: OI_correction}). \textit{Top left}: The original science spectrum is shown in black, the telluric spectrum in magenta. \textit{Top right}: The telluric-corrected science spectrum (dotted black line) is shifted to the stellar frame (solid black line); the telluric-corrected photospheric template (dotted red line) is shifted, broadened, and veiled to match the science spectrum (solid red line). \textit{Bottom left}: The telluric- and photospheric-corrected spectrum of the 6300 \AA\ line is shown in solid black line; the telluric and photospheric spectra are plotted with dashed lines for reference. The break at $\sim$400--500 km/s is due to the echelle orders on the detector. \textit{Bottom right}: The same procedure is applied to the 5577 \AA\ line.}
\label{fig: spectra_corr}
\end{figure*}

\section{DATA ANALYSIS} \label{sec: analysis}
\subsection{Correction of [OI] spectra} \label{sec: OI_correction}
The emission lines analyzed in this work are affected by absorption features produced in stellar atmospheres and by emission and absorption in the Earth’s atmosphere, also called telluric contamination. Telluric emission is removed during the spectral extraction process while telluric absorption, only present in the vicinity of the [OI] 6300 line, is removed by using spectra of early type standard stars with no stellar absorption at these wavelengths. Correction of the line profiles for stellar photospheric features, required for both [OI] 6300 and [OI] 5577, follows a long established procedure \citep{hartig89,hartig91,basri90,hartig95}, which relies on spectral templates matched in temperature and surface gravity to the target star coupled with an estimate of the continuum excess emission, known as veiling, that dilutes photospheric features. The veiling is defined as the ratio of the excess to the photospheric flux as: $r_{\lambda} = F_{exc}/F_{phot}$.
Figure \ref{fig: spectra_corr} shows one example to illustrate the main steps of this procedure, which has been performed with custom IDL procedures developed in previous work for correction of infrared spectra \citep{banz14,banz17}, and here adapted to this wavelength range. 

First, the original science spectrum is corrected for telluric absorption, by division to the telluric standard spectrum after correction for small differences in airmass (Fig.\ref{fig: spectra_corr}a). Second, a photospheric template is shifted by cross-correlation with the science spectrum, and broadened and veiled to match the stellar photospheric lines observed in the science spectrum (Fig.\ref{fig: spectra_corr}b). This is obtained by minimizing the residuals after subtraction of the telluric-corrected science and photospheric spectra. The residuals are measured over two spectral ranges at each side of the [OI] lines, and a grid of broadening and veiling values are explored to minimize the residuals. We also explore a range in spectral types, around the literature spectral type of each science target, to determine which one best represents the photospheric spectrum near the [OI] lines.

The stellocentric frame for the 6300 \AA\ [OI] line is determined by cross-correlation of the photospheric Calcium line at 6439.07 \AA\ with a Phoenix stellar model \citep{phoenix}. In some Keck-HIRES spectra where this line was not covered, we adopt the radial velocity estimate from \citet{fang18}. For the 5577 \AA\ [OI] line we instead use five strong photospheric lines between 5565 and 5590 \AA\ (Fig.\ref{fig: spectra_corr}d). By measuring the velocity shift independently for these two spectral ranges, which fall on different echelle orders, we estimated an uncertainty in wavelength calibration for this dataset and therefore in our estimates of stellar radial velocities and [OI] line centroids (the uncertainties given in Table \ref{tab: sample}). The distribution of the radial velocity differences between the two estimates is peaked on zero and has a standard deviation of $\sim1$\,km/s, consistent with the estimate by \citet{pasc15} for Keck-HIRES spectra. Further details on the radial velocity uncertainties in the HIRES spectra are included in \citet{pasc15} and \citet{fang18}. We do not find any significant difference between the wavelength precision as estimated from the different orders in spectra taken with MIKE or HIRES in this sample. 

The grid of photospheric standards used in this work is included in Table \ref{tab: phot_stds}; these are all weak-lined TTauri stars (WTTS) observed with HIRES. This extended grid of standards enabled us to improve the photospheric correction in roughly one third (10/30) of the spectra published in \citet{simon16}. We remark that a finely-sampled grid of spectral type templates is important to properly retrieve [OI] emission, and therefore for a reliable analysis of the emission properties.
Photospheric residuals at each side of the [OI] lines can affect the spectral shape (and therefore the centroid and width) of any [OI] emission at velocities approximately between -200 and -80 km/s (blue side of the 6300 \AA\ [OI] line) and between +30 and +200 km/s (red side of the line), depending of the width of photospheric lines and on the strength of [OI] emission (e.g. Figure \ref{fig: spectra_corr}). The impact of photospheric correction is therefore higher in the [OI] emission components that are broader than $\gtrsim $\,50--60\,km/s in FWHM, while narrower emission components centered or close to zero velocity are less affected (see more details in Appendix \ref{App: phot_corr}).

We also conducted a thorough comparison of the effects of using WTTS versus Main Sequence (MS) stars as photospheric templates, since bright MS standards can be acquired with shorter exposure times and have higher signal to noise, and a grid of MS standards was acquired as part of the Magellan-MIKE part of this sample (Section \ref{sec: data}). However, we found that, in pre-MS stars of K and M type, an artificial emission feature on the red side of the 6300 \AA\ [OI] line is produced at the 5-10\% level over the continuum, probably due to the sensitivity of these photospheric lines to stellar gravity. We therefore stress that WTTS should be used to properly retrieve the [OI] 6300 \AA\ emission from pre-MS stars, especially for K and M types.

\subsection{Composite Gaussian fits to [OI] lines} \label{sec: OI_fits}
\begin{figure*}%[ht]
\centering
\includegraphics[width=1\textwidth]{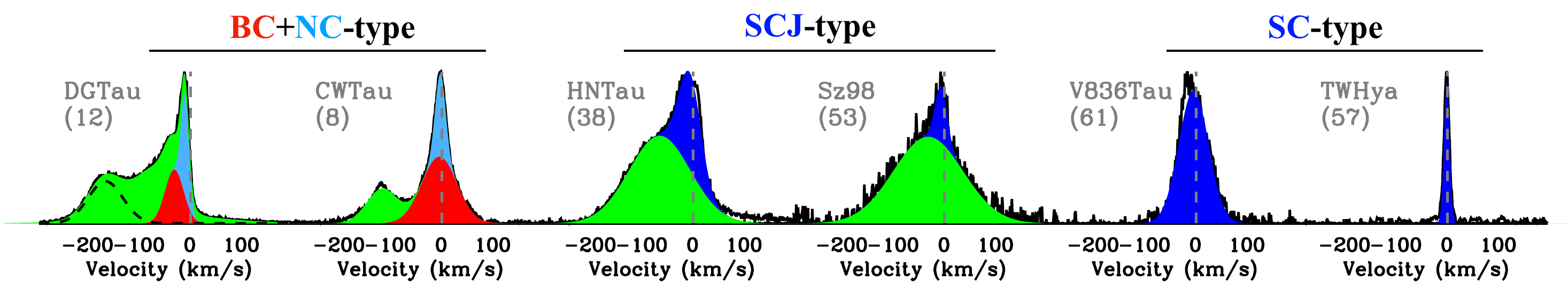} 
\caption{Representative examples of [OI] line profiles, showing ``BC+NC"-type LVC (DGTau and CWTau), ``SCJ"-type LVC (HNTau and Sz98), and ``SC"-type LVC (V836Tau and TWHya) by color-coding their HVC and LVC components as described in Section \ref{sec: OI_class}: HVCs are in green, LVC BC are in red, LVC NC in light blue, and LVC SC and SCJ in dark blue. Line profiles for the entire sample are shown in Appendix \ref{App: linefits_figs}. Where multiple are present, we mark with a dashed black line the most blueshifted HVC component used in the analysis.}
\label{fig: representative_lines}
\end{figure*}

Once telluric- and photospheric-corrected, [OI] emission at 6300 \AA\ often shows structured line profiles with multiple distinct peaks. We follow previous work and model each [OI] spectral profile as the composite of Gaussian functions \citep{natta14,rigl13,simon16}. In this work we adopt a line fitting IDL procedure developed in previous work to fit multi-component infrared CO spectra \citep{banz15}, which embeds the {\tt MPFIT} fitting code written by \citet{mpfit}. This procedure allows for up to six Gaussians with independent centroids, widths, and peak values to reproduce the observed line profile. The best fit composite model for [OI] emission and the number of necessary Gaussian components are determined by $\chi ^2$ minimization, by adding an additional Gaussian in a given line fit only if it improves the reduced $\chi ^2$ by more than 20\%. A composite of Gaussian functions provides good fits (reduced $\chi ^2 < 1.5$) in most spectra in this sample. The majority of spectra ($\sim 80$\%) requires a total of $\leq 3$ Gaussian components to be fitted, the rest requires more. LVC components require only 1 or 2 Gaussians;  HVC components require up to 3 components in 2/32 objects. The fitting procedure provides measurements and uncertainties of [OI] line widths, centroids, 5577/6300 ratios, and EWs for each velocity component. Figure \ref{fig: representative_lines} shows examples of line fits, and all [OI] lines and Gaussian fits are included in Appendix \ref{App: linefits_figs}, separated into classes as explained in Section \ref{sec: OI_class}.

\begin{figure*}%[ht]
\includegraphics[width=1\textwidth]{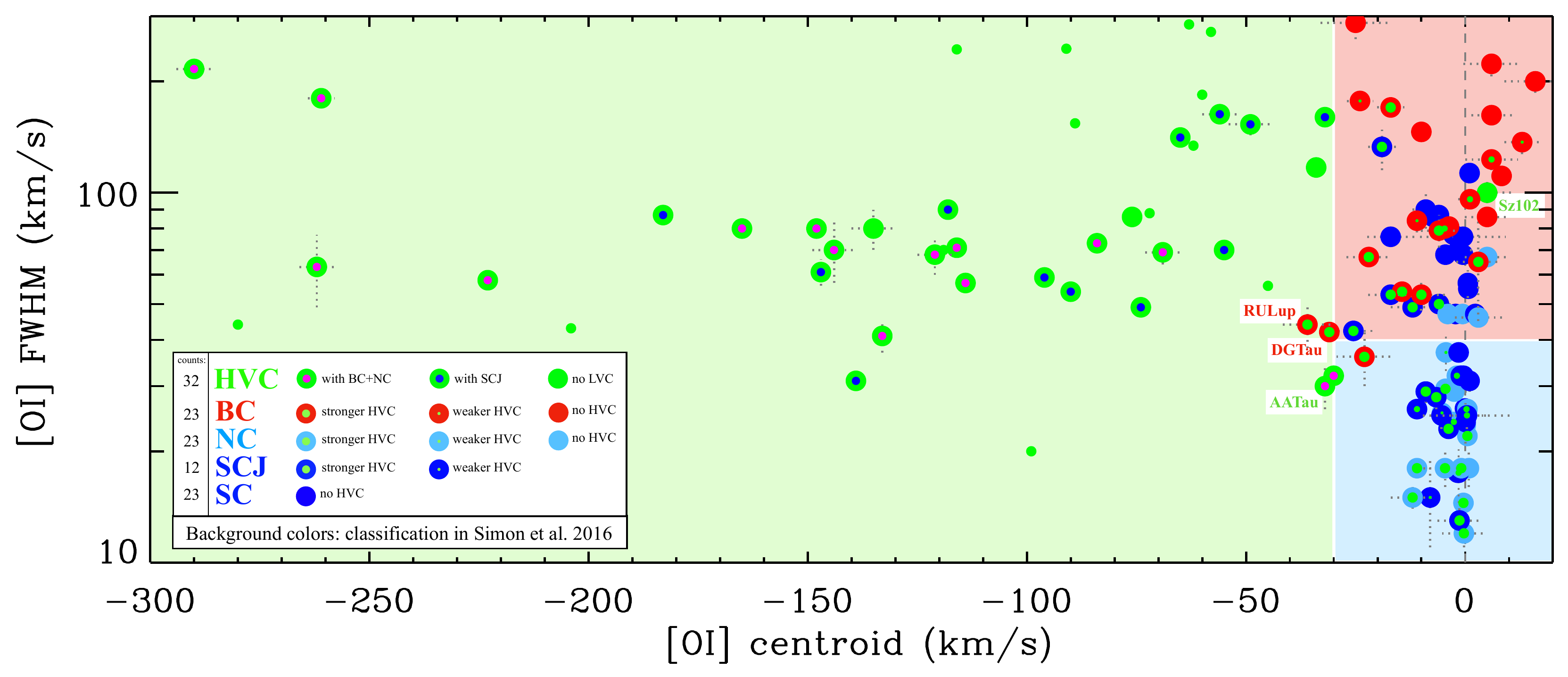} 
\caption{Parameter space of measured centroids and FWHM of [OI] Gaussian components, using color coding as explained in the text and displayed in the inset in this figure. A green interior indicates that a given LVC also has an HVC at larger blueshifts, and it is sized proportional to HVC EW$_{\rm tot,corr}$ (this color-coding is applied to all figures in the paper). The most blueshifted HVC Gaussian component in each object is showed as a large green dot, with interior color to distinguish if they have an LVC-SCJ (blue) or LVC-BC+NC (pink). Smaller green dots show HVC components at intermediate velocities detected in some objects and excluded from the analysis (see Section \ref{sec: OI_params}). AATau and Sz102 are mentioned in Section \ref{sec: OI_class} as HVCs projected to low velocities by highly-inclined systems.}
\label{fig: OI_parspace}
\end{figure*}

\subsection{Empirical classification of [OI] components} \label{sec: OI_class}
As summarized in Section \ref{sec: OIhistory}, previous work has identified two main velocity components in the [OI] emission spectra observed from TTauri stars, which have been interpreted as two physically distinct components due to collimated jets (the high-velocity component, HVC) and disk winds (the low-velocity component, LVC). 
Recently, the analysis and interpretation of LVC has been refined using higher resolution spectroscopy and Gaussian fitting of the observed line profiles. \citet{simon16}, with part of the dataset included in this work, set the separation between HVC and LVC at observed line centroid velocities $|v_c| = 30$\,km/s, based on the lowest-velocity Gaussian component attributed to a jet within the sample (in AA Tau). Within the LVC, they also identified two distinct Gaussian components in many spectra \citep[as initially found in two objects by][]{rigl13}, a broader component dominating the line wings and a narrower component dominating the central peak of the emission line (color-coded respectively in red and light blue in Figure \ref{fig: representative_lines}). The presence of these two distinct Gaussian components has been supported by measuring higher 5577/6300 line ratios in BC as compared to NC, consistent with emission from disk regions that have different densities/temperatures (\citealt{simon16}, and more recently \citealt{fang18}). 

Spectral decomposition into Gaussian components is a convenient and simple procedure to interpret spatially-unresolved [OI] spectra that include emission from different regions/phenomena (see more discussion in Appendix \ref{App: OI_caution}). In this work, we perform a comparative analysis of the properties of these empirically-defined Gaussian components, to investigate their origin and links.
We adopt the following empirical classification scheme based on, as main criteria, i) the observed velocity centroids, which we interpret as the intrinsic velocity of a given component projected along the line of sight, and ii) the number of Gaussian components found by the fitting procedure. This classification is also illustrated in Figures \ref{fig: representative_lines} and \ref{fig: OI_parspace}.

\begin{description}
\item[High-velocity component (HVC)] These components are defined by line centroids more blueshifted than $\lesssim -30$\,km/s; HVCs are detected in 32/61 spectra, and they also typically have a LVC\footnote{There are three exceptions: V409Tau and IPTau do not show any Gaussian centroid less blueshifted than $\gtrsim -30$\,km/s, and Sz102 is dominated by a jet in the plane of the sky.}. HVCs are color-coded in green in all figures, with a blue interior if they have a SCJ-type LVC and pink interior if they have a BC+NC-type LVC.

\item[Low-velocity component (LVC)] These components are defined by line centroids less blueshifted than $\gtrsim -30$\,km/s, they are found in 58/61 spectra in this sample, and depending on the number of Gaussian components that fall in this velocity range they are further divided into:

\item[ - Double LVC component (BC+NC)] These LVC require a combination of a broad (\textbf{BC}, color coded in red) and a narrow (\textbf{NC}, color coded in light blue) gaussian for a good fit; they are found in 23 objects and 80\% of them have an HVC.

\item[ - Single LVC component (SC and SCJ)] These LVCs are well fit by a single Gaussian, and we color code them in dark blue. Most of them ($\sim70$\%) do not have an HVC and we name them \textbf{SC}; these are detected in 23 objects. A smaller fraction ($\sim30$\%) do have an HVC, and we name them \textbf{SCJ}; these are detected in 12 objects.
\end{description}

This classification scheme is based on the same Gaussian-fit criteria adopted in \citet{simon16}, in terms of separation between HVC and LVC (at a centroid velocity of -30 km/s) and of number of Gaussian components within the LVC (a maximum of 2). The difference in this analysis is that, rather than assigning each single-Gaussian LVC to BC or NC depending on the FWHM, as done in \citealt{simon16} \citep[and][]{mcginnis18,fang18}, we keep track of the single-component LVC profiles as a distinct category, and we use the notation SC or SCJ depending on whether they are accompanied by an HVC. One reason for this approach is that this paper is exploring kinematic properties of [OI] components and their dependence on disk and accretion properties, and as will be shown below, the SC and SCJ show some differences from the BC/NC components. Another reason is that we now have a large enough sample of LVC requiring two gaussians that we find overlap in the narrowest BC and broadest NC, precluding a clean separation based only on the FWHM. We do not separately track the BC+NC with or without HVC because the sample size is too small, with only 6/23 BC+NC profiles where HVC is not detected\footnote{These can be identified visually in all figures, by not having a green dot at their center. For up to 4/6 of these (DHTau, FMTau, GHTau, and RNO90), it cannot be excluded that a weak HVC might be present (see Appendix \ref{App: linefits_figs}).}.
In the parallel work by \citet{fang18}, single-component LVCs are instead separated into BC or NC depending on their FWHM as in \citet{simon16}, because it is the disk emitting region that is used to determine the mass outflow rates, regardless of whether 1-2 Gaussian components are found in the LVC and on whether HVC is present or not.

\begin{figure*}%[ht]
\centering
\includegraphics[width=1\textwidth]{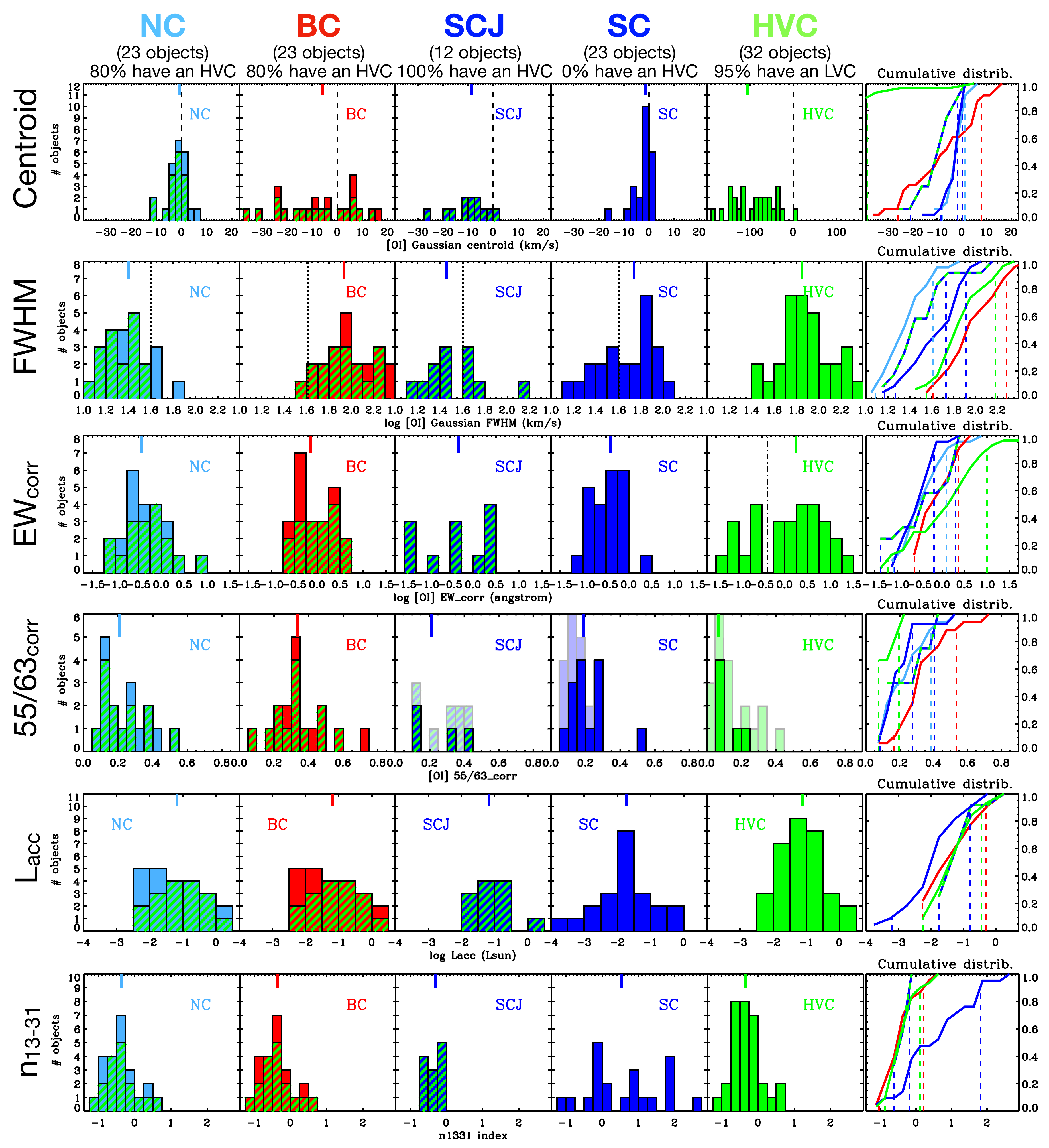} 
\caption{Histograms of the distributions of measured properties of [OI] components. Median values are marked by vertical solid bars at the top of each plot. The fraction of LVC components that have HVC are marked in dashed green in each distribution. The dashed lines in the cumulative distributions show the intervals that include 80\% of the total sample of each component (excluding 10\% at both ends of the distributions). Dotted lines in the FWHM plots indicate the separation between BC and NC as adopted in \citet{simon16}. A dashed-dotted line in the EW distribution marks the separation between ``strong" and ``weak" HVCs as adopted in Figure \ref{fig: LVC_distributions}. Upper limits in 55/63 are shown in lighter colors.}
\label{fig: OI_histos}
\end{figure*}

Figure \ref{fig: OI_parspace} shows the kinematic parameter space covered by the [OI] 6300 \AA\ components for the 61 spectra with [OI] detections, tracking centroids and FWHM for a total of 90 individual components (32 HVC, 23 BC+NC, 12 SCJ, 23 SC). The line profiles and fit parameters for the entire sample are shown in Figures \ref{fig: BC+NC_lines} to \ref{fig: SC_lines} and in Tables \ref{tab: OIdouble_param} to \ref{tab: HVC_param} in Appendix \ref{App: linefits_figs}, separated into the empirical LVC classes. Components are identified in each figure throughout this paper by the color coding described above; in addition, LVC with/without HVC are distinguished by the presence/absence of a green interior, whose size is related to the HVC EW$_{tot,corr}$ (Section \ref{sec: OI_params}). For comparison, the lighter-color background in Figure \ref{fig: OI_parspace} shows where the HVC/LVC assignments from \citet{simon16} would fall, with green for HVC, red for BC and light blue for NC (note the location of AA~Tau, which marked the HVC/LVC boundary in that work). According to this previous classification, 9/23 SC and 5/12 SCJ would have been assigned to BC, the rest to NC (see also Figure \ref{fig: OI_histos}).

A few remarks about Figure \ref{fig: OI_parspace}. The figure shows the kinematic parameter space of [OI] components as observed, i.e. as projected along the line of sight with us. Line-of-sight projection effects increase/decrease the true gas kinematic properties depending on the viewing angle (these effects will be tested in Section \ref{sec: results} and discussed in Section \ref{sec: discuss}). Therefore, boundaries between HVC and LVC, and between individual LVC classes, may show some overlap in the kinematic parameter space as observed. The observed overlap between FWHM of BC and NC noted above is one example of that. Another example is the region at coordinates of approximately -30 and 40 km/s, where four LVC and two HVC components are clustered together. The two BC components that have measured centroids at -31 and -36 km/s (DGTau and RULup) are classified as LVC instead of HVC because their centroids are consistent with -30 km/s within the uncertainties, and they have only one other LVC component at lower velocity; also, they have only moderate disk inclinations of 30--35$^{\circ}$. The two HVC components (in AATau and V853Oph) are classified as HVC because they are the third component found by the Gaussian fits, i.e. there are already two LVC components at lower velocity, and they are detected only in the 6300 \AA\ line, consistently with the low 5577/6300 line flux ratios found in HVCs; also, their disks are highly inclined (50--70$^{\circ}$), therefore projecting any HVC to lower observed velocities.
An extreme example of projection effects is provided by the strong jet from the almost edge-on disk around Sz102 \citep[e.g.][]{louvet16}, also marked in Figure \ref{fig: OI_parspace}. \citet{fang18} have recently showed that the [SII] 4068/[OI] 6300 ratio can be used as an additional parameter to separate HVC from LVC and that the [OI] line from Sz102, in spite of centroids close to the stellar velocity, is consistent with shocks, and hence should be classified as HVC.

\subsection{Parameters used in the analysis of [OI] emission} \label{sec: OI_params}
Here, we describe the parameters used in the next section to illustrate the main results from this work. All the [OI] properties used in the analysis refer to what measured in the 6300 \AA\ line, apart from the 5577/6300 line flux ratio that clearly includes both lines.
For each Gaussian component the equivalent width measured against the veiled continuum (EW$_{meas}$) is corrected for the veiling as:

\begin{equation} \label{eqn: EW_corr}
{\rm EW}_{corr} = {\rm EW}_{meas} \times (1+r_{63}) \, .
\end{equation}
where $r_{63}$ is the veiling measured around the 6300 \AA\ line.

The [OI] line flux ratio 5577/6300 (hereafter 55/63$_{meas}$) is measured by fitting the 5577 \AA\ with the 6300 \AA\ line profile, to determine the flux scaling factor between the two lines (we do that for the individual components, when multiple are present). This procedure takes full advantage of the velocity-resolved line profiles, and provides simultaneous verification that their line centroids and widths match, especially when multiple components are present. As in \citet{simon16}, we generally find that the 5577 \AA\ lines, when detected, show only LVC emission. In most cases, the LVC has similar Gaussian fits at 6300 and 5577 \AA\, but in some objects the 6300 line is blueshifted from the 5577 line (IPTau, DOTau, DKTau, CWTau, DFTau, AS205, DGTau); this was previously found in a few objects also by \citet{hartig95} and \citet{simon16}.
Upper limits for 55/63$_{meas}$ are estimated by scaling down the 6300 \AA\ line until it is consistent with the continuum + noise measured at 5577 \AA . In seven objects we cannot determine an upper limit for the 5577/6300 ratio because of i) the presence of strong and broad emission lines next to and blended with the 5577 \AA\ line (in AS353A, EXLup08, RULup, VVCrAS, all having multiple and strong HVC components and high accretion rate), or ii) complex photospheric residuals left in the range of the 5577 \AA\ line (in EXLup, ITTau, V773Tau).

Estimating the flux ratio between the 5577 \AA\ and 6300 \AA\ [OI] lines requires correcting the ratio between the measured equivalent widths for differences in both the veiling and photospheric continuum level at the two wavelengths, as follows:

\begin{equation} \label{eqn: 55/63_corr} 
55/63_{corr} = 55/63_{meas} \times \frac{\rm Cont_{55}}{\rm Cont_{63}} \times \frac{1+r_{55}}{1+r_{63}} \, ,
\end{equation}
where Cont$_{\lambda}$ is the photospheric continuum at 5577 \AA\ and 6300 \AA\ that we measured from stellar spectra from the Pickles Atlas \citep{pickles}. The second and third terms on the right side of the equation are illustrated in Appendix \ref{App: EWcorr_plots}.

In analyzing the HVC we use only the most blueshifted Gaussian component for the kinematics (centroid and FWHM) but the sum of all the blueshifted Gaussian components as a measure of the total strength of the HVC emission toward the observer (EW$_{\rm tot}$; corrected for veiling as discussed earlier in this section). Selecting only the most blueshifted component for the kinematics is a proxy for the highest velocity gas in each outﬂow, an approach similar to the analysis of \citet{appenz13}\footnote{\citet{appenz13} also considered the blue-edge wing of the observed forbidden emission, and report a velocity $\approx 20\%$ higher than that of the line centroid, on average.}. In 20/32 objects a single Gaussian describes the entire blueshifted HVC emission; in 10/32 objects the HVC emission requires an additional one or two Gaussian components at intermediate velocities, sometimes with a broad wing and narrow center (DOTau and AS353A). Red-shifted HVC components are detected in 10/32 objects, and in 2/10 of these the HVC is only red-shifted (BPTau and FZTau); red-shifted HVCs are not included in this analysis. All the blueshifted HVC components are shown in Figure \ref{fig: OI_parspace}, with the most blueshifted components as large green points and the additional components, if present, as smaller green points. The latter are not included in the subsequent analysis of the HVC kinematics. 

The accretion luminosities, $L_{\rm acc}$, are taken from the literature (see Appendix \ref{App: sample_refs}). Other studies have shown good correlations with various forbidden line component luminosities with accretion luminosity, but since our spectra are not flux calibrated, we do no include forbidden line luminosities in our analysis but rely on the line equivalent widths, corrected as shown above, to compare emission strengths among components. 

The infrared $n_{13-31}$ index is used as a proxy for dust in the inner circumstellar disk region, where values larger than $\sim$0 point to dust depletion and inner cavities \citep{furlan09}. We have measured the $n_{13-31}$ index from archival infrared spectra taken with the medium-resolution ($R \sim 700$) provided by the Spitzer IRS spectrograph \citep{IRS}, as reduced in previous work \citep{pontopp10} or available from the online CASSIS database \citep{cassis}. The measured $n_{13-31}$ values\footnote{Note that \citet{furlan09} defined the  $n_{13-31}$ index from low-resolution ($R \sim 100$) IRS spectra using wavelengths between 12.8--14\,$\mu$m and 30.3--32\,$\mu$m. However, these ranges include strong molecular emission easily detected in medium-resolution IRS spectra \citep[e.g.][]{pontopp10}. To avoid the strongest molecular lines we narrow the ranges to 13--13.2\,$\mu$m and 30--30.15\,$\mu$m.} are reported in Table \ref{tab: sample} and are available for 95\% of the sample.

Another property that affects the observed properties of [OI] emission is the viewing angle. As a proxy for this quantity we take the disk inclination with respect to the line of sight. Disk inclinations are available for 80\% of the sample and reported in Table \ref{tab: sample}. 
For each object we adopt values from the most reliable data available from previous work, i.e. we prefer inclinations derived from spatially-resolved disk images. Most of the disk inclinations used here are from spatially-resolved millimeter imagery tracing the outer disk, while in some cases there is availability of inner disk tracers, e.g. near-infrared CO spectro-astrometry (see Appendix \ref{App: sample_refs} for individual references). Evidence for an inner disk misaligned with respect to the outer disk is reported for a few objects in Appendix \ref{App: object_notes}.

\begin{figure*}%[ht]
\centering
\includegraphics[width=0.32\textwidth]{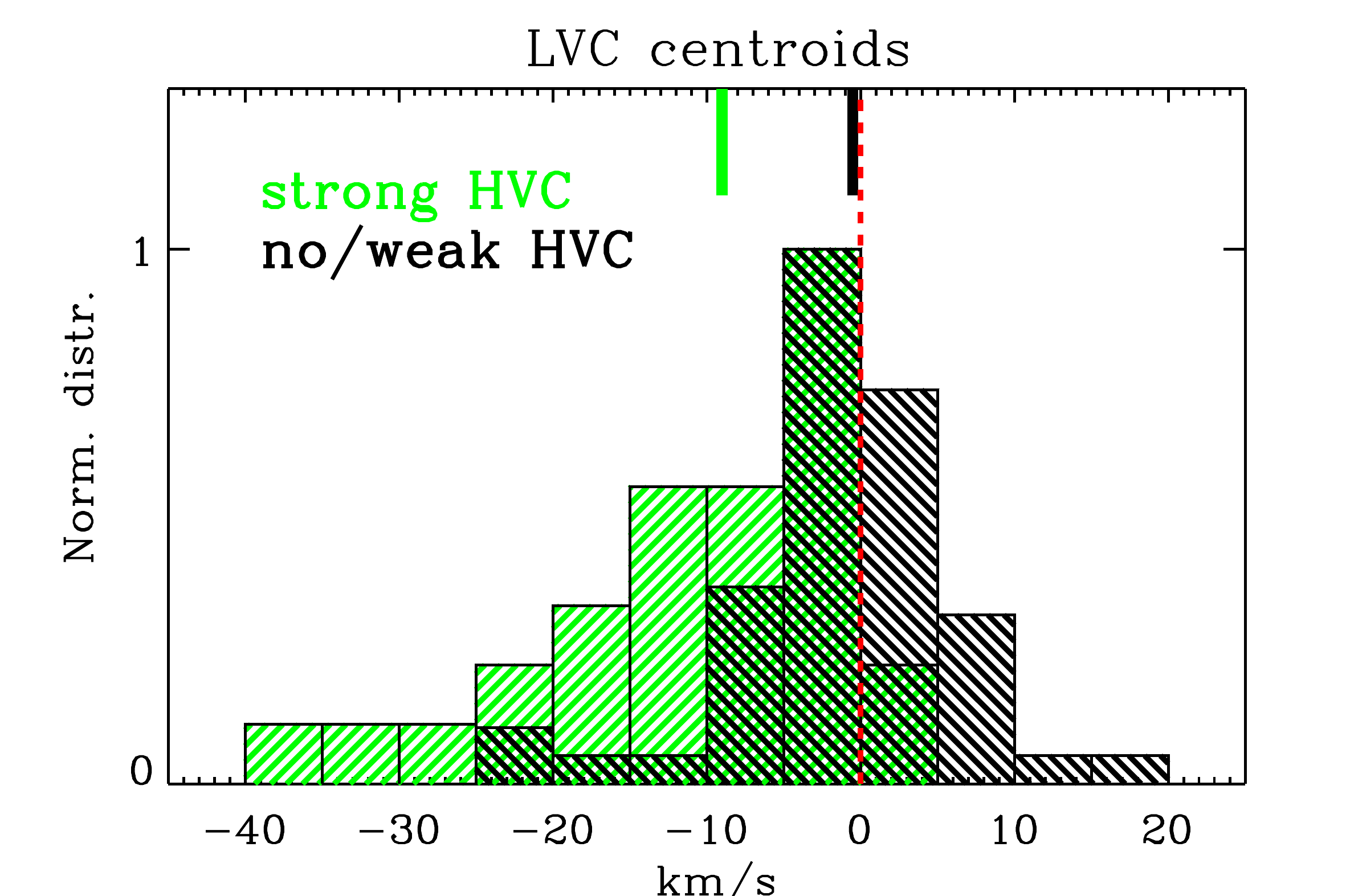} 
\includegraphics[width=0.32\textwidth]{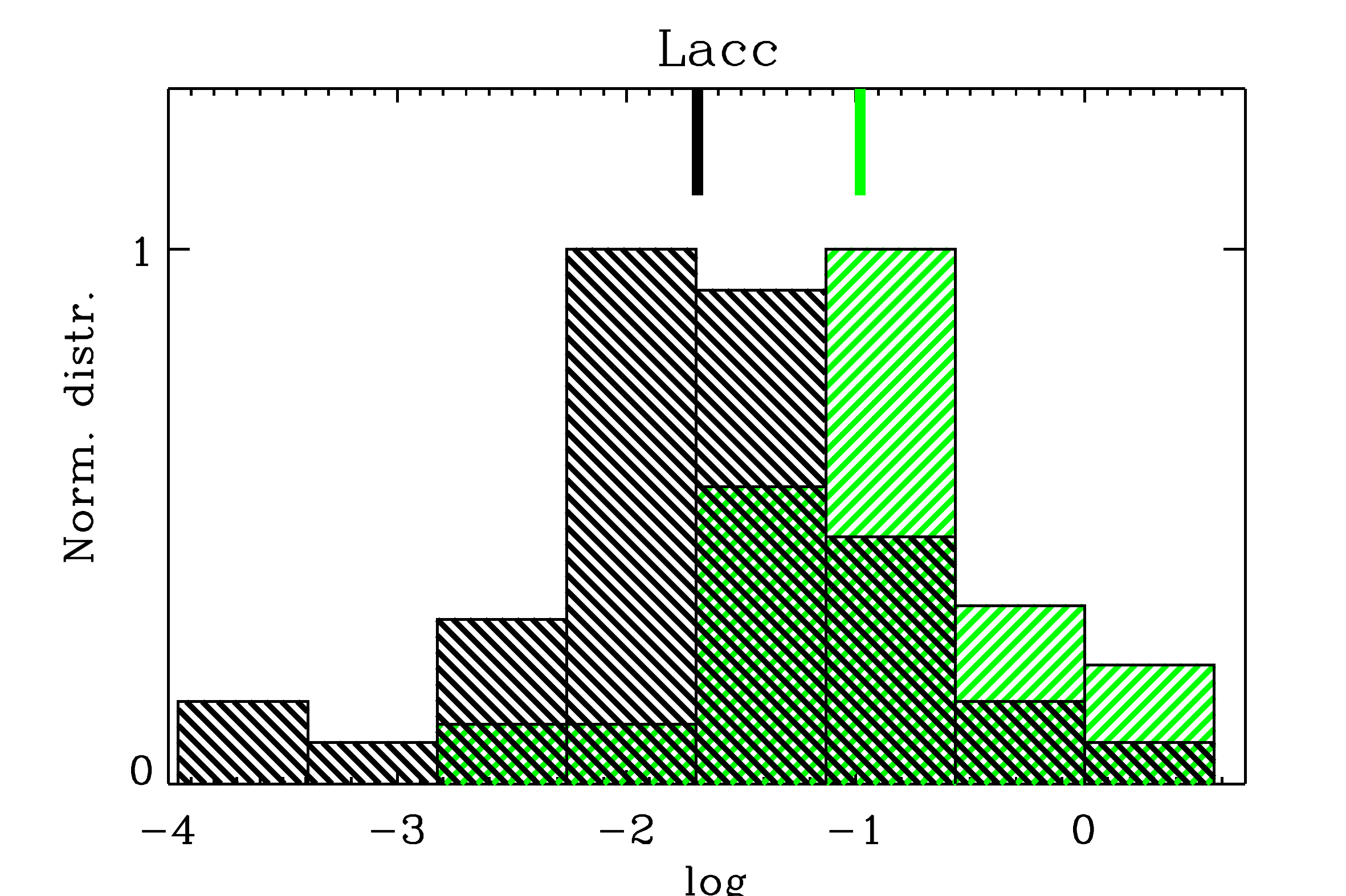} 
\includegraphics[width=0.32\textwidth]{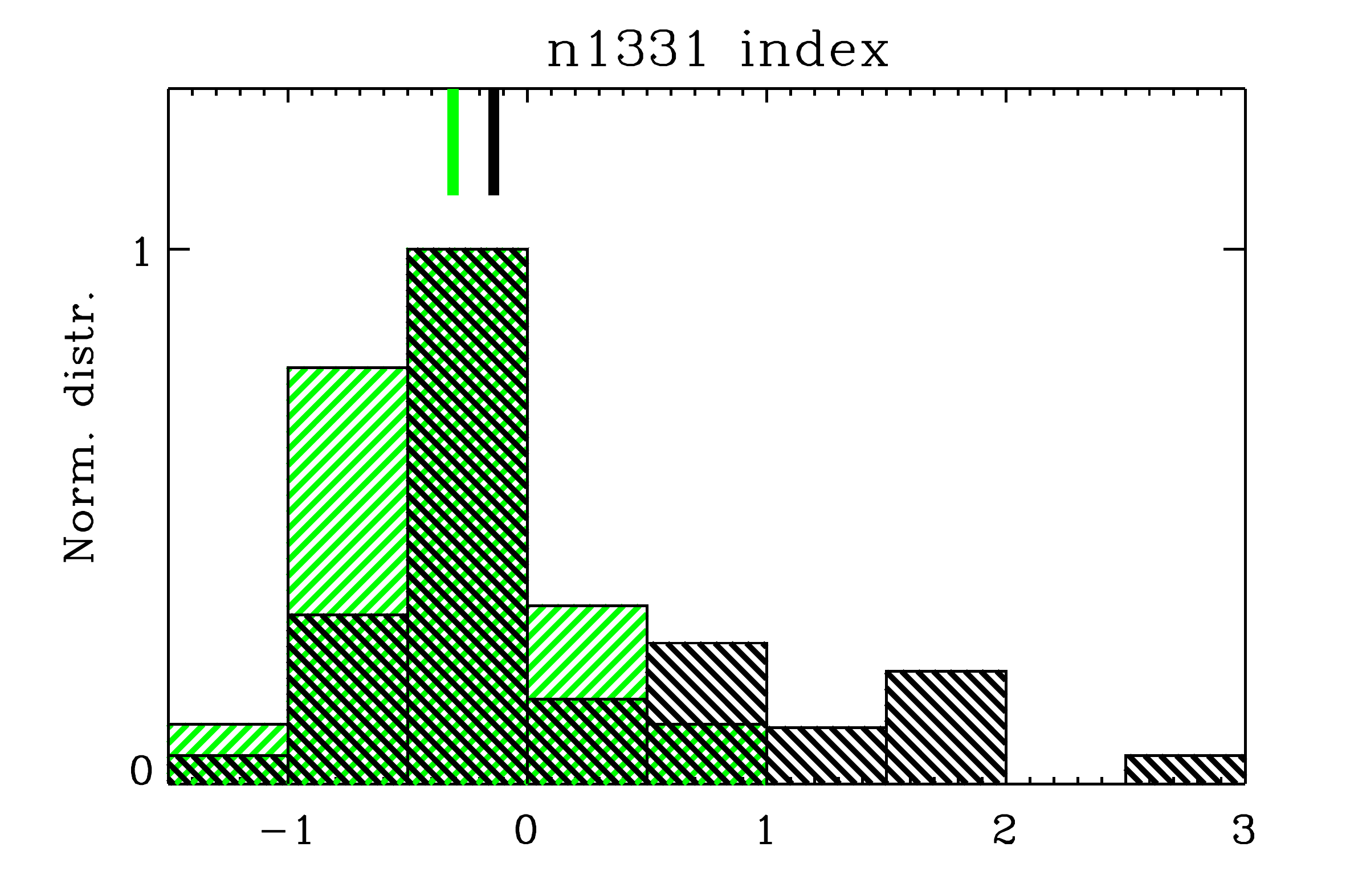} 
\caption{Different behavior of objects that have a strong HVC from those that do not have it or only weak (see text for details). The LVC is typically blueshifted when a HVC is present and strong, while it is rather symmetric around zero in the opposite case. Objects with HVC typically have higher $L_{\rm acc}$ and lower $n_{13-31}$ index than those without HVC. Median values are marked by vertical solid bars at the top of each plot.}
\label{fig: LVC_distributions}
\end{figure*}

\section{RESULTS} \label{sec: results}
\subsection{Overview of [OI] emission properties} \label{sec: OI_prop_histos}
Figure \ref{fig: OI_histos} provides an overview of the properties of all the [OI] components identified in this sample, including their kinematics (centroid and FWHM), EW$_{corr}$ (Eqn. \ref{eqn: EW_corr}), and 55/63$_{corr}$ line ratios (Eqn. \ref{eqn: 55/63_corr}). In addition, Figure \ref{fig: OI_histos} shows how the [OI] components are distributed in terms of disk infrared index $n_{13-31}$ and $L_{\rm acc}$. 

Multiple interesting aspects are revealed by this systematic comparison of [OI] components, and we highlight here some important ones. 
In terms of kinematics, all centroid distributions peak in the blue, with some notable difference between classes. The most blueshifted components after the HVC are the BC and SCJ, with median centroids of -6 and -9\,km/s, respectively. The distributions of NC and SC centroids, while less blueshifted overall, also peak at negative velocities, with median and mean at $\sim$\,-1.3 and -2 km/s for both componts. This suggests an origin in outflowing gas, albeit slower than that traced by the BC and SCJ. Only the BC, among LVC components, includes some red-shifted centroids, up to $\approx +15$\,km/s; in Appendix \ref{App: red_BCs} we explain why these red-shifted centroids are in most cases affected by contamination, and should be taken with extra caution.

In terms of FWHM, the distributions of BC and NC separate rather well, with a small overlap region at $\sim$40--60\,km/s. SCJs and SCs span the full range of FWHMs found in NCs and extend into the BC range. In relation to the 55/63 ratio, low values (and often non detections in the 5577\,\AA\ line) dominate in HVC components, as previously found by \citet{hartig95} and confirmed by \citet{fang18}. The LVC 55/63 ratios are typically higher in the BC than in the NC, in agreeent with \citet{simon16} and \citet{fang18}. SC ratios are mostly similar to NC ratios.

A striking difference is found in the distributions of infrared indices. HVCs, and therefore all LVCs that are found with them, are mostly found in objects with $n_{13-31}<0$ and their median value is -0.35. In contrast, 65\% of SCs have $n_{13-31} > 0$, up to values of $\sim 3$, and their median is +0.5.

In the following sections, we describe the most significant correlations identified within the multi-dimensional parameter space of [OI] emission properties. We use the Pearson coefficient to evaluate if a correlation is statistically significant. If so, we fit the data with an outlier-resistant linear regression routine and report the best-fit parameters in Table \ref{tab: fit_params} in the Appendix. The critical value above which a correlation is significant at the 5\% level depends on the sample size: it is 0.50 for a sample of 12 objects as in the SCJ; 0.35 for a sample of 23 objects as in the BC+NC and SC alone; and 0.30 for a sample of 32 objects as in the HVC. These values are given here for guidance; the individual correlations presented below are considered on the basis of the specific sample size available in each case, which can be different depending on the parameters considered (e.g. disk inclinations and $n_{13-31}$ are not available for 100\% of the sample, see Section \ref{sec: OI_params}).
The interpretation of these correlations is discussed in Section \ref{sec: discuss}, not in this section.

\begin{figure*}%[ht]
\centering
\includegraphics[width=1\textwidth]{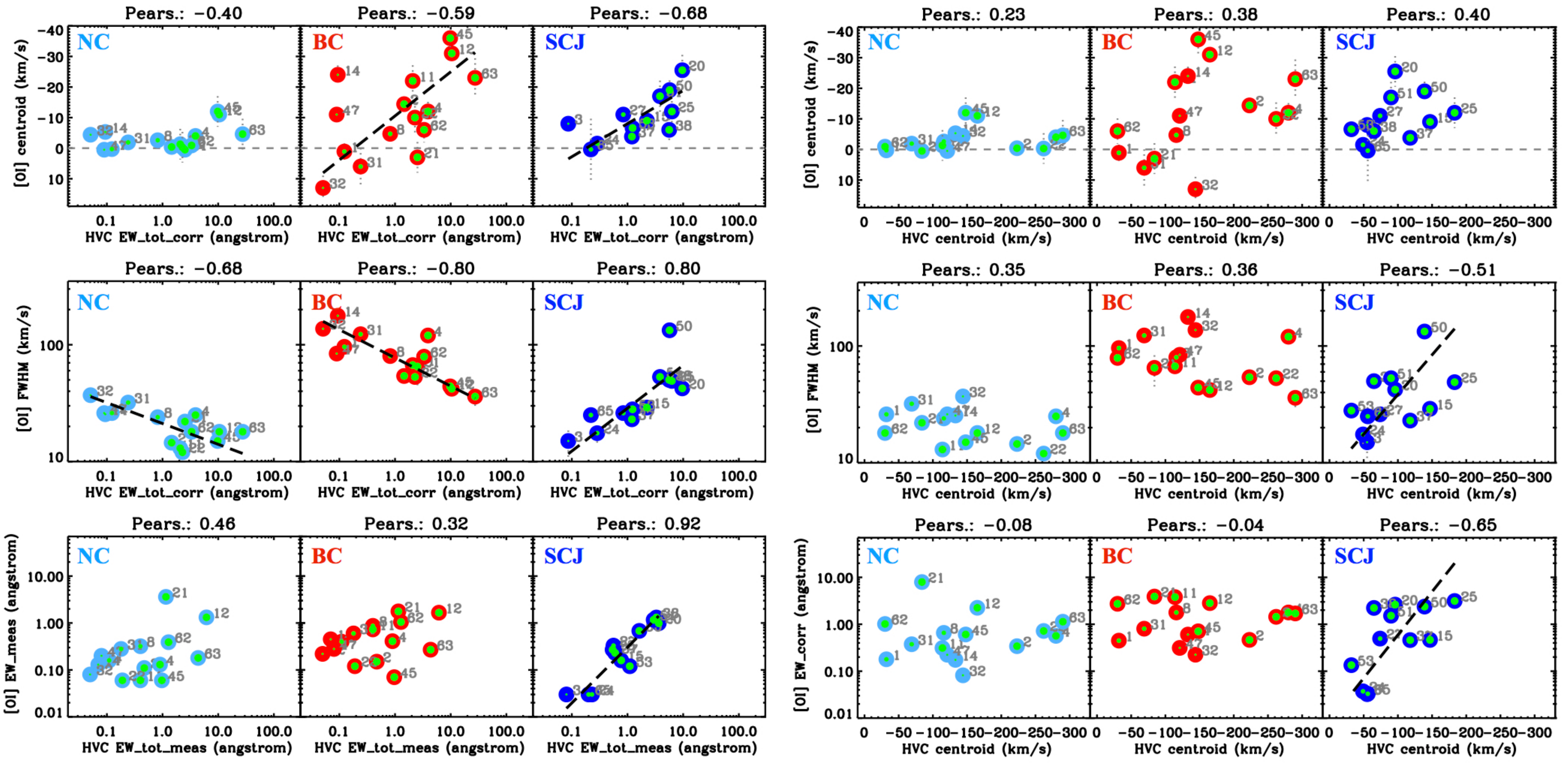} 
\caption{Correlations between LVC and HVC properties. Pearson correlation coefficients are reported at the top of each plot. Linear fits are shown as black dashed lines, their best-fit parameters reported in Table \ref{tab: fit_params}. Color-coding as in Figure \ref{fig: OI_parspace}.}
\label{fig: BCNCSCJ_HVC_corr}
\end{figure*}

\begin{figure*}%[ht]
\centering
\includegraphics[width=1\textwidth]{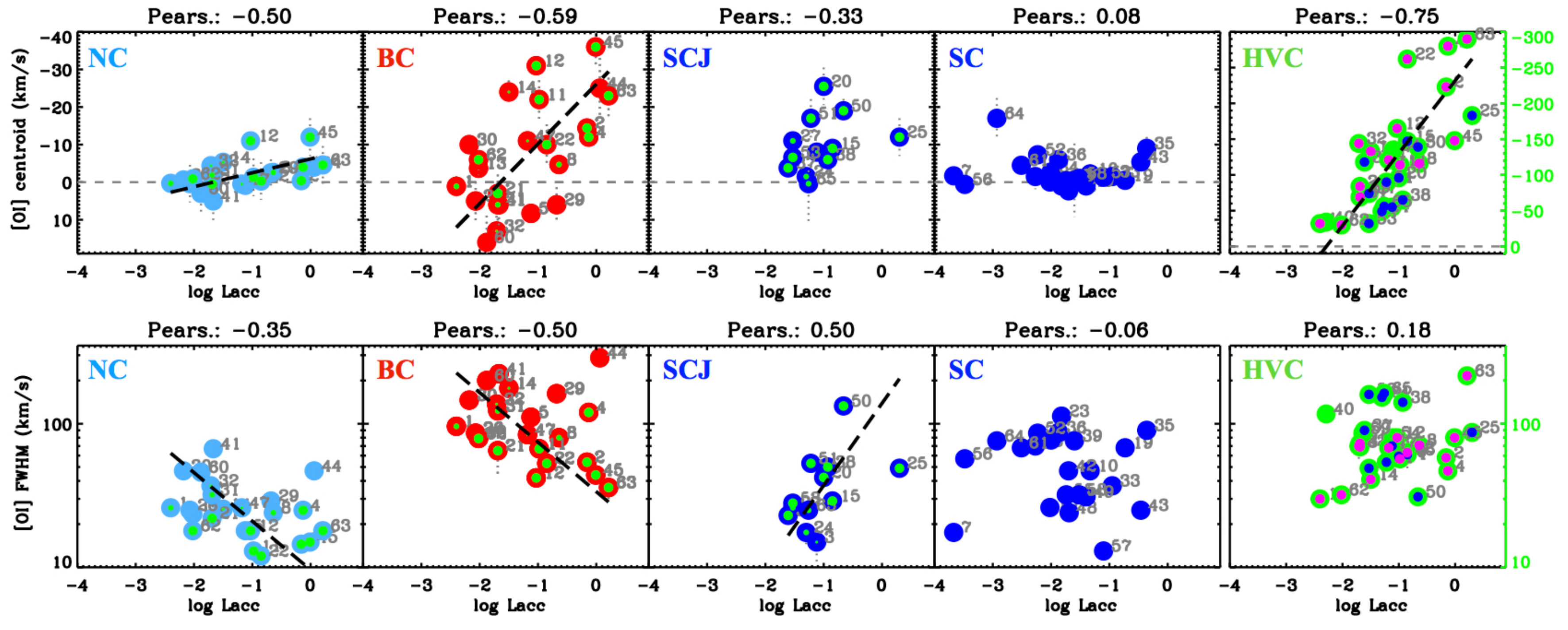} 
\caption{Correlations between [OI] properties and accretion luminosity. A similar figure displaying correlations with the veiling is included in the Appendix (Figure \ref{fig: veiling_correl}).  Linear fits are shown as black dashed lines, their best-fit parameters reported in Table \ref{tab: fit_params}. Color-coding as in Figure \ref{fig: OI_parspace}. Note the different y axis range for the HVC centroids, in the top right plot.}
\label{fig: OI_Lacc_corr}
\end{figure*}

\begin{figure*}%[ht]
\centering
\includegraphics[width=1\textwidth]{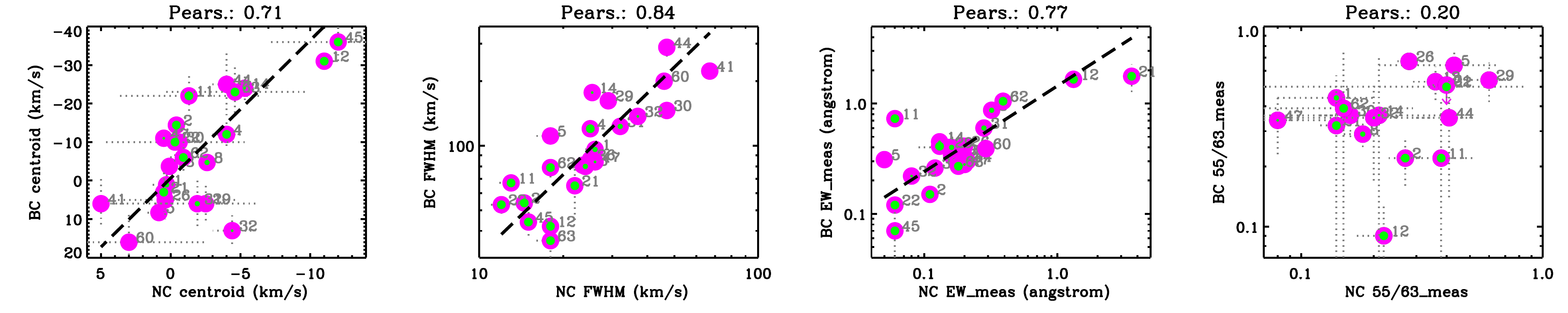} 
\caption{Correlations between BC and NC properties.  Linear fits are shown as black dashed lines, their best-fit parameters reported in Table \ref{tab: fit_params}.}
\label{fig: BCNC_correlations}
\end{figure*}

\subsection{Correlations between LVC and HVC} \label{sec: res_LVC_HVC_corr}
A new contribution of this study is to systematically investigate the dependence between the kinematic properties of LVC components (centroid, FWHM) and the HVC properties. A difference in the distribution of LVC centroid velocities is shown in Figure \ref{fig: LVC_distributions} between sources with strong HVC compared to those without or only a weak HVC (log EW$_{\rm{tot},corr} < -0.4$, see Figure \ref{fig: OI_histos}). LVC components tend to be more blueshifted in objects with strong HVC, independent of the classification as double or single Gaussian components \citep[as recently noticed also by][]{mcginnis18}. Moreover, HVCs are typically found in objects that have high accretion and low $n_{13-31}$ (middle and right panels of Figure \ref{fig: LVC_distributions}). 

A detailed exploration of the relation between LVC and HVC properties is presented in Figure \ref{fig: BCNCSCJ_HVC_corr}. In this figure the left panel shows the LVC centroid, FWHM, and EW against the HVC equivalent width while the right panel shows the same properties against the HVC centroid velocity. Here, LVCs are broken out into kinematic classes with the left column of each panel for the NC (light blue), the middle column for the BC (red), and the right column for the SCJ (dark blue with green interior, to remind the reader of the color-coding presented above).

It is the strength of the HVC (left panel), more than the velocity of the HVC (right panel), that is systematically linked to the LVC properties. First, the BC and SCJ centroid velocities correlate with the HVC equivalent width\footnote{The NC centroids show a similar trend.}, a correlation that explains the finding of more blueshifted LVC centroids in objects with strong HVC (Figure \ref{fig: LVC_distributions}). Second, all LVC FWHMs correlate with the HVC equivalent width, but while the FWHM of the NC and BC decreases with increasing HVC EW, the FWHM of the SCJ increases. Third, only the SCJ shows a correlation between its EW and that of the HVC, and between both its FWHM and EW with the HVC centroid velocity (right panel), in the sense that more blueshifted HVC is linked to broader and stronger SCJs.

\subsection{Correlations with accretion} \label{sec: res_accr}
It is well documented that the [OI] HVC and LVC luminosities are well correlated with the accretion luminosity (Section \ref{sec: OIhistory}). Here, we explore for the first time whether the [OI] kinematic properties also correlate with the accretion luminosity.
Figure \ref{fig: OI_Lacc_corr} shows the relations between the accretion luminosity and the centroid velocity, FWHM and equivalent width for all 5 forbidden line components broken out into kinematic classes by column as NC, BC, SCJ, SC and the most blueshifted HVC (green, with interior colors designating their LVC as either BC+NC in pink or SCJ in dark blue, to remind the reader of the color-coding presented above).   

A new finding from this work is that the centroid velocity of 3 out of 5 kinematic components (BC, NC, HVC) is correlated with the accretion luminosity, including the HVC itself (upper panel of Figure \ref{fig: OI_Lacc_corr}). SCJ centroids are not significantly correlated at the 5\% level for this small sample size, but still clearly show a similar trend of increasing blueshift for higher accretors. We note that the datapoint that brings the correlation under significant level in this case is EXLupi during the 2008 accretion outburst (\#25); still, in this figure, as in Figure \ref{fig: BCNCSCJ_HVC_corr} and, later, in Figure \ref{fig: JETvelocities}, the relative changes measured in [OI] emission in EXLupi between quiescence (\#24) and outburst (\#25) align very well with the correlations with accretion discovered in this work from the rest of the sample.
Only the centroids of the SC, which have no HVC, are not correlated and do not show any trend with $L_{\rm acc}$.
Thus, except for the SC and before accounting for either viewing angle or stellar mass, the velocity of the outflowing gas, whether in the HVC or LVC, is higher when the accretion luminosity is higher.

The lower row of Figure \ref{fig: OI_Lacc_corr} shows that the FWHM of the NC, BC, and SCJ correlate with the accretion luminosity. As found with the HVC equivalent width (left panel of Figure \ref{fig: BCNCSCJ_HVC_corr}, middle plots), the sense of the correlation for the NC and BC with $L_{\rm acc}$ is opposite that of the SCJ: while the former show decreasing line width with increasing $L_{\rm acc}$, the SCJ FWHM increase with increasing $L_{\rm acc}$. No correlation is present between the accretion luminosity and the FWHM of the SC in this sample.

\subsection{Correlations between BC and NC} \label{sec: res_BCNC_corr}
Twenty three out of 61 objects in our sample have two LVC components \citep[38\%, consistent with the fraction found in half of the sample by][]{simon16}, hence the kinematic properties of the BC and NC can be compared to each other on a star to star basis. As shown in Figure \ref{fig: BCNC_correlations}, the centroid velocities, FWHM, and equivalent widths of the NC and BC are correlated with each other. The centroids are significantly correlated also when excluding the two most blueshifted objects, DGTau \#12 and RULup \#45 (the Pearson coefficient goes from 0.71 to 0.55). We also compare the [OI]5577/[OI]6300 line ratios between the two components and find the BC has a more constant, and typically higher, [OI] line ratio than the NC, as found in \citet{simon16} and \citet{fang18}.

\begin{figure*}%[ht]
\centering
\includegraphics[width=1\textwidth]{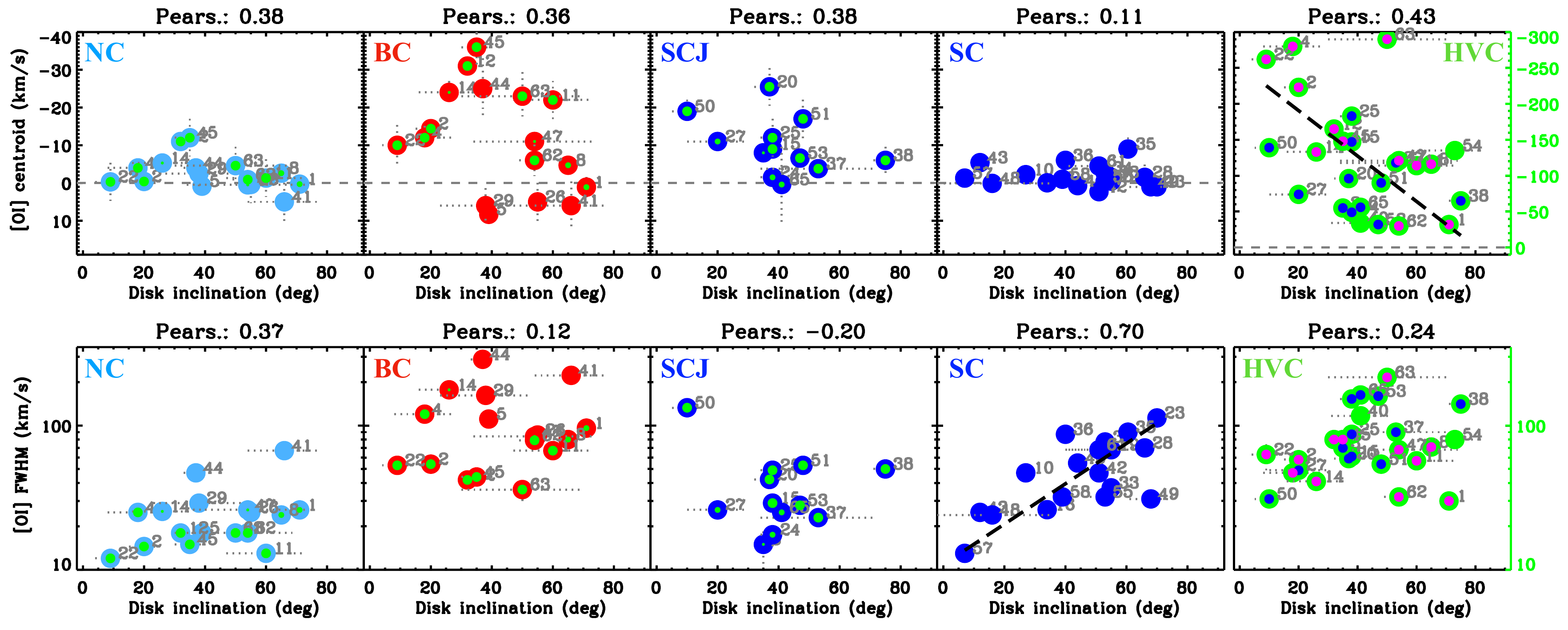} 
\caption{Correlations between [OI] emission properties and disk inclinations. The interpretation of the observed line widths for LVC as broadened by Keplerian rotation in the disk is discussed in Section \ref{sec: disc_kepl}. The correlation between HVC centroids and disk inclination is discussed in detail in Section \ref{sec: disc_jetvels}. Linear fits are shown as black dashed lines, their best-fit parameters reported in Table \ref{tab: fit_params}. Color-coding as in Figure \ref{fig: OI_parspace}. Note the different y axis range for the HVC centroids, in the top right plot.}
\label{fig: OI_incl_corr}
\end{figure*}

\begin{figure*}%[ht]
\centering
\includegraphics[width=1\textwidth]{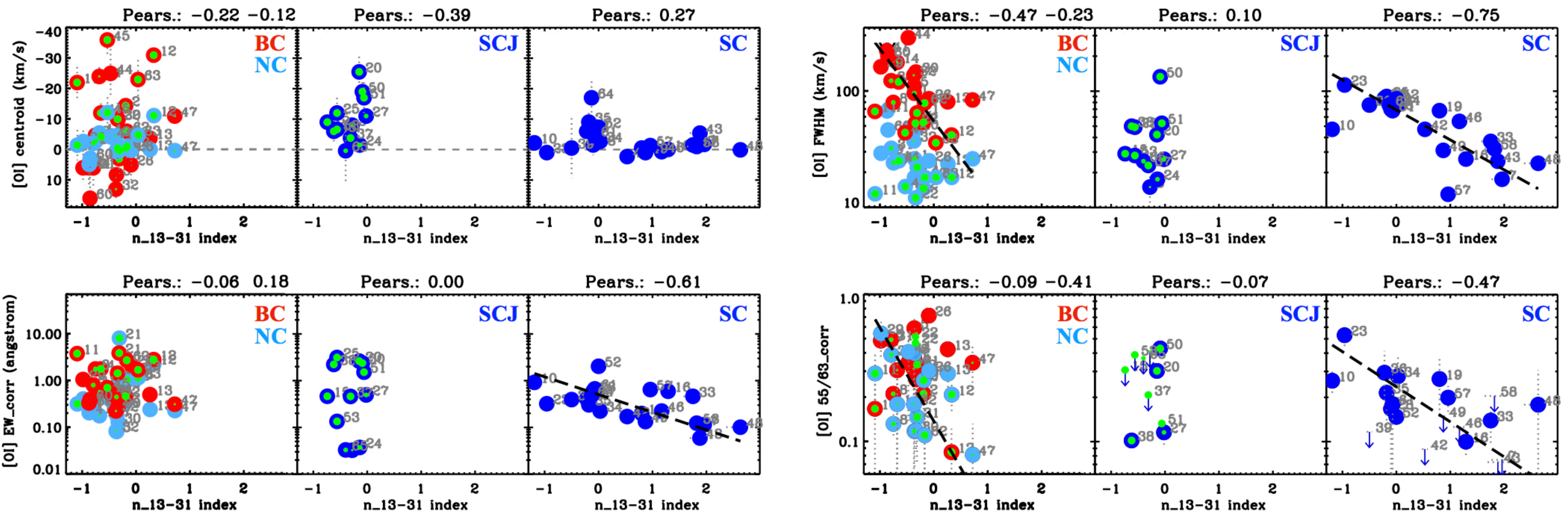} 
\caption{Correlations with $n_{13-31}$ index. BC and NC are overlapped for a more compact visualization; above each plot, the first Pearson coefficient refers to BC, the second to NC. Linear fits are shown as black dashed lines, their best-fit parameters reported in Table \ref{tab: fit_params}. Color-coding as in Figure \ref{fig: OI_parspace}.}
\label{fig: SC_n1331_corr}
\end{figure*}

\subsection{Correlations with viewing angle} \label{sec: res_incl}
In Figure \ref{fig: OI_incl_corr}, we explore correlations between disk inclination and the centroid velocity and FWHM for all [OI] components. 
The relation between the observed centroid velocity and the disk inclination follows the pattern for an outflow perpendicular to the disk for the (most blueshifted) HVC, in agreement with the findings of \citet{appenz13}. LVC components, excluding the SC, show only weak trends: SCJ in the same sense as HVC while BC, and to a less extent NC, show the largest blueshifts at intermediate inclinations of $\sim 35^{\circ}$. However, as shown in Figures \ref{fig: BCNCSCJ_HVC_corr} and \ref{fig: OI_Lacc_corr}, the observed LVC centroids are well correlated with the accretion luminosity and the strength of the HVC, so that any inclination effect is likely secondary compared to what is presumably intrinsically higher velocity outflows in systems with higher accretion rates. 

In terms of line widths, only the SC is correlated with disk inclination (the interpretation of these trends in terms of Keplerian rotation in the disk will be discussed in Section \ref{sec: discuss}). Again, the NC, BC, and SCJ line widths are instead primarily linked to the accretion luminosity and the HVC equivalent width (Figures \ref{fig: BCNCSCJ_HVC_corr} and \ref{fig: OI_Lacc_corr}).

\subsection{Correlations with infrared index} \label{sec: res_infrindx}
Figure \ref{fig: SC_n1331_corr} summarizes the relations between the infrared index and the LVC properties: centroids, FWHM, equivalent width, and the [OI] 5577/6300 ratio. As already shown in Figure \ref{fig: LVC_distributions}, HVC sources (thus including most NC, BC, and all SCJ) have $n_{13-31} \lesssim 0$ pointing to  optically thick inner disks. Only the SC component includes many disks with positive $n_{13-31}$, i.e. disks with dust-depleted inner regions. A new finding of this work is that all SC properties (but their centroids) correlate with the infrared index $n_{13-31}$.

The inferred correlations show that as the optical depth of the inner disk decreases ($n_{13-31}$ increases), the SC EW, FWHM, and 55/63 ratio also decrease.
Similar trends have been presented in the literature when comparing [OI] emission in full and transition disks, using different tracers for the dust content in inner disks (e.g., \citealt{simon16,mcginnis18,fang18}).
Within the smaller range of infrared indices spanned by NC+BC and SCJ, the only emerging correlation is the one between the BC line width and $n_{13-31}$. The sense of the correlation is the same as for the SC but the slope is different\footnote{The trend is also shared by the NC but the Pearson coefficient is brought under a significant correlation level by two outliers (\#11 and \#47).}. In addition, the NC shows a significant correlation in the 55/63 ratios, which decrease with increasing $n_{13-31}$, again same trend as the SC but with a different slope.
None of the LVC centroid velocities are correlated with the infrared spectral index.

\section{DISCUSSION} \label{sec: discuss}
We have systematically investigated five empirical classes of [OI] emission at 6300 \AA\ and explored whether they are linked to each other and to disk properties that probe the evolution of inner disk gas ($L_{\rm acc}$) and dust ($n_{13-31}$). Four of these classes are LVC: BC and NC (from LVC profiles that require two Gaussians for a good fit), and the single-Gaussian SCJ (those that also have an HVC) and SC (those that do not have any HVC). The main results from the analysis described in Section \ref{sec: results} can be summarized as follows:

\begin{itemize}
\item LVCs that have an HVC show strong correlations that tie LVC kinematics to the accretion and jet properties (Figures \ref{fig: BCNCSCJ_HVC_corr} and \ref{fig: OI_Lacc_corr}): LVC FWHMs correlate with the HVC EW and $L_{acc}$, with the peculiarity that BC and NC FWHM decreases together with $L_{acc}$ (or HVC EW), while SCJ FWHM increase. The blueshifts of BC are positively correlated with both HVC EW and $L_{acc}$, and those of NC and SCJ with either HVC EW or $L_{acc}$; HVC blueshifts are positively correlated with $L_{acc}$.

\item BC and NC centroids, FWHMs, and EWs are positively correlated (Figure \ref{fig: BCNC_correlations}). The largest BC and NC blueshifts are for disk inclinations of $\sim 35^{\circ}$; in contrast, HVCs present the largest blueshifts for close to face-on ($incl = 0^{\circ}$) disks (Figure \ref{fig: OI_incl_corr}).

\item The EW, FWHM, and 55/63 ratios of SCs, which do not have HVC and include many disks with inner dust cavities, are negatively correlated with the $n_{13-31}$ index, i.e. the depletion of inner dust results in weaker and narrower lines and lower 55/63 ratios (Figure \ref{fig: SC_n1331_corr}).
\end{itemize} 

Thus, three LVC kinematic components (BC, NC, SCJ) have properties that overall correlate with the strength of the HVC and the accretion luminosity, while one (SC) has properties that overall correlate with the infrared spectral index. 

In the next sections, we discuss these findings focusing on the possible origin of these kinematic components. In particular, we address: rotation and outflow motions imprinted in the observed LVC profiles (Section \ref{sec: disc_kepl}), the origin in MHD winds for LVCs that also have an HVC (Section \ref{sec: disc_wind}), the link between HVC velocities and accretion, which is found to be associated to the two types of LVC profile (BC+NC and SCJ, Section \ref{sec: disc_jetvels}), some open questions associated to these two types of LVC profiles (Section \ref{sec: disc_questions}), and the evolution of inner disk winds as traced by SC-type LVC emission (Section \ref{sec: disc_evol}).

\begin{figure*}%[ht]
\includegraphics[width=1\textwidth]{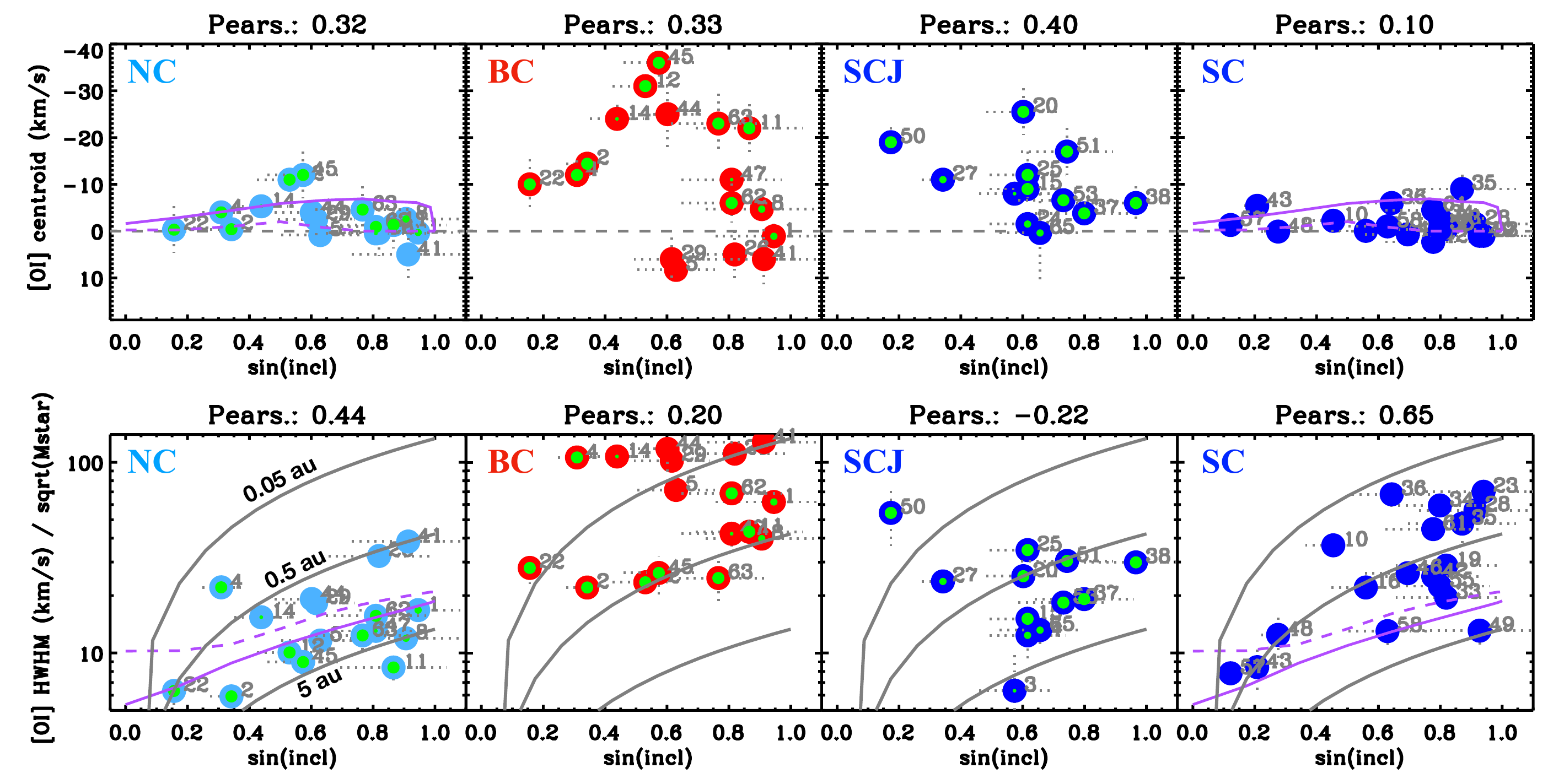} 
\caption{Similar to Figure \ref{fig: OI_incl_corr}, but using the sine of the inclination and the HWHM divided by the square root of the stellar mass. Keplerian models for gas emitting from disk radii of 0.05, 0.5, and 5 au are shown as solid grey lines in the bottom plots. Purple lines show the kinematics of [OI] 6300 \AA\ lines as produced in photoevaporative wind models for a full disk \citep[solid line,][]{EO16} and a disk with an inner cavity of 8.3~au \citep[dashed line,][]{EO10}.}
\label{fig: OI_incl_corr+MODELS}
\end{figure*}

\subsection{Keplerian and outflow imprints on [OI] emission} \label{sec: disc_kepl}
Coupling system inclination to kinematic properties from high resolution spectra provides physical insight into the origin of the gas emission. 
For the HVC, the linear relation between blueshift and disk inclination (Figure \ref{fig: OI_incl_corr}) is consistent with HVCs tracing jets launched perpendicular to their disks, in agreement with previous work (Section \ref{sec: OIhistory}). In contrast, in the same objects, LVCs show no significant linear trends; instead, the centroids of the BC, and to a lesser extent the NC, show a maximum blueshift at a viewing angle of $\sim 35^{\circ}$. This suggests that this angle maximizes the velocity projected toward the observer, i.e. the wind is launched at that angle with respect to the disk rotation axis. This value provides the first observational match from forbidden line spectra to the minimum angle between the rotation axis and magnetic field lines needed to launch an MHD disk wind \citep{bland82}, and remarkably matches predictions for conical-shell MHD winds \citep{kurosawa12}. This angle also remarkably matches the opening angle of cone-shaped CO emission that has been measured around the jet in HH30 from recent ALMA data \citep{louvet18}. While these other works analyzed different tracers that do not necessarily trace a similar portion of the outflow as [OI] emission, it is remarkable that they could overall trace the same basic underlying wind geometry, as supported by the similar semi-opening angles. The evidence for this maximum-projection angle in BC and NC will be combined in Section \ref{sec: disc_wind} to the other correlations found in this analysis, to support an MHD origin for both BC and NC.

Further examination of the relations between the disk inclination and the [OI] centroids and line widths are shown in Figure \ref{fig: OI_incl_corr+MODELS}. The upper panel repeats the relation between the LVC centroid and disk inclination shown in Figure \ref{fig: OI_incl_corr}, while the lower panel explores the relation between the disk inclination and an associated Keplerian radius for the LVC based on the line half width at half maximum (HWHM). Specifically, the HWHM divided by the square root of the stellar mass is plotted against the sine of the disk inclination, and lines of constant Keplerian radii, $\rm{R}_{kepl} = (\rm sin \,\textit{i}\, /\, {\rm HWHM})^2 \, G\,M_{\star}$, are shown at 0.05, 0.5 and 5 au. If the LVCs are broadened by Keplerian rotation, the implied emitting radii for the BC are between $\approx 0.02$--0.5 au while for the NC they are further out, between $\approx 0.5$--5 au, in good agreement with other studies \citep{simon16,mcginnis18,fang18}. 

If Keplerian rotation is the dominant source of the line broadening, and a particular component is emitted from a similar disk radius across all the sources, then a correlation would be expected in the lower panel of Figure \ref{fig: OI_incl_corr+MODELS}.
The correlation is significant in the SC and NC, which also have the lowest blueshifts (upper panel). In contrast, the line widths of BC and SCJ, which have the largest blueshifts, are not correlated with the viewing angle. This interestingly different behavior could be interpreted in the framework of disk winds if the emitting gas extends well above the disk midplane, where the line widths will be affected by relative contributions between the poloidal and toroidal velocity. As gas moves along a streamline, away from the disk surface, the poloidal velocity increases while the toroidal velocity decreases \citep[e.g.][]{KP00,kurosawa06}. In this framework, NC and SC might trace gas relatively close to the disk surface, at lower poloidal velocities and rotating close to the Keplerian speed at larger disk radii. In contrast, at smaller disk radii, the BC and SCJ might trace gas in a higher-velocity part of the wind with a large poloidal velocity but a toroidal velocity lower than the Keplerian speed at the disk surface where it was launched. In this case, the range of BC/SCJ line widths could correspond to a range of heights above the disk surface, instead of a range of disk radii inferred from a sole Keplerian interpretation.

These scenarios need dedicated investigations with wind models applied to the kinematics of [OI] emission. In the rest of the discussion, we associate a [OI] emitting radius to individual sources only for SC components (Section \ref{sec: disc_evol} and Table \ref{tab: OIsingle_param}), because of their low outflow velocity and evidence for the most highly significant correlation, in this sample, between FWHM and disk inclination as expected by Keplerian rotation (Figures \ref{fig: OI_incl_corr} and \ref{fig: OI_incl_corr+MODELS}). However, even in this case we should note that the observed [OI] kinematics might not trace the wind footpoint in a simple way, if [OI] is emitting from a wind that has a large vertical extension above the disk midplane as in existing photoeavaporative models \citep[up to 35~au above the disk,][]{EO10,EO16}. For the other LVC components, the potential range of emitting radii from a sole Keplerian interpretation of the line widths is displayed in Figure \ref{fig: OI_incl_corr+MODELS}.

\subsection{BC+NC: MHD winds kinematically linked to accretion and jets} \label{sec: disc_wind}
Evidence for an origin in an MHD wind was recently found for one LVC component, the BC, since its emitting radii at $< 1$~au indicate a disk wind within the gravitational radius, implying a non-thermal launching process \citep{simon16}.
By finding a tight kinematic connection between BC and NC, and their correlation with HVC and accretion (Figures \ref{fig: BCNCSCJ_HVC_corr}, \ref{fig: OI_Lacc_corr}, and \ref{fig: BCNC_correlations}), our analysis now helps to clarify also the origin of the NC, which remained elusive in recent works and was still debated between a photoevaporative or an MHD wind. The ambiguity comes from NC line widths and centroids, which agree with model predictions by \citealt{EO10,EO16} for disks with and without inner cavities (the purple lines in Figure \ref{fig: OI_incl_corr+MODELS}), suggesting that the NC kinematics are consistent with those expected for a photoevaporative wind (the case of SC components is discussed in Section \ref{sec: disc_evol}). However, our analysis now shows that i) the presence of BC and the low infrared indeces of these disks do not support the existence of an inner cavity in the gas or dust (which in these models produces the lowest [OI] blueshifts near zero velocity), and that ii) it needs to be demonstrated why such a tight kinematic connection between BC and NC (Figure \ref{fig: BCNC_correlations}) should exist if the two components arise in two different types of winds (MHD for BC and photoevaporative for NC). 

Rather, the tight correlations between BC and NC kinematics, together with their correlations with HVC EW and L$_{acc}$, suggest that both BC and NC are part of a same MHD wind. Previous work suggested a common origin for the line excitation, as the line luminosity of BC, NC, and HVC were all found to correlate with L$_{acc}$ \citep{simon16,nisini18,mcginnis18,fang18}. Our discovery of a link between the LVC kinematics and L$_{acc}$ demonstrates that, in addition to a common excitation, these lines share a physical origin in a common outflow process that is connected to accretion. These links favor an MHD origin for all LVC components that also have an HVC, including SCJs, a scenario where the wind/LVC is collimated into a jet/HVC by magnetic stresses \citep[e.g.][]{cabrit99,ferreira06,roman09}. The broad range of wind launching radii possibly implied by this interpretation, extending from inside 0.1 au (the BC) out to several au (the NC), could be explained by the radially-extended ``D-winds" and their re-collimation beyond several Alfv\'en radii \citep[e.g.][]{pudr07}. These winds might also naturally explain a link between the wind velocity and the jet strength and accretion, where angular momentum is extracted from the disk over a range of radii beyond co-rotation \citep[see also the Discussion in][]{fang18}. A key contribution from future modeling work could be to test the [OI] kinematics and their link to jets and accretion across different possible scenarios of MHD winds.

\begin{figure*}%[ht]
\includegraphics[width=1\textwidth]{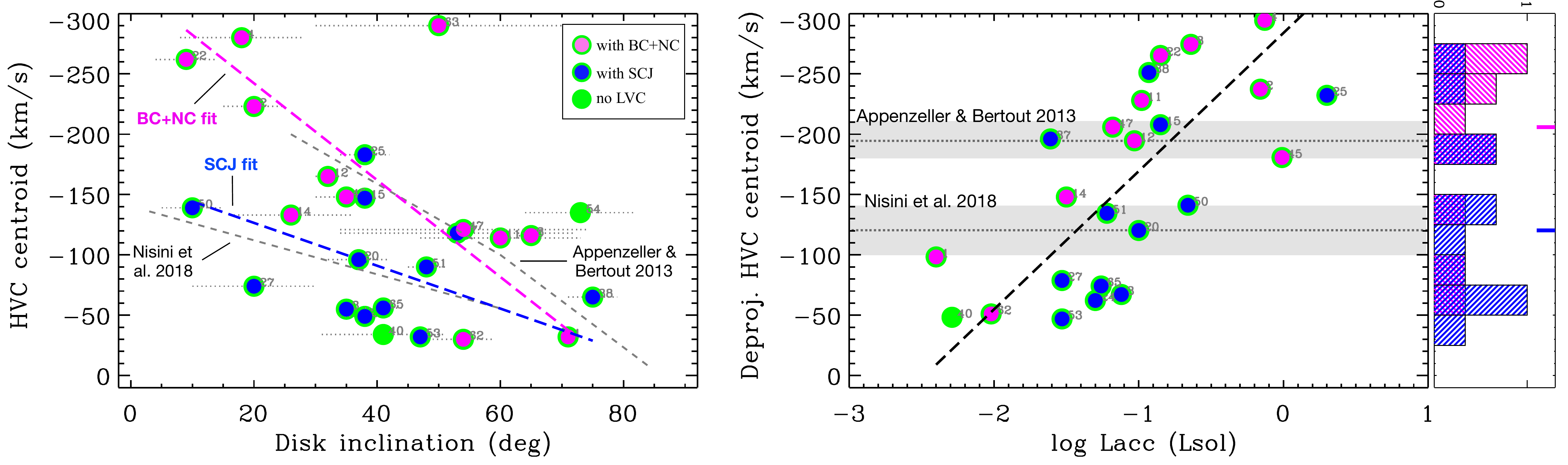} 
\caption{\textit{Left:} Same as top right of Figure \ref{fig: OI_incl_corr}. The trends reported by \citet{appenz13} and \citet{nisini18} are reported with grey dashed lines. Linear fits to the HVCs that have BC+NC/SCJ are shown with magenta/blue dashed lines. \textit{Right:} Same as top right of Figure \ref{fig: OI_Lacc_corr}, but showing the deprojected HVC velocity; a linear fit is shown in black. Average and standard deviation of jet velocities estimated by \citet{appenz13} and \citet{nisini18} are shown as grey shaded regions. The histogram to the right highlights the dichotomy in LVC type discussed in the text; median values for SCJ and BC+NC are marked by the two solid horizontal bars.}
\label{fig: JETvelocities}
\end{figure*}

\subsection{HVC velocities are linked to accretion and to the LVC velocity structure} \label{sec: disc_jetvels}

Previous work provided evidence that jets are connected to accretion, and that the HVC of [OI] emission probes jets (Section \ref{sec: OIhistory}). In addition to the already known correlation between line luminosity and accretion, our analysis finds for the first time that also the HVC centroid velocity correlates with $L_{acc}$ (Figure \ref{fig: OI_Lacc_corr}), suggesting that higher accretion comes with faster jets.

The observed relation between the centroid of the most blueshifted component of the HVC and the disk inclination was shown in Figure \ref{fig: OI_incl_corr}, where the largest blueshifts are seen for the most face-on disks ($incl = 0^o$). We repeat this plot in Figure \ref{fig: JETvelocities} (left panel), but also superpose two relations (gray lines) reported in \citet{appenz13} and \citet{nisini18}: both lines show the same trend but have different slopes and offsets. After deprojection, \citet{appenz13} found an average jet velocity of $\sim$200 km/s for their sample of bright sources mostly in Taurus, while \citet{nisini18} found $\sim$120 km/s for Lupus sources. \citet{nisini18} commented on this discrepancy and suggested the deprojected HVC/jet velocity might depend on the mass of the central stars. 
 
In our sample, the SCJ sources follow the relation of  \citet{nisini18} while the BC+NC follow that of \citet{appenz13}, see Figure \ref{fig: JETvelocities} (left plot). The right panel of Figure \ref{fig: JETvelocities} further reveals a tight relation between the deprojected HVC velocity and the accretion luminosity\footnote{Two objects with high accretion and uncertain disk inclination fall outside the plot: VVCrA and Sz102; their de-projected HVC velocities appear to be at $\sim 450$km/s, but are consistent with $\sim 300$km/s within 1 sigma of their inclination estimates.}:  SCJ sources are primarily associated with lower velocity HVC and lower accretion luminosities while BC+NC with higher HVC velocities and higher accretion luminosities, although there is significant overlap between the two. The average terminal velocity and the standard deviation from \citet{appenz13} and \citet{nisini18} are also indicated on the figure. Although these earlier works did not classify their LVC profiles, it is possible that the sample in \citet{appenz13} was dominated by higher accretion rate sources with BC+NC profiles while the sample in \citet{nisini18} was primarily composed of lower accretion rate sources with SCJ profiles. In support of this scenario, we note that of the 10 Lupus objects we have in common with \citet{nisini18}, 9 are SC/SCJ and only 1 is BC+NC (RULup), and they have median $M_{\star}$ and log$L_{acc}$ of 0.5 M$_\odot$ and -1.53 L$_\odot$. Moreover, of the 8 objects in common with \citet{appenz13} with available disk inclinations, 6 are BC+NC and only 2 are SC/SCJ (CYTau and DLTau), and they have median $M_{\star}$ and log$L_{acc}$ of 0.8 M$_\odot$ and -0.85 L$_\odot$. 
 
In conclusion, our dataset provides a bridge between these previous studies, suggesting that there is a continuum of HVC terminal velocities, from $\approx$ 50 to 300 km/s, that depends on accretion. The lower velocity range ($< 100$~km/s) is thought to be too low for a bonafide jet \citep[e.g.][]{frank14} and might be due to a transition in the outflow properties (e.g. the recollimation) from a fast wind to a jet in lower versus higher accretion rate sources. Why the type of LVC emission, whether SCJ or BC+NC, should depend on $L_{acc}$ and the terminal outflow velocity is yet to be fully understood. We do note, however, that if LVC truly MHD winds that feed the jet (Section \ref{sec: disc_wind}), and if line widths roughly trace their launching radius (Section \ref{sec: disc_kepl}), a different launching region for BC and SCJ (Figure \ref{fig: OI_incl_corr+MODELS}) could provide a natural explanation for faster jets fed by a wind at smaller disk radii (the BC) relative to slower jets fed by a wind at larger disk radii (the SCJ).

\subsection{On the dichotomy in the LVC velocity structure} \label{sec: disc_questions}
A puzzling aspect of the LVC structure is why it sometimes has a double component while in others it appears as a single component. The double component structure, if indeed signifying formation at two different disk radii, could imply a radial region around $\approx 0.5$~au where little or no forbidden emission arises, providing a potential similarity to recent MHD models that find discontinuities in the wind radial structure \citep{nolan17,suriano18}. In contrast, those LVC with a single component structure would suggest no break in the disk surface brightness, and the emission could come from a wide range of radii in order not to show a double peak rotational profile \citep{hartig95,simon16}. In the case of the SCJ, the associated Keplerian radii primarily lie somewhat in between those for the BC and NC (Figure \ref{fig: OI_incl_corr+MODELS}), while for the SC there is a wide range of associated Keplerian radii, extending from $\approx$ 0.1 to 2 AU, almost the full range of both the BC and NC. Here we focus the attention on those sources that also have HVC and discuss those without HVC in the next section.

As seen in Section \ref{sec: disc_jetvels}, the BC+NC structure is most prevalent in sources with higher jet terminal velocities. The tendency for the line widths of both the BC and NC to decrease as the accretion luminosity increases could suggest that their formation region, if broadening is primarily Keplerian, moves outward as their blueshifts increase, a relation unexpected in wind models. However, in an MHD wind scenario where the toroidal velocity decreases as the poloidal velocity increases along the streamlines (Section \ref{sec: disc_kepl}), a reduction in FWHM could correspond to tracing the wind at higher altitudes from the midplane, rather than at larger disk radii. In this case, the BC might form in a narrow range of inner disk radii in all objects ($\approx 0.05$~au, from the BC that do not have strong HVC), arising in an MHD wind that directly feeds the jet. Still, the kinematic correlation with NC suggests that this MHD wind has a conical-shell structure and could be radially extended (Section \ref{sec: disc_kepl}), where it is the inner wind region that is mostly feeding the jet, rather than the wind region at larger distance from the star \citep{fang18}.

The SCJ structure is instead most prevalent in sources with lower HVC velocities (Section \ref{sec: disc_jetvels}). Otherwise they show similar trends with the HVC emission strength and the accretion luminosity as the BC+NC sources with the exception that their FWHM increases with increasing accretion and outflow diagnostics. If line broadening is primarily Keplerian, this behavior is consistent with the expectation from disk wind models that smaller launch radii result in a faster outflow, in contrast to what found for BC as discussed above. If both BC and SCJ are tracing winds feeding the jets, an emitting region of SCJ at larger disk radii, relative to the smaller disk radii probed by BC (Figure \ref{fig: OI_incl_corr+MODELS}), might provide an explanation for the lower HVC velocities found in these objects (Figure \ref{fig: JETvelocities}), as proposed in Section \ref{sec: disc_jetvels}.
While some of these findings are therefore suggestive of distinctive behaviors between the BC+NC and SCJ, possibly due to different wind structures, larger sample sizes would be helpful to better characterize these LVC classes in the future. Even with the sample available for this work, the number of objects in each of the LVC classes is not large, with the SCJ class the least well sampled and where unresolved shocks may contaminate or produce the observed LVC emission in at least 3/12 of these objects (see Appendix \ref{App: disc_shock}).

\begin{figure*}%[ht]
\includegraphics[width=1\textwidth]{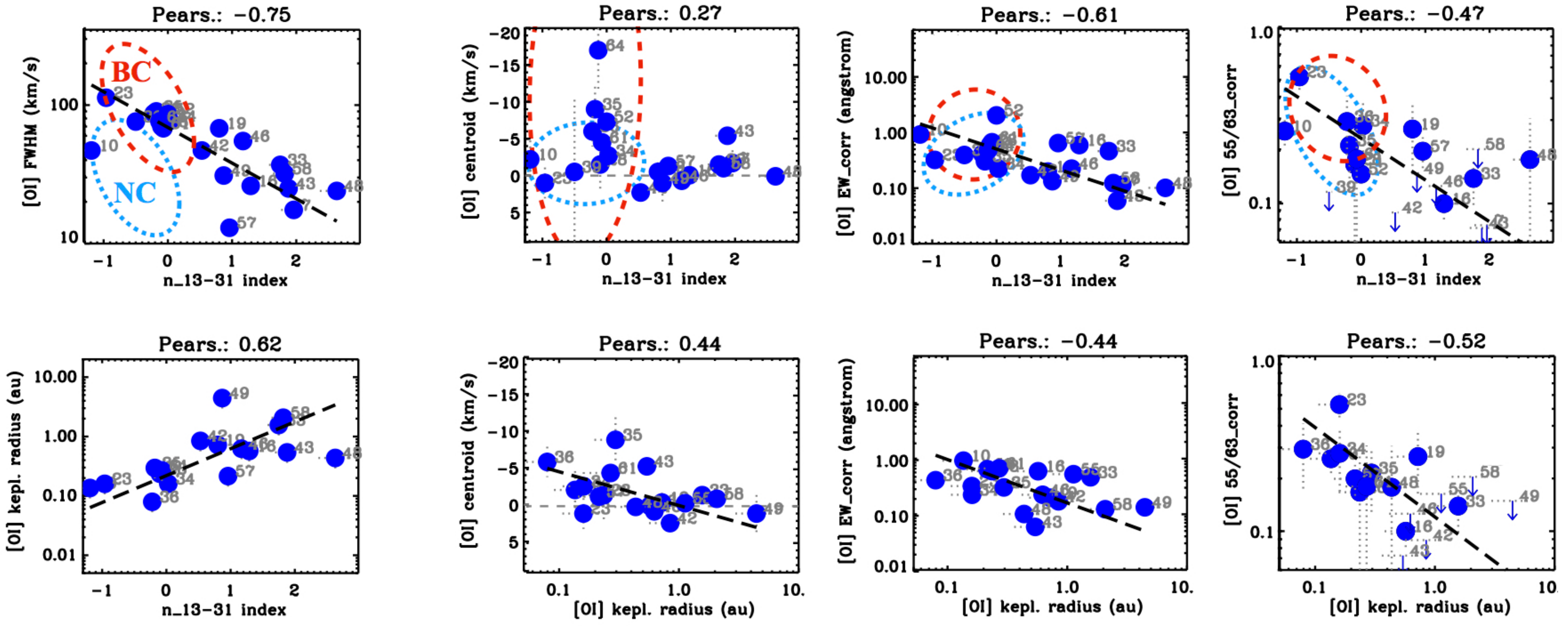} 
\caption{Summary of SC correlations with the infrared index $n_{13-31}$ and the Keplerian radius (note that some sources are missing from some plots, depending on the availability of these parameters, see Table \ref{tab: sample}). Red dashed and light-blue dotted ellipses in the upper plots mark the regions spanned by BC and NC components, for comparison.}
\label{fig: SC_discussion}
\end{figure*}

\subsection{SC: tracing the evolution of inner disk winds} \label{sec: disc_evol}

\citet{EP17} proposed that the LVC emission could trace the evolution of inner disk winds, as based on the prevalence of single-component and narrow [OI] emission lines reported by \citet{simon16} in disks classified as ``transitional". However, they pointed out that the sample was still small, especially in the later evolutionary stages when inner disks are dispersing. With a larger sample and the inclusion of infrared index values, here we use SC components to explore how the gas traced by [OI] emission evolves with the dust in inner disks.

As shown in Section \ref{sec: results}, SC sources span the largest range in $n_{13-31}$ as well as a wide range of accretion luminosities. They thus include sources with optically thick inner disks and high accretion luminosities along with a significant number of sources (13/21) with $n_{13-31} > 0$, indicative of dust-depleted inner disks, which also have lower accretion luminosities. Thus, the SC sources include objects in later evolutionary phases than any other [OI] component, where accretion onto the star has decreased and the inner disk has become more optically thin, in some cases forming an inner disk cavity that is detected through millimeter-wave imaging (e.g. TWHya, LkCa15, RYLup, and several others).
In Figure \ref{fig: SC_discussion}, we summarize the statistically significant correlations found in the SC components that are illustrative of their behavior together with dust depletion and the SC emitting radius.

As the SC sources transition to larger infrared indices, their inferred Keplerian radii become larger (i.e. their FWHM become narrower) and their centroid velocities become less blueshifted (Figure \ref{fig: SC_discussion}).
In addition, as $n_{13-31}$ increases, the SC EW$_{corr}$ and 55/63$_{corr}$ decrease indicating the [OI] emitting region traces cooler/less dense gas further away from the star as the inner disk dust is depleted. This in turn suggests that [OI] emission co-evolves with inner disks where both dust and gas are being depleted possibly in an inside-out fashion. These results support suggestions in other recent works on [OI] emission \citep{simon16,mcginnis18,fang18}, and align well with evidence from other tracers that probe the evolution of molecular gas in inner disks \citep{bp15,hoadl15,banz17,banz18}.

The full distribution of the SC centroid velocities shows predominantly blueshifted emission, with a median at -1.2 km/s very similar to that for the NC (Section \ref{sec: results} and Figure \ref{fig: OI_histos}). It is thus likely that overall the SC also traces a disk wind, and for 12/18 of those with available disk inclination the FWHM is too large to be explained by existing photoevaporative models (Figure \ref{fig: OI_incl_corr+MODELS}). Therefore, for at least these 12 objects it is probably an MHD disk wind formed inside the gravitational radius for thermal escape. Moreover, we note that an additional discrepancy with existing photoevaporative models is that they produce larger [OI] FWHM when an inner disk cavity forms \citep{EO10}, rather than smaller as the data show (Figure \ref{fig: SC_discussion}). It is therefore possible that overall all SC components trace the same MHD wind that evolves together with the dust content in the inner disk. In an MHD wind scenario, it is interesting that no HVC is detected. Maybe these sources have a slow wind with a large opening angle, larger than that indicated by the BC+NC, which does not recollimate into a jet. 

The SC objects with the largest $n_{13-31}$ index ($> 0.5$) and largest Keplerian radii are all known to be transition disks with inner dust cavities, and their centroid velocities are overall consistent with the stellar velocity. In fact, the blueshifts become progressively smaller as the inner dust clears and the emission comes from further out in the disk (Figure \ref{fig: SC_discussion}), consistent with expectations from wind models. As discussed in the case of TW~Hya \citep{pasc11}, profiles centered at the stellar velocity could arise in a disk wind if the [OI] emitting gas originates within the dust cavity, as the receding wind is not occulted, or in alternative from truly bound disk gas extending at large radii beyond the cavity.

\begin{figure*}%[ht]
\centering
\includegraphics[width=1\textwidth]{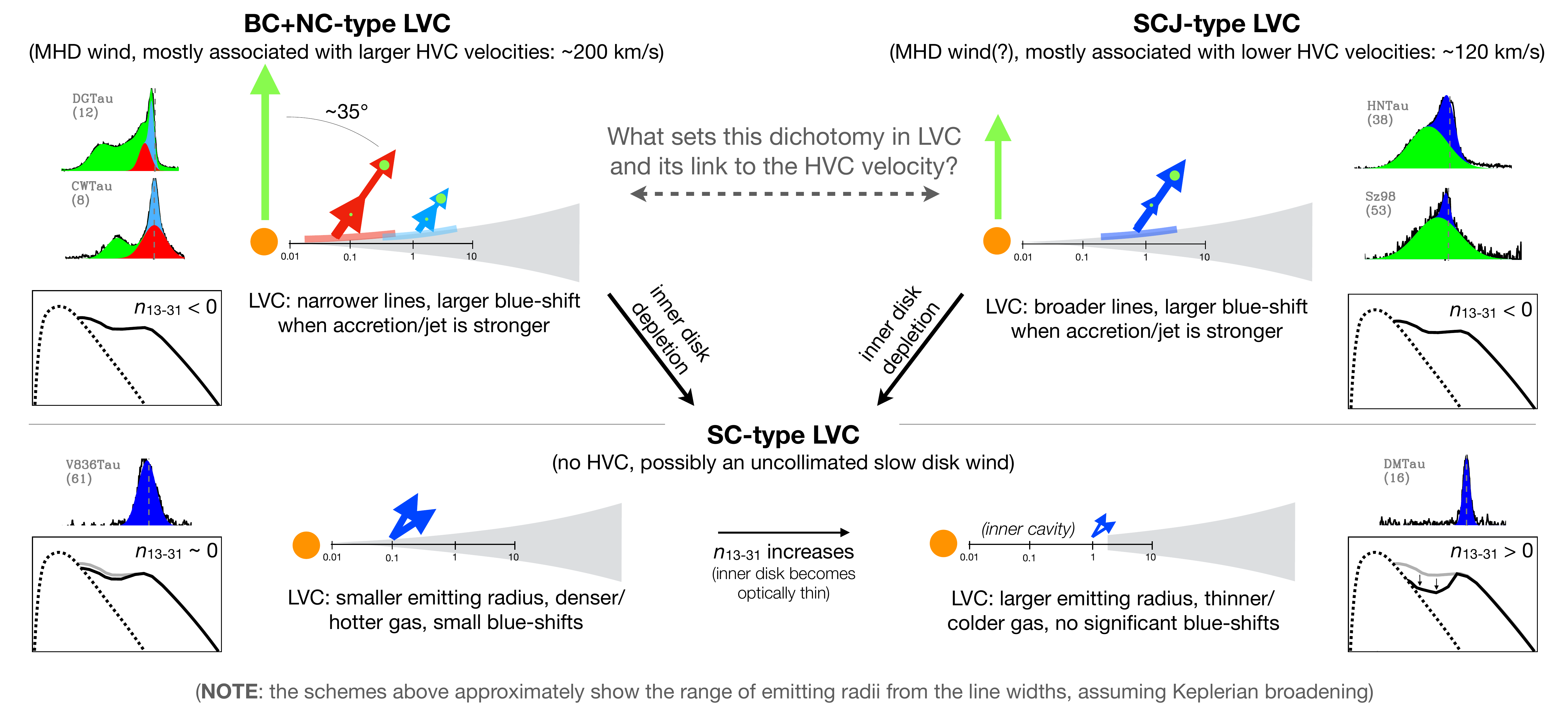} 
\caption{Cartoon summarizing [OI] emission properties and their evolution. Arrows visualize [OI] components properties: the centroid velocity (arrow length), the line width (arrow thickness), and the direction of the jets/winds. The behavior of different LVC components as a function of the HVC EW (Figure \ref{fig: BCNCSCJ_HVC_corr}) is shown as in previous plots, by the size of a green dot inside the arrow. Insets showing SEDs aim at representing the measured values of $n_{13-31}$; the change of $n_{13-31}$ in SCs is exaggerated for better visualization.}
\label{fig: OI_cartoons}
\end{figure*}

\section{Summary and Conclusions}
We have investigated the properties of [OI] emission at 6300 \AA\ and 5577 \AA\ in T~Tauri stars. Key points of this analysis include: 1) a sample size of 65 objects (61 with detected [OI] emission) spanning disks from ``full" to ``transitional"; 2) a spectral resolution of $\sim7$\,km/s, which reveals in detail the kinematics of [OI] emission; 3) the use of a dense grid of photospheric standard stars, to recover [OI] emission where affected by photospheric absorption; and 4) the systematic analysis of [OI] properties and correlations in conjunction with the accretion luminosity and disk properties (specifically the disk inclination and $n_{13-31}$ index). This work largely follows the analysis of \citet{simon16}, and contributes to improvements in terms of sample size, quality of the photospheric-corrected [OI] line profiles, and a systematic analysis of the multi-dimensional parameter space of [OI] properties.

We analyzed and compared the properties of five types of velocity components, identified empirically in the observed emission through multi-Gaussian fit decomposition: a high-velocity component (HVC) with line blueshifts larger than -30\,km/s found in 32 objects, and four types of low-velocity (LVC) components with line blueshifts lower than -30\,km/s: BC and NC (found in 23 objects) which describe the broader wings and narrower center of LVC profiles, and single-Gaussian LVC components that we separated into those that have an HVC (called SCJ, found in 12 objects) and those that do not (called SC, found in 23 objects). This approach allowed us to track kinematic properties of the LVC that had not been explored previously, where all single-component LVCs were assigned to either BC or NC depending on their FWHM \citep{simon16,mcginnis18,fang18}. The analyses by \citet{simon16} and \citet{fang18} are complemented by this work, and our findings support in particular that BC must be an MHD wind \citep{simon16} where most of the mass outflow rate is its inner part (the BC) radii rather than in its outer part \citep[the NC,][]{fang18}.

The main focus of this work has been the kinematic behavior of the LVC, as a contribution to ongoing investigations about its origin and its connection to jets and disk dispersal. This analysis identified several correlations demonstrating the connection between the wind kinematics and accretion, the presence/absence and strength of jets, and the degree of inner disk dust depletion. The main findings and conclusions that we draw from this analysis are listed as follows, and summarized in Figure \ref{fig: OI_cartoons}.

\begin{list}{-}{}
\item LVC kinematics (centroid and FWHM) overall correlate with HVC EW, when an HVC is present, and with the accretion luminosity; these correlations strongly suggest a physical link between LVC, HVC, and accretion, beyond the excitation of [OI] lines by accretion as proposed from luminosity correlations found in previous work (Sections \ref{sec: res_LVC_HVC_corr} and \ref{sec: res_accr}).

\item BC and NC kinematics are correlated, and they show the largest blueshifts at a disk inclination of $35^{\circ}$; this suggests that they are part of a same radially-extended conical-shell wind launched at this angle from the disk rotation axis (Sections \ref{sec: res_BCNC_corr} and \ref{sec: res_incl}).

\item Together with recent/parallel analyses of the emitting region and line ratios \citep{simon16,fang18}, the two points above reinforce the interpretation of LVC as tracing MHD winds that feed jets, when also an HVC is present; moreover, this new analysis supports an MHD origin for NC too, in addition to BC (Sections \ref{sec: disc_kepl} and \ref{sec: disc_wind}).

\item The velocity of the outflowing gas, whether in the HVC or LVC, is higher when the accretion luminosity is higher; a faster wind, higher accretion, and a faster and stronger jet all go together. The positive correlation between HVC velocity and accretion luminosity suggests that stronger accretion drives faster jets, and that the jet may be faster/slower when the MHD wind is launched at smaller/larger radii as traced by the BC/SCJ (Section \ref{sec: disc_jetvels}).

\item SC properties systematically correlate with the infrared index $n_{13-31}$, supporting a scenario where winds and inner disks evolve together: [OI] emission moves to larger disk radii, loses velocity, and traces less dense gas and/or lower temperatures, as inner disks become more optically thin (Section \ref{sec: disc_evol}). These new findings add increasing details to an evolutionary scenario for [OI] emission that is emerging from previous work \citep{hartig95,simon16,EP17,mcginnis18}.

\end{list}

Some aspects of [OI] emission still challenge our current understanding: why LVC has sometimes one sometimes two Gaussian components, why this dicothomy may be linked to the jet velocity, and why BC+NC line widths correlate in opposite sense to the SCJ line width, with the HVC equivalent width and accretion luminosity (Section \ref{sec: disc_questions}). 
Yet, the fundamental relations found in this work demonstrate that accretion, the jet velocity, and the kinematics of disk winds are all connected. A scenario that might comprise all the observed trends is one where a radially extended MHD wind launches and feeds jets, where the wind launching region and velocity are directly linked to the jet and to accretion onto the star. Dedicated modeling will determine which types of MHD winds can reproduce the properties and correlations revealed by [OI] kinematics, so as to better understand the role of these winds as potential drivers of accretion and of the evolution and dispersal of inner disks.

\acknowledgements 
The authors are grateful to James Keane for observing the Magellan-MIKE spectra used in this paper and to Elisabetta Rigliaco for helping with the observing strategy. The authors wish to thank several colleagues for helpful conversations on the findings of this work, in particular Pauline McGinnis, Sylvie Cabrit, Barbara Ercolano, and Antonella Natta; the authors also thank J\'ozsef Varga for providing a measurement of the inner disk inclination in RULup from MIDI data.
A.B. is grateful to Min Fang for providing the grid of photospheric standards, part of the sample included in the analysis, and for several helpful discussions during the parallel realization of this work and of the work presented in \citet{fang18}.
I.P., U.G., and S.E. acknowledge support from a Collaborative NSF Astronomy \& Astrophysics Research Grant (ID: 1715022, ID:1713780, and
ID:1714229). This material is based upon work supported
by the National Aeronautics and Space Administration under
Agreement No. NNX15AD94G for the program ``Earths in
Other Solar Systems". The results reported herein benefitted from collaborations and/or information exchange within NASA Nexus for Exoplanet System Science (NExSS) research coordination network sponsored by NASA’s Science
Mission Directorate.
This paper includes data obtained at the W.M. Keck Observatory, which is operated as a scientific partnership among the California Institute of Technology, the University of California and the National Aeronautics and Space Administration. The Observatory was made possible by the generous financial support of the W.M. Keck Foundation. The authors wish to recognize and acknowledge the very significant cultural role and reverence that the summit of Mauna Kea has always had within the indigenous Hawaiian community. We are grateful to the Hawaiian community to have the opportunity to conduct observations from this mountain.
This paper includes data gathered with the 6.5 meter Magellan Telescopes located at Las Campanas Observatory, Chile.

\appendix

\section{[OI] emission lines and fits} \label{App: linefits_figs}
\begin{figure*}%[ht]
\includegraphics[width=1\textwidth]{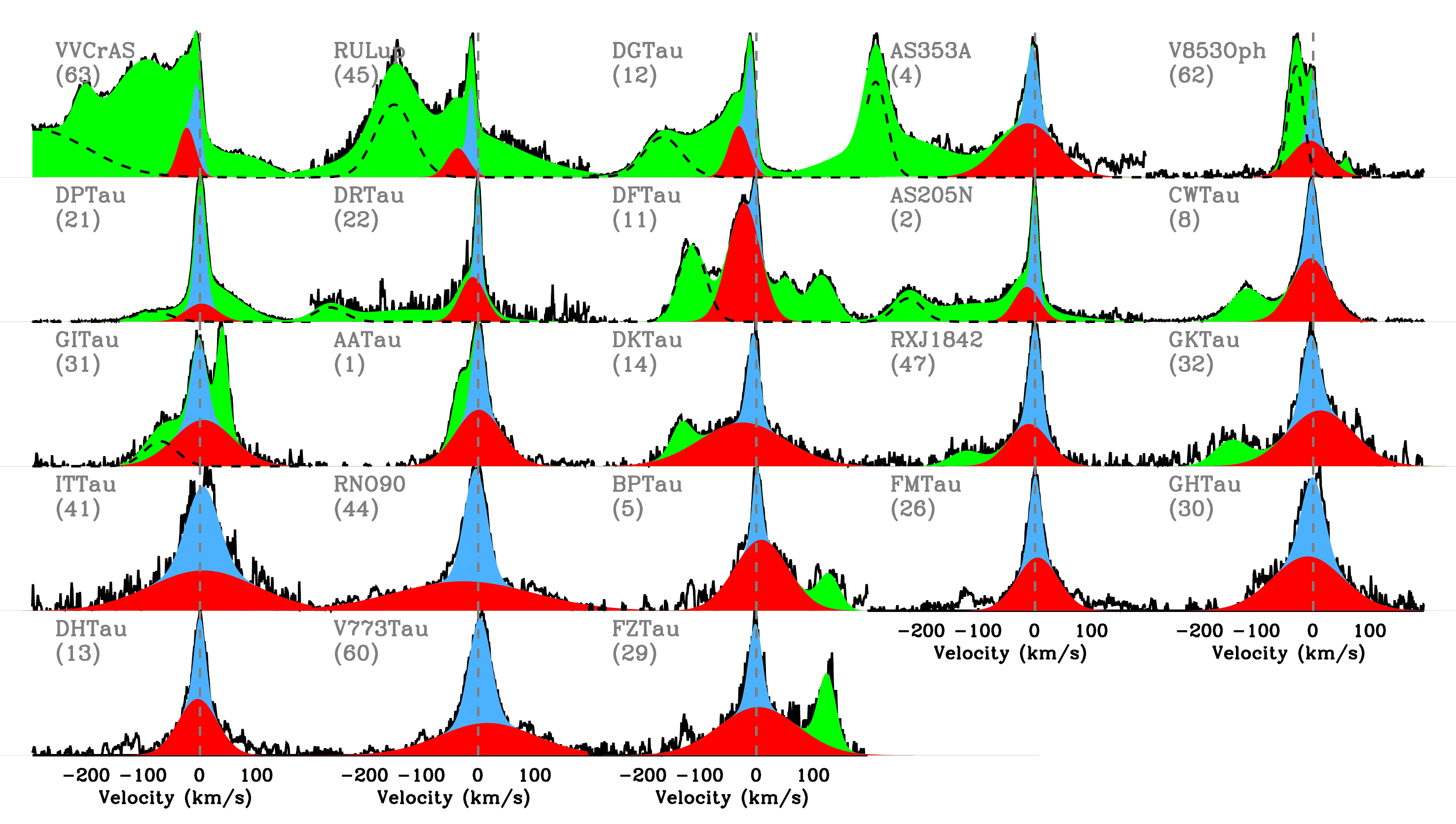} 
\caption{[OI] line profiles and velocity components for ``BC+NC" objects, ordered by the HVC EW$_{tot,corr}$ to visualize the correlations shown in Figure \ref{fig: BCNCSCJ_HVC_corr}.}
\label{fig: BC+NC_lines}
\end{figure*}

\begin{figure*}%[ht]
\includegraphics[width=1\textwidth]{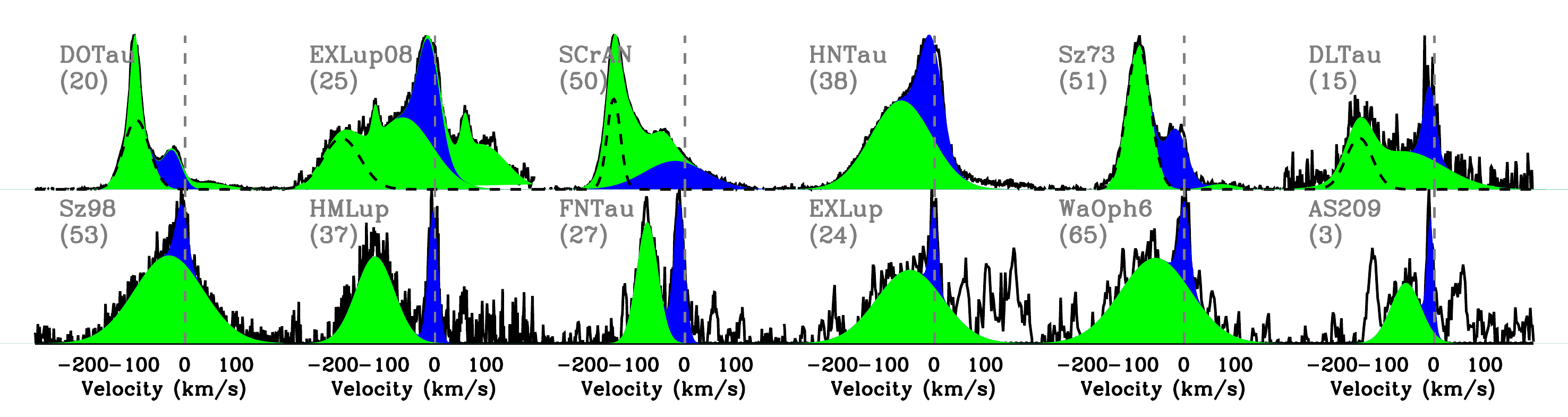} 
\caption{[OI] line profiles and velocity components for ``SCJ" objects, ordered by the HVC EW$_{tot,corr}$ to visualize the correlations shown in Figure \ref{fig: BCNCSCJ_HVC_corr}. }
\label{fig: SCjets_lines}
\end{figure*}

\begin{figure*}%[ht]
\includegraphics[width=1\textwidth]{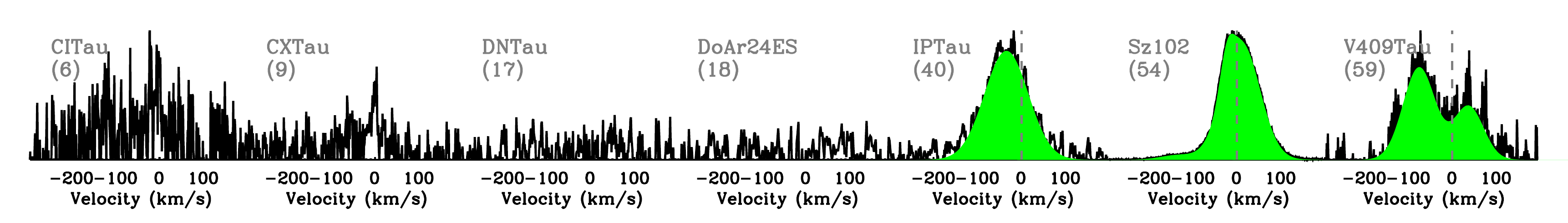} 
\caption{[OI] line profiles and velocity components for objects with no detection or only HVC. CITau and CXTau tentatively show some [OI] emission, but their low S/N does not allow to measure reliably its kinematic properties. }
\label{fig: LO_lines}
\end{figure*}

\begin{figure*}%[ht]
\includegraphics[width=1\textwidth]{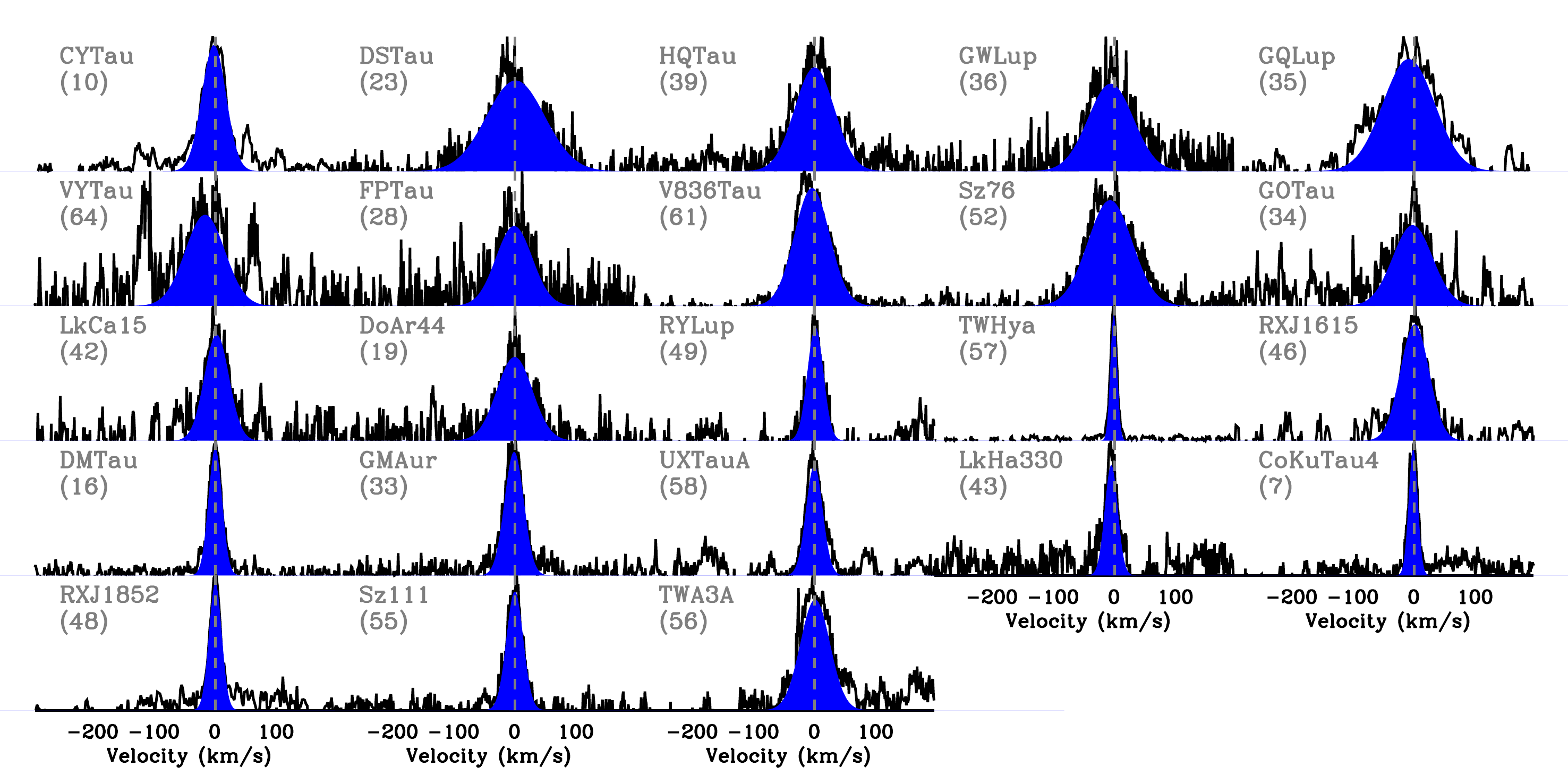} 
\caption{[OI] line profiles and velocity components for ``SC" objects, ordered by $n_{13-31}$ index (Figure \ref{fig: SC_n1331_corr}). Sz111 and TWA3A do not have $n_{13-31}$ measurements, and are put at the end. The two symmetric features at the sides of [OI] emission in VYTau are considered spurious (see Figure \ref{fig: Susp_phot_resid}).}
\label{fig: SC_lines}
\end{figure*}

Figures \ref{fig: BC+NC_lines} to \ref{fig: SC_lines} and Tables \ref{tab: OIdouble_param} to \ref{tab: HVC_param} report the whole sample of [OI] lines and their Gaussian fits, separated into LVC classes as described in the main text (Section \ref{sec: OI_class}).

\section{References for Table 1} \label{App: sample_refs}
Stellar masses are taken from \citet{HH14} for Taurus sources and from the compilation in \citet{salyk13} for sources in other regions, apart from AS353A \citep{rigl15}, DFTau \citep{allen17}, EXLup \citep{galli15}, RXJ1615 \citep{andrews11}, RYLup \citep{alcala17}, and V853Oph \citep{sart03}.

Accretion luminosities for our sample are taken from \citet{simon16} for 30 objects, as estimated from the H$\alpha$ luminosity, and from \citet{fang18} for 29 objects, as estimated from the average of several permitted line luminosities (including H$\alpha$ where available). The value for LkHa330 is taken from \citet{salyk13}, for EXLup in quiescence from \citet{sicagu12}, and during the 2008 outburst from \citet{aspin10}.

Disk inclinations references (reference numbers given in Table \ref{tab: sample}): (i1) \citet{tripathi17}, (i2) \citet{simon17}, (i3) \citet{pontopp11}, (i4) \citet{ansdell16}, (i5) \citet{guillo11}, (i6) \citet{pietu14}, (i7) \citet{vdmarel18}, (i8) \citet{hales18}, (i9) \citet{hug10}, (i10) \citet{bp15}, (i11) \citet{andrews16}, (i12) \citet{casass18}, (i13) \citet{fedele18}, (i14) \citet{tazzari17},  (i15) \citet{macgreg17},  (i16) \citet{scicluna16},  (i17) \citet{kudo08},  (i18) \citet{louvet16},  (i19) \citet{curiel97},  (i20) \citet{cox13}. When the estimated disk inclinations are uncertain, but error-bars are not provided in the original works, we assume here $20^{\circ}$ to mark them as more uncertain (as in the case of references i9 and i16).

\section{Notes on individual objects} \label{App: object_notes}
\subsection{CY\,Tau}
CY\,Tau is potentially a double component LVC with a strong NC and a weak BC, but the photospheric residuals at both sides of the line do not allow to reliably confirm the presence of a BC. We therefore fit a single component and include this object in the SC class, but we warrant that higher S/N and better photospheric correction may reveal a BC in the future. 

\subsection{IP\,Tau}
There is only one [OI] component detected in IP\,Tau, and it presents properties that are intermediate between HVC and LVC. The [OI] line profile varies over time, as reported previously by \citet{simon16}. In this work we analyze the 2012 spectrum, while we also re-corrected the 2006 spectrum; the sawtooth-like residuals shown in \citet{simon16} have disappeared. Here we classify its [OI] emission as a HVC, although it is unclear why no LVC component would be detected in this object. In Sz102, an LVC is possibly embedded into the HVC, which is projected at low velocities due to the close to edge-on orientation of the disk; however, the disk inclination in IPTau is estimated to be $35^{\circ}$, and so this does not seem a viable explanation. IPTau also has an inner disk cavity detected at millimeter wavelengths \citep{Long18}. 

\subsection{LkHa\,330}
This is an object where the inner and outer disk may be misaligned: CO spectro-astrometry gives an inclination of $12^{\circ}$ \citep[the value we use in this analysis, from][]{pontopp11}, while mm imaging gives $39^{\circ}$ \citep{tripathi17}; the presence of spirals in the outer disk has also been proposed to possibly be accompanied by a misaligned inner disk, as seen in other disks \citep[see discussion in][]{banz18}. 

\subsection{RU\,Lup}
The BC in RU\,Lup is embedded in a broad multi-component HVC emission, and given the S/N of this spectrum its detection is tentative. In support of its detection, we find that in the parameter space shown in Figure \ref{fig: OI_parspace} the BC lies next to two other BCs that are detected at high S/N in DG\,Tau and VV\,CrA\,S (with centroids at $\sim - 30$\,km/s and FWHM at $\sim 40$\,km/s). Additional support may come from the H$_2$ line profiles as observed in the UV by \citet{Herczeg05}, which show two components that closely resemble the kinematic of the [OI] LVC, with centroids at -12 and -30 km/s \citep[Fig. 12 in][]{Herczeg05}. 

The disk inclination in RULup is uncertain, and there may be a misalignment between inner and outer disk. The outer disk is well spatially resolved in existing ALMA images, and suggests a face-on view of only $3^{\circ}$ \citep{tazzari17}. On the other hand, spectro-astrometry of NIR CO emission suggested a larger inclination of $35^{\circ}$ \citep{pontopp11}, supported also by the rotation period and rotational broadening of stellar lines that suggest $24^{\circ}$ \citep{Herczeg05}, and by fits to MIDI visibilities at 10.7 $\mu$m that find $55^{\circ}\pm10^{\circ}$ \citep[J. Varga, private communication as based on data published in][]{varga18}. We therefore adopt an inclination of $35^{\circ}$ here, and point out that the inner and outer disk in RULup may be misaligned.

\subsection{RY\,Lup}
We adopt the inclination from the ALMA dust continuum image \citep{tazzari17}, but note that the [OI] FWHM of this object would agree better with the correlation in SC FWHM vs inclination (Figure \ref{fig: OI_incl_corr}) if the inner disk is misaligned as compared to the outer disk, and has a lower inclination closer to face-on orientation ($< 38^{\circ}$) as suggested by \citet{vdmarel18} from SED fitting. 

\subsection{VV\,CrA}
VVCrAS and N are part of a complex binary system with separation of $\approx 2"$ and two circumstellar disks, possibly misaligned, where one star is possibly viewed through the disk of the other one \citep{smith09}. The two disks are as of yet not spatially resolved at millimeter wavelengths. \citet{scicluna16} modeled the two disks using SMA, ATCA, and MIDI data, and proposed a disk inclination of $\approx 50^{\circ}$ for both disks, but warning that the value is uncertain and needs direct measurement from higher-resolution millimeter interferometry data. We note here that by adopting a disk inclination of $\sim 50^{\circ}$ for VVCrAS (and we assume an error-bar of $20^{\circ}$ to mark its high uncertainty), its HVC centroid looks like an outlier in Figure \ref{fig: OI_incl_corr}, and that a lower disk inclination of $\sim 20^{\circ}$ would bring it close to other observed components.

\begin{figure*}%[ht]
\includegraphics[width=1\textwidth]{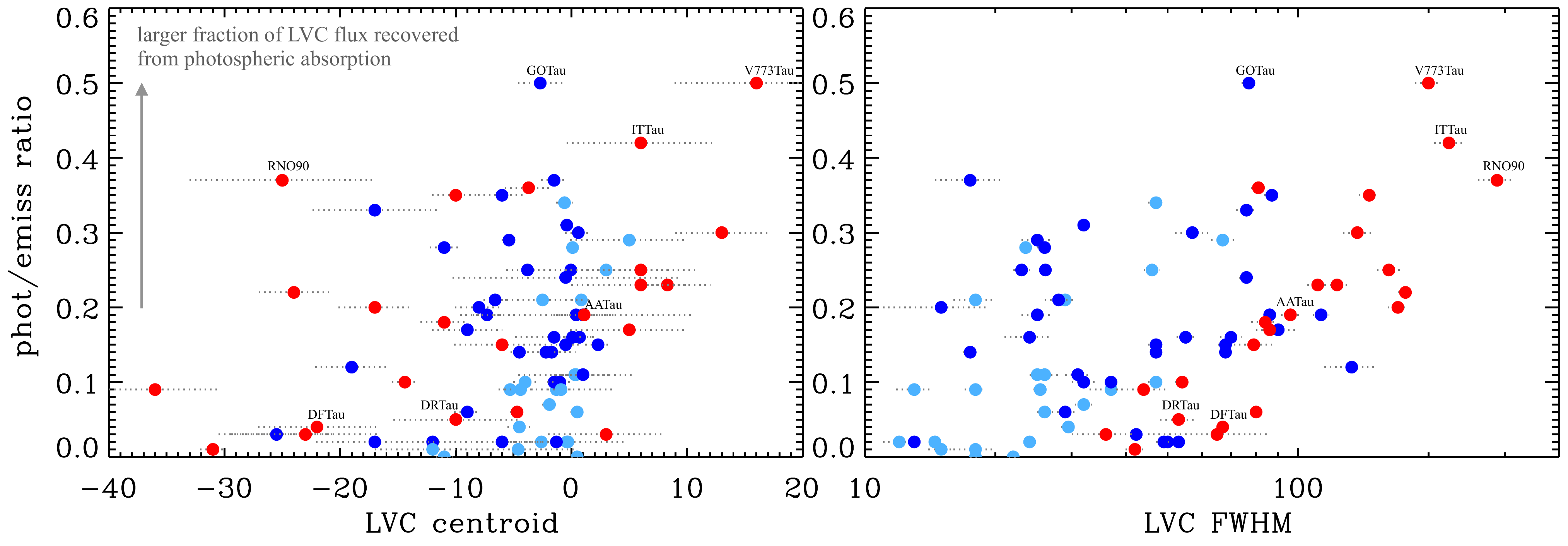} 
\caption{Impact of photospheric correction (Section \ref{sec: OI_correction}) on the measured properties of LVC Gaussian components (Section \ref{sec: OI_fits}), estimated as a measure of the fraction of the LVC flux that has been recovered from photoshperic absorption (see examples in Figure \ref{fig: CompMerit_lines}).}
\label{fig: CompMerit_figure}
\end{figure*}

\begin{figure*}%[ht]
\includegraphics[width=1\textwidth]{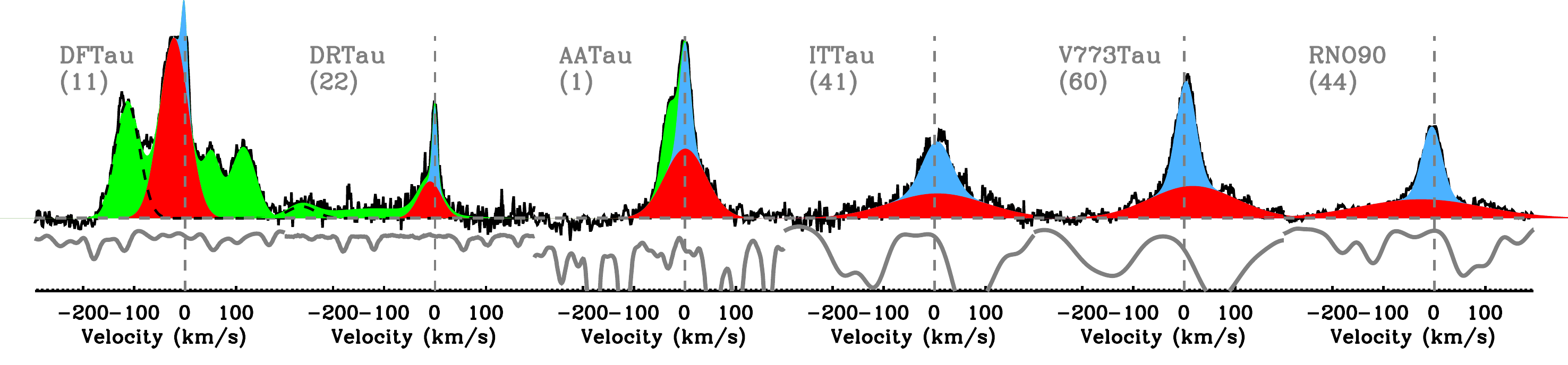} 
\caption{Examples to illustrate the y axis values in Figure \ref{fig: CompMerit_figure}: DFTau and AATau are examples of a low fraction of LVC recovered from photospheric lines (the photosphere is highly veiled), AATau an intermediate case, and ITTau, V773Tau, and RNO90 are the most extreme cases where a large fraction of the BC component is recovered from the broad and deep photospheric lines. The photospheric spectrum is shown below each spectrum for reference.}
\label{fig: CompMerit_lines}
\end{figure*}

\begin{figure*}%[ht]
\centering
\includegraphics[width=1\textwidth]{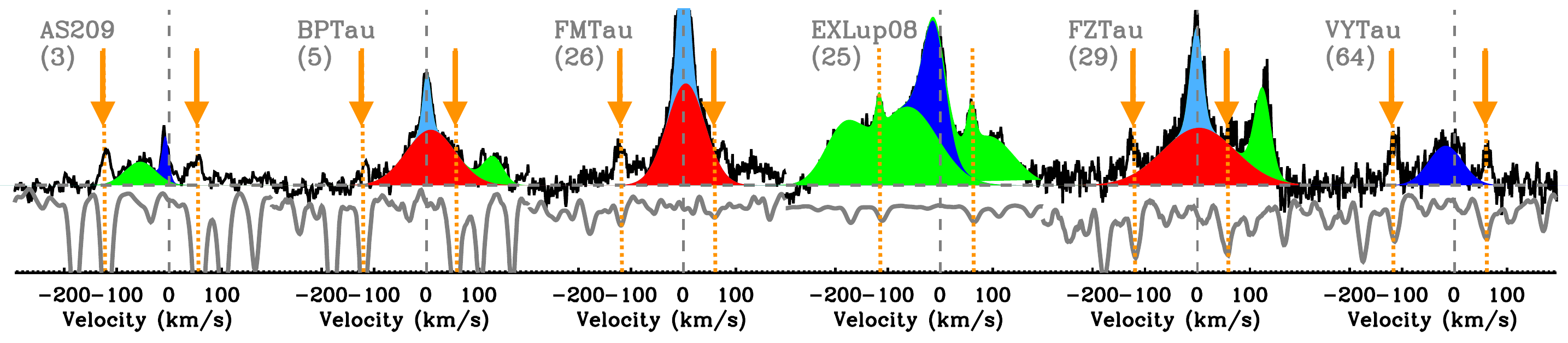} 
\caption{Spurious HVC components (marked with an arrow), produced possibly by photospheric residuals (Appendix \ref{App: phot_corr}) or by FeI emission lines at each side of the 6300 \AA\ [OI] line (Appendix \ref{App: FeI_contamin}).}
\label{fig: Susp_phot_resid}
\end{figure*}

\section{Impact of photospheric correction on [OI] Gaussian components} \label{App: phot_corr}
In Figure \ref{fig: CompMerit_figure}, we show an estimate of the impact of photospheric correction on the flux of the recovered [OI] LVC components. The parameter on the y axis is a proxy for how much LVC flux has been recovered from the absorption caused by the stellar photosphere. The larger its value, the larger fraction of a given LVC component has been recovered from photospheric absorption. The deeper and broader the photospheric lines, the larger is the fraction of LVC flux recovered, depending on how much a given LVC overlaps to photospheric lines (examples are included in Figure \ref{fig: CompMerit_lines}). Figure \ref{fig: CompMerit_figure} illustrates two important things: LVCs are typically more affected when they have some emission on the red side of the 6300 \AA\ (i.e. typically the broad LVC lines), because of two (sometimes deep) photospheric lines on the red side of the 6300 \AA\ line. This is best seen in the BC components: they show a trend in both their centroids and FWHM. The three BCs that stand out as the highest fraction of flux recovered from photospheric lines are ITTau, V773Tau, and RNO90, because they all have deep and broad photospheric absorption (Figure \ref{fig: CompMerit_lines}). In some cases, the BC FWHM of these objects shows up as an obvious outlier in correlations analyzed in Section \ref{sec: results} (RNO90, \#44, in the bottom panel of Figure \ref{fig: OI_Lacc_corr}), and is therefore excluded from the linear fit due to the higher uncertainty of their photospheric correction.

This analysis also highlights the uncertainty of some weak and narrow HVC components, shown in Figure \ref{fig: Susp_phot_resid}. These components can be easily recognized from all other HVC components by their narrow line width and symmetry with the photospheric lines. These narrow features are excluded from the analysis presented in this work. We include also EXLup08 in this figure, to highlight that at least in this object these two features, which are present even before photospheric correction, correspond to emission from FeI lines at each side of the [OI] line (see Appendix \ref{App: FeI_contamin} and \citealt[][]{fang18}).

\section{Contamination from permitted iron transitions} \label{App: FeI_contamin}
\begin{figure*}%[ht]
\centering
\includegraphics[width=1\textwidth]{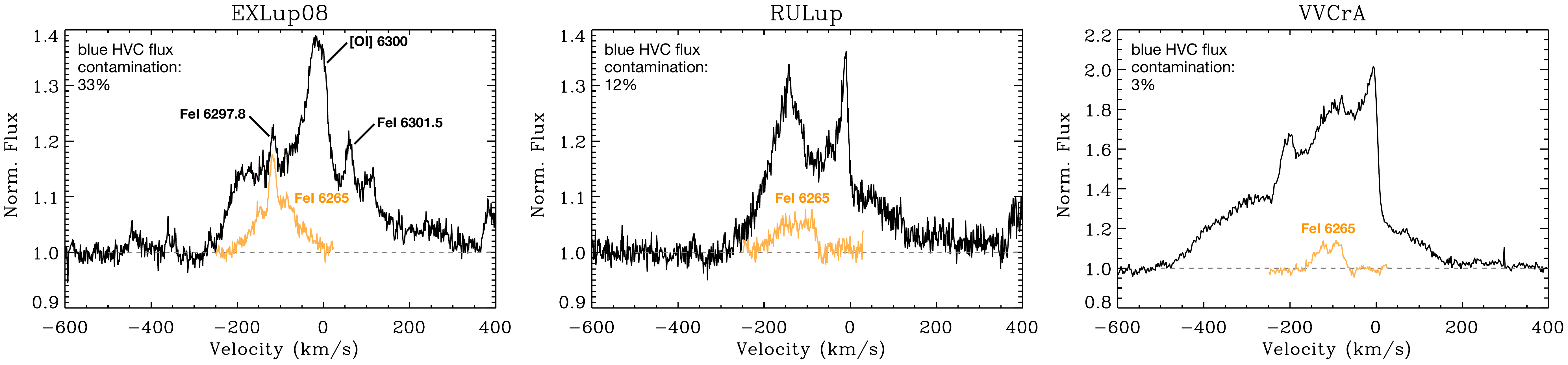} 
\caption{Contamination of [OI] HVC emission from FeI lines. The FeI 6265 \AA\ line, similar in excitation to the 6297 \AA\ line, is shifted to the rest frequency of the 6297 \AA\ line for visual comparison.}
\label{fig: FeI_contamin}
\end{figure*}

The analysis in this work and in the parallel analysis by \citet{fang18} found that [OI] HVC components can be contaminated by FeI emission (previously found in another case by \citealt{petrov14}). Figure \ref{fig: FeI_contamin} shows the three cases where FeI emission has been identified in this sample, and displays the fraction of flux contamination for HVC by using an unblended FeI line of similar excitation. We have subtracted these contamination estimates from the measured HVC equivalent widths used in the analysis for these three objects (Table \ref{tab: HVC_param}).

\section{A note on red-shifted BC components} \label{App: red_BCs}
A fraction of double-component objects (7/23) show red-shifted BCs (Figure \ref{fig: OI_parspace}). These components are among the broadest measured BC components (FWHM $\gtrsim 100$\,km/s), and the significance of their red-shifted centroids is at the 1--2 sigma level only, apart for BPTau and GKTau where it is 3--4 sigma. There are a few reasons to take these red-shifted BCs with extra caution, before concluding that they have real red-shifts. BPTau, FMTau, and FZTau include potential emission from FeI or from photospheric residuals (Figure \ref{fig: Susp_phot_resid}). The two broadest ones (FWHM $\gtrsim 200$\,km/s), in ITTau and V773Tau, are affected by deep and broad photospheric line on the red side of the [OI] 6300 \AA\ line (Figure \ref{fig: CompMerit_lines}). GITau has a strong narrow red-shifted HVC component, which may affect the shape of the BC.

\section{A note on [OI] spectral classifications} \label{App: OI_caution}
The data used in this and in similar recent/parallel analyses lack spatial information on emission that may come from physically different phenomena or spatial regions. With a slit width of $\approx 1"$, we pick up any emission within a radius of $\approx 70$ au from a star at the distance of Taurus. Spectro-imaging studies, which have been possible so far only for a few bright objects \citep[including CWTau and DGTau in this sample, e.g.][]{bacc00,coffey07}, have found that outflows from TTauri stars show complex structures both spatially and kinematically within $1"$. Separating the observed [OI] emission into the sum of Gaussian components is an attempt to identify physically-distinct parts of outflows, and in the absence of spatially-resolved data or a model to guide profile decomposition this is the best approach that can be taken at the present time with this and similar datasets. The fact that the LVC can always be described either by a single Gaussian or by a combination of a broad and a narrow Gaussian, the latter often with different centroids and [OI] line ratios, suggests that this is a reasonable approach until additional information is available. A key contribution of future work could be to test how the spectrally-defined Gaussian components may separate spatially over the extent of the complex observed outflows, especially when they may include multiple phenomena (jets, winds, and/or shocks). In this regard, current challenges of spectro-imaging studies include the small samples observed to date, and the inability to resolve the emission on spatial scales smaller than $\approx 10-20$\,au, the most important scale to study inner disk winds.

With what said above in mind, none of the empirical classifications proposed so far, whether based on kinematics \citep[as in][and this work]{simon16,mcginnis18} and on line ratios \citep[as in][]{fang18}, should be considered strictly definitive or free from uncertain classifications of some components/objects, especially those that show strong variability \citep[e.g. in DK~Tau,][]{simon16,fang18} or those at the edges of any sharp boundaries adopted in individual parameters (e.g. line centroids or 55/63 ratios) that may be unable to capture the real, higher complexity of the data. One lesson learned from this and recent analyses of high-resolution [OI] optical spectra is that these classifications are helpful to identify and analyze specific aspects of the data, yet without excluding other possible choices that in the future may allow to highlight additional aspects.

\section{Veiling and photospheric correction for EW values} \label{App: EWcorr_plots}
\begin{figure*}%[ht]
\includegraphics[width=1\textwidth]{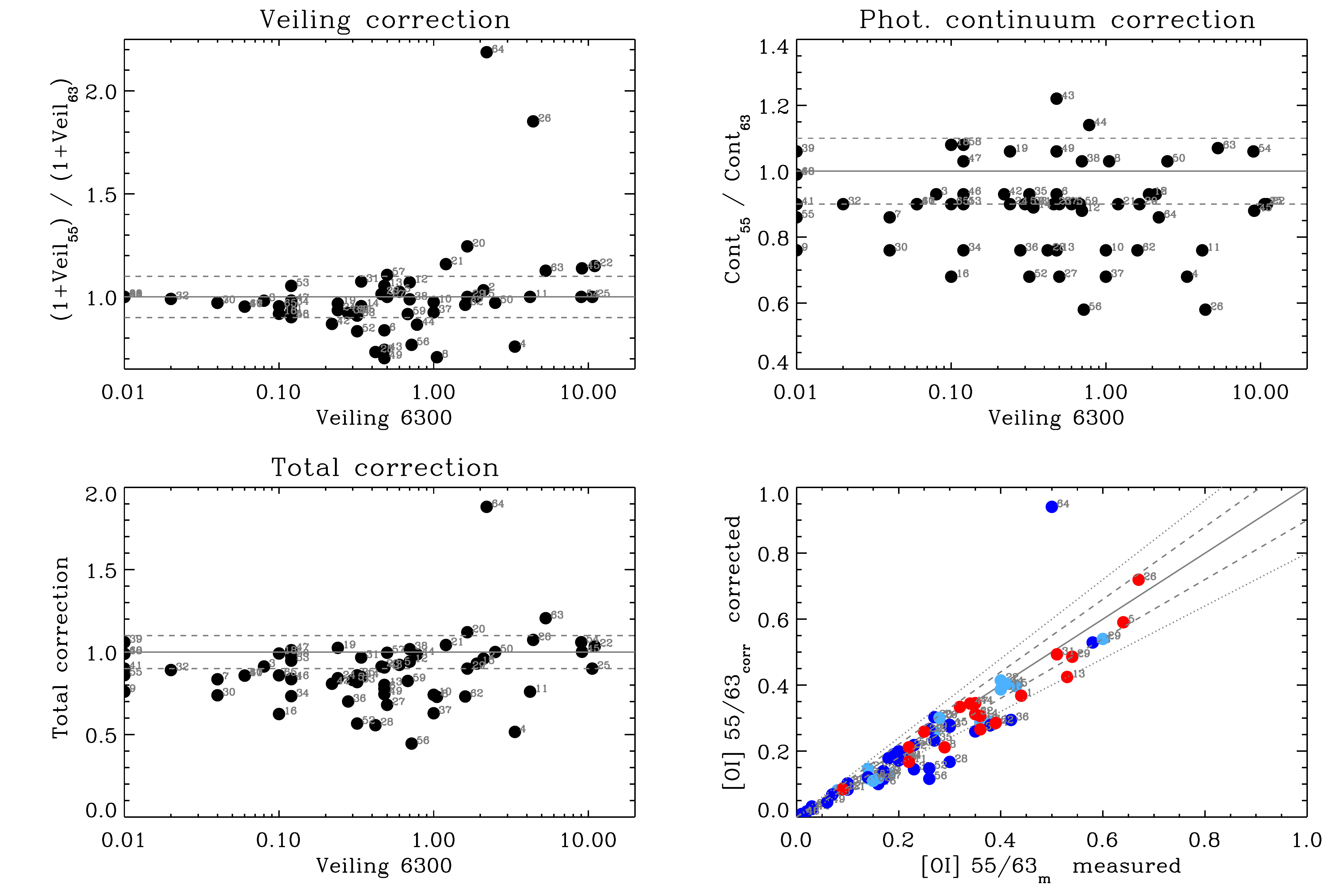} 
\caption{Correction of 55/63$_{m}$ for veiling and photospheric continuum difference between 6300 and 5577 \AA . The two plots at the top show the two terms in Equation \ref{eqn: 55/63_corr}. Dashed and dotted lines mark a 10\% and 20\% change in the plotted values.}
\label{fig: EWcorr_plots}
\end{figure*}

Figure \ref{fig: EWcorr_plots} shows the two terms in Equation \ref{eqn: 55/63_corr}, used to correct the measured equivalent widths for the veiling and photospheric continuum difference between 6300 and 5577 \AA . The one object that stands out is VYTau (\#64), where the measured veiling at 6300 and 5577 \AA\ differs the most (possibly due to a contamination from FeI lines, see Figure \ref{fig: Susp_phot_resid}).

\section{Veiling correlations} \label{App: veiling_correl}
\begin{figure*}%[ht]
\centering
\includegraphics[width=1\textwidth]{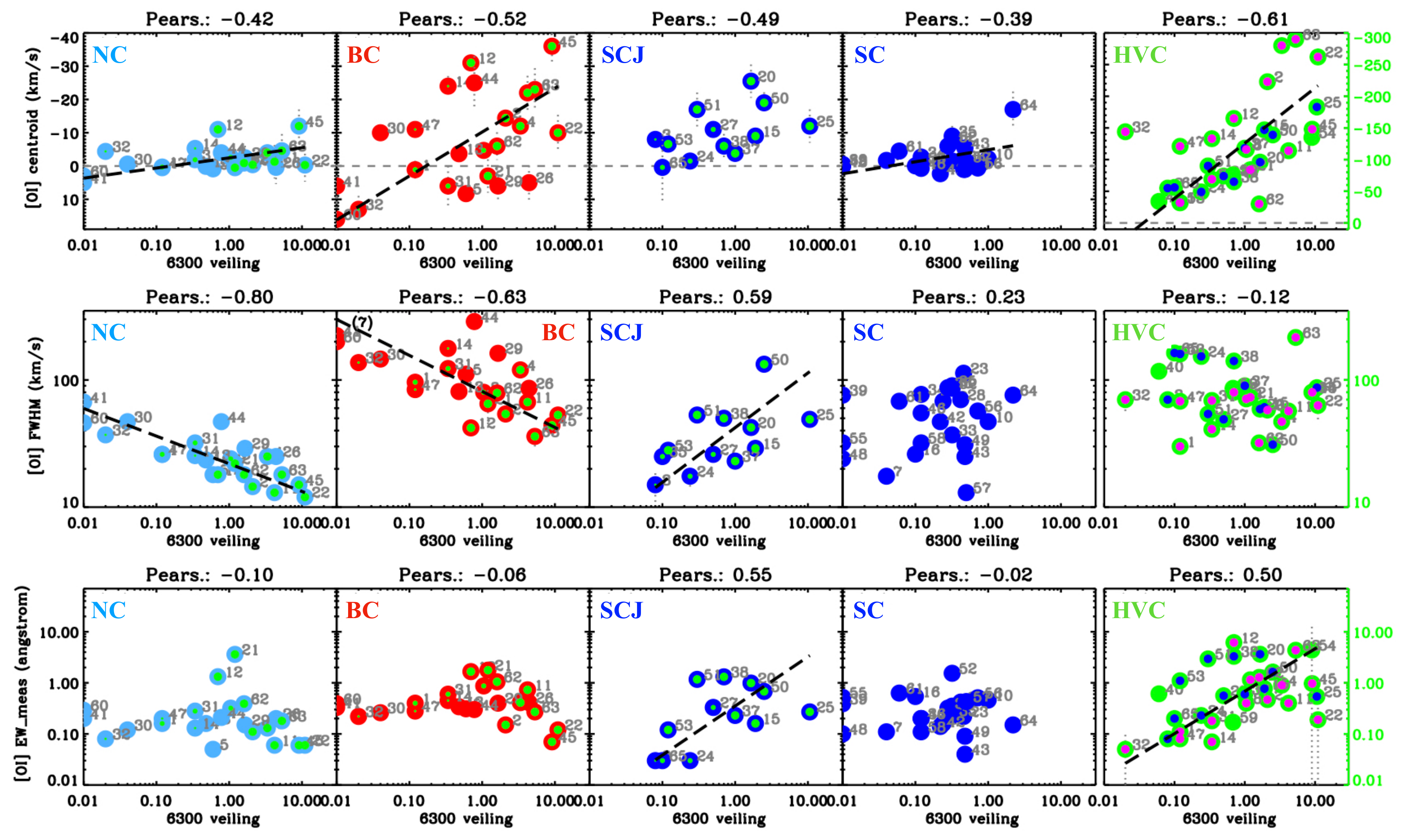} 
\caption{Correlations between [OI] properties and veiling as measured at 6300 \AA . Linear fits are shown as black dashed lines.}
\label{fig: veiling_correl}
\end{figure*}

\begin{figure*}%[ht]
\centering
\includegraphics[width=0.49\textwidth]{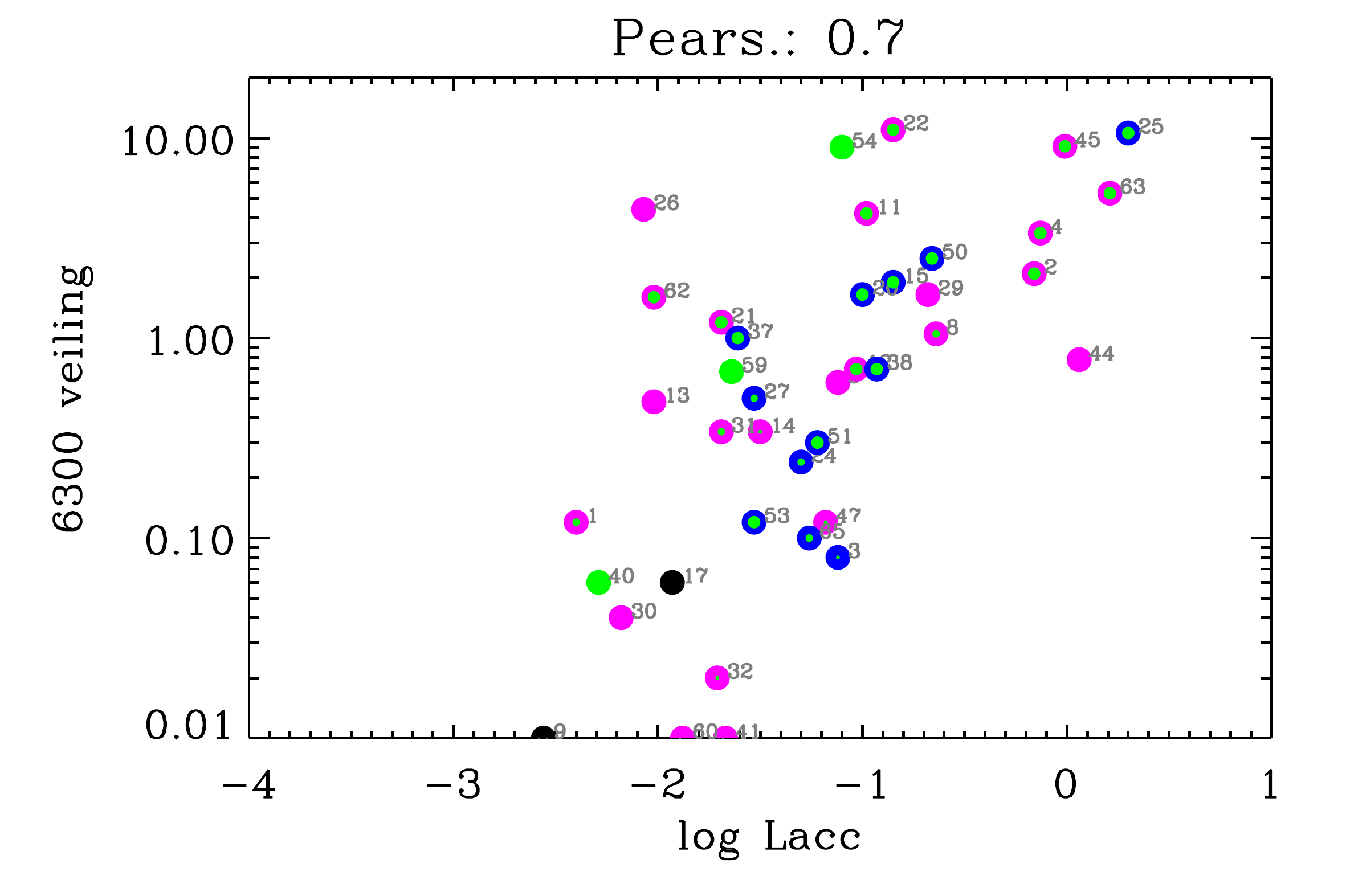} 
\includegraphics[width=0.49\textwidth]{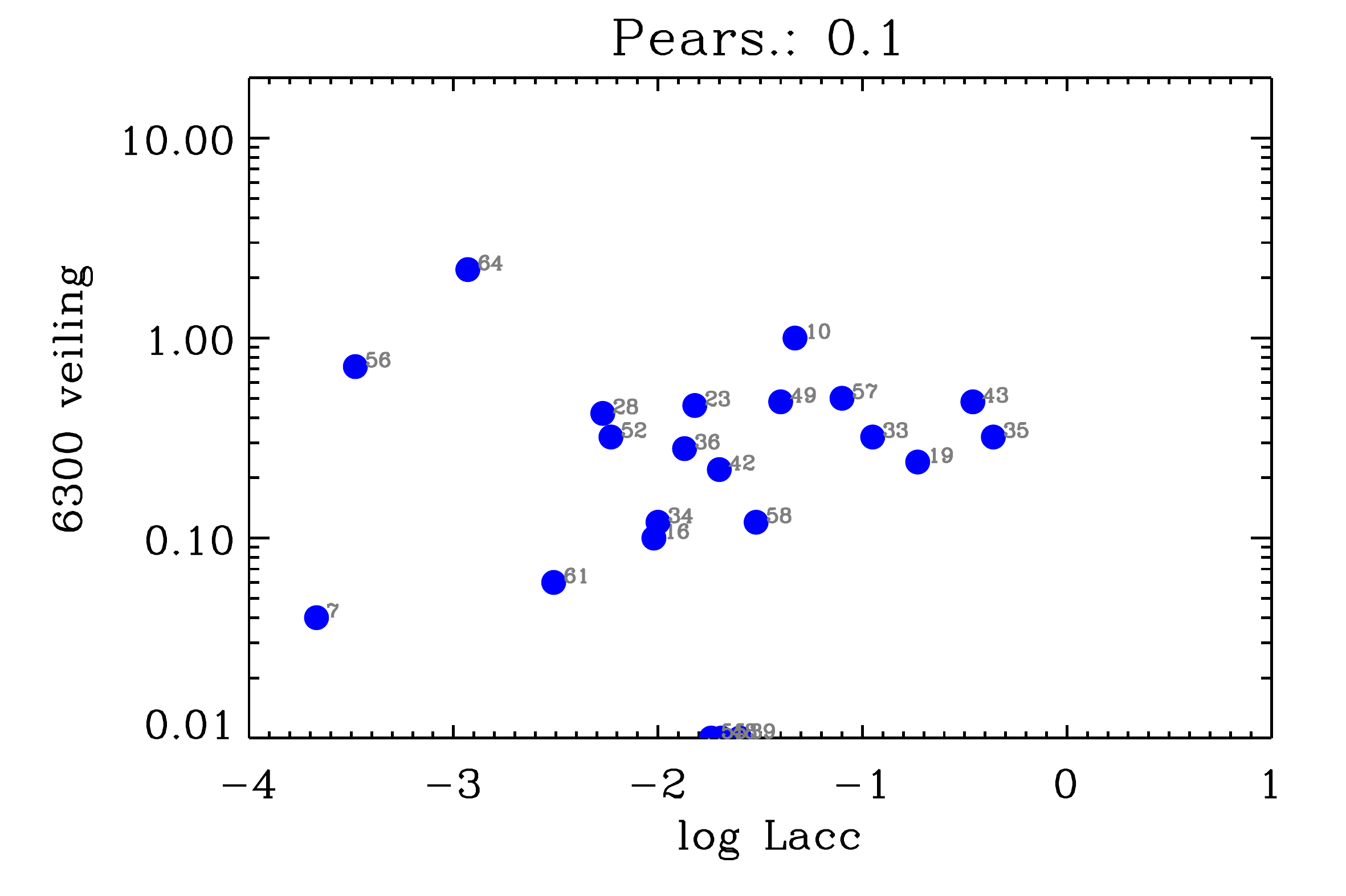} 
\caption{Correlation between accretion luminosity and veiling measured at 6300 \AA\ for the sample included in this work. SCs are shown in the plot to the right, all other components to the left.}
\label{fig: Lacc_veiling}
\end{figure*}

Figure \ref{fig: veiling_correl} shows the correlations found between [OI] emission properties and the veiling measured at 6300 \AA . This figure is analogous to Figure \ref{fig: OI_Lacc_corr} that showed the correlations with accretion luminosity. Figure \ref{fig: Lacc_veiling} shows the correlation between veiling and accretion luminosity for this sample. The figure is separated in two panels, showing SC to the right, and all others to the left, to highlight the notably different correlation. Note that the stronger correlation seen in the plot to the left is supported by the relative change measured in EXLup between quiescence and outburst (\#24 and \#25).
About some notable outliers: VYTau (\#64) has the largest deviation between the veiling measured at 6300 and 5577 \AA\ (Appendix \ref{App: EWcorr_plots}); TWA3A (\#56) has been classified in the literature as an M4, but the latest spectral template we have available is M3, and the measured veiling could be overestimated.

\section{Potential contamination from shocks in SCJ} \label{App: disc_shock}
\begin{figure*}%[ht]
\centering
\includegraphics[width=1\textwidth]{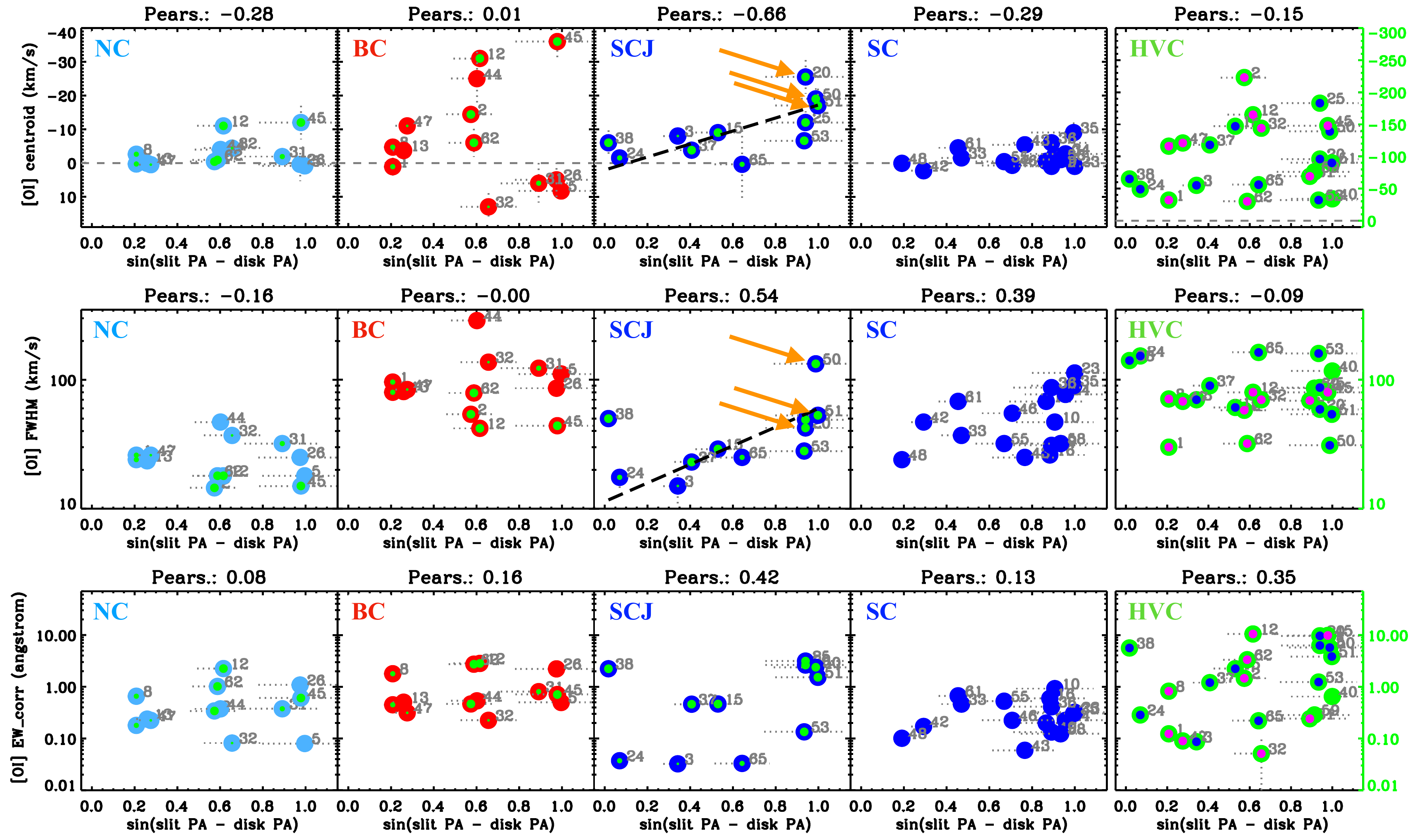} 
\caption{Correlations between [OI] properties and slit PA alignment to the jet PA. If jets are launched perpendicular to their disk, the slit PA is more aligned to the jet for x axis values closer to 1. Orange arrows indicate the three peculiar SCJ profiles discussed in the text (Appendix \ref{App: disc_shock}).}
\label{fig: OI_slitPA+fits}
\end{figure*}

To test the potential spatial extent of the observed [OI] emission, we have checked for correlations between its properties and the slit alignment to the direction of the jet. Figure \ref{fig: OI_slitPA+fits} shows the correlations found between [OI] emission properties and the slit PA alignment to the jet PA. If jets are typically launched perpendicular to their disk, the slit PA is more aligned to the jet for x axis values closer to 1, in this figure. 

Only weak trends, which yet could be expected if a larger portion of the jet falls into the slit, are found in HVC blue-shifts and equivalent widths, showing some increase as the slit is more aligned parallel to the jet (but without a significant correlation coefficient). The strongest correlations are found in SCJ blue-shifts and FWHM, showing an increase for both as the slit is more aligned to the jet. This could be interpreted as a larger portion of a shock (or faster shocks along the jet) being included within the slit. However, the absence of these correlations in HVCs, and the small sample size in SCJs, warrant some caution. There are three peculiar line profiles in the SCJ class that show a strong HVC, the largest blue-shifts within the SCJ class, and a peak-to-continuum ratio increasing from the LVC to the HVC, unlike any other line profile in the whole sample (\#20 DOTau, \#50 SCrAN, and \#51 Sz73, Figure \ref{fig: SCjets_lines}). If these three are excluded, the correlations shown in Figure \ref{fig: OI_slitPA+fits} become statistically insignificant (Pearson coefficients of 0.39 and 0.25). It is therefore unclear whether these three objects trace the same wind as the other SCJs, or rather shocks. A good test for future work would be to obtain multiple slit PAs for a sample of SCJ objects (at least two positions, at $0^{\circ}$ and $90^{\circ}$ from the disk PA). In one case, EXLup, we do have two spectra observed at different PA, but they were also observed in two very different accretion regimes for the central star, so that any PA alignment effects are combined with the intrinsic change in the accretion-outflow phenomena.

The 55/63 ratios could also help distinguishing shocks from winds, as shown in \citet{fang18}. However, 55/63 measured in SCJs are mostly upper limits, and are consistent both with LVC and HVC as measured in other objects, so this parameter alone does not help to distinguish their origin in this sample. In only one SCJ, HNTau (\#38), \citet{fang18} measures a [SII] 4068/ [OI] 6300 ratio as low as found in HVCs, suggesting that at least in this case the SCJ may be tracing similar excitation conditions as those that excite HVCs.

\section{Testing SCJ as potential NC+BC profiles} \label{App: SCJ_test}
\begin{figure*}%[ht]
\centering
\includegraphics[width=1\textwidth]{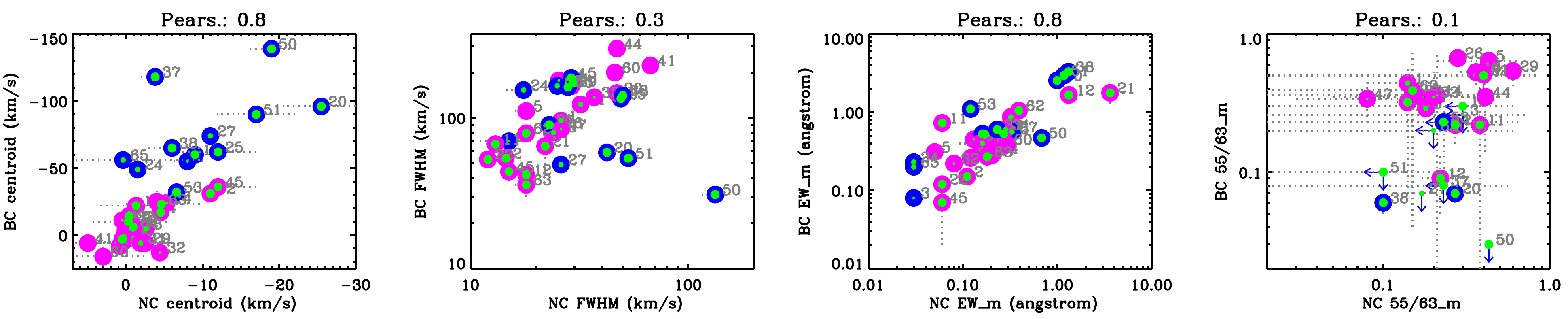} 
\caption{Similar to Figure \ref{fig: BCNC_correlations}, but including also SCJs by assuming that their LVC is NC and their HVC is a highly blue-shifted BC.}
\label{fig: LVC_corr}
\end{figure*}

Figure \ref{fig: LVC_corr} shows the correlations found between BC and NC properties (as in Figure \ref{fig: BCNC_correlations}) in comparison to the properties of the SCJ class. In this comparison, we assume that all SCJs are NCs and that their HVC is a highly blue-shifted BC, to test the scenario that the SCJ class may trace a similar outflow process than the BC+NC profiles, although having larger velocity in the BC. In terms of centroids and EWs, SCJs line up well with the correlations found between BC and NC. In terms of FWHM too, apart from three outliers already identified above for their peculiar line profile (\#20 DOTau, \#50 SCrAN, and \#51 Sz73). In terms of 55/63 ratios, the values measured in SCJ objects populate a lower region on the y axis, supporting a jet origin for their HVC components (which have lower ratios than the BC), but they are consistent with the same region on the x axis, i.e. without constraining the origin of their LVC components as different from the NC (at least in terms of 55/63 ratios, as already pointed out in Appendix \ref{App: disc_shock}). This comparison highlights the possibility that some properties of different velocity components within the [OI] profiles may behave similarly, beyond their empirical classification as HVC or different classes of LVC. This is something that may deserve further investigation in future work, possibly by obtaining spatial information on the emission.

\begin{deluxetable*}{l c c c c c c c c c c}
\tabletypesize{\small}
\tablewidth{0pt}
\tablecaption{\label{tab: OIdouble_param} Properties of [OI] NC and BC emission.}
\tablehead{ & \multicolumn{4}{|c|}{NC} & \multicolumn{4}{|c|}{BC} & & 
\\ %\cline{2-5}
\colhead{Name} & \colhead{FWHM} & \colhead{Centroid} & \colhead{55/63$_{m}$} & \colhead{EW$_{m}$} & \colhead{FWHM} & \colhead{centroid} & \colhead{55/63$_{m}$} & \colhead{EW$_{m}$} & \colhead{Cont} & \colhead{Veil}
\\ 
  & (km/s) & (km/s) &  & (\AA) & (km/s) & (km/s) &  & (\AA) & corr & corr }
\tablecolumns{11}
\startdata
AATau & 26.0 $\pm$ 2.0 & 0.3 $\pm$ 1.1 & 0.14 $\pm$ 0.06 & 0.16 & 96.0 $\pm$ 4.6 & 1.1 $\pm$ 2.7 & 0.44 $\pm$ 0.10 & 0.40 & 0.90 & 0.93 \\
AS205N & 14.5 $\pm$ 0.4 & -0.4 $\pm$ 0.5 & 0.27 $\pm$ 0.04 & 0.11 & 54.0 $\pm$ 2.0 & -14.4 $\pm$ 1.1 & 0.22 $\pm$ 0.06 & 0.15 & 0.93 & 1.03 \\
AS353A & 25.0 $\pm$ 1.1 & -4.0 $\pm$ 1.0 & -- & 0.13 & 120.0 $\pm$ 9.0 & -12.0 $\pm$ 3.1 & -- & 0.41 & 0.68 & 0.76 \\
BPTau & 18.0 $\pm$ 2.0 & 0.9 $\pm$ 0.9 & 0.43 $\pm$ 0.22 & 0.05 & 111.0 $\pm$ 6.0 & 8.3 $\pm$ 1.8 & 0.64 $\pm$ 0.11 & 0.31 & 0.90 & 1.02 \\
CWTau & 24.0 $\pm$ 0.6 & -2.6 $\pm$ 0.5 & 0.18 $\pm$ 0.03 & 0.32 & 80.0 $\pm$ 2.0 & -4.7 $\pm$ 0.6 & 0.29 $\pm$ 0.04 & 0.87 & 1.03 & 0.71 \\
DFTau & 13.0 $\pm$ 1.2 & -1.3 $\pm$ 5.0 & 0.38 $\pm$ 0.15 & 0.06 & 67.0 $\pm$ 2.0 & -22.0 $\pm$ 5.1 & 0.22 $\pm$ 0.20 & 0.73 & 0.76 & 1.00 \\
DGTau & 18.0 $\pm$ 0.5 & -11.0 $\pm$ 0.7 & 0.22 $\pm$ 0.05 & 1.32 & 42.0 $\pm$ 2.0 & -31.0 $\pm$ 1.1 & 0.09 $\pm$ 0.04 & 1.66 & 0.88 & 1.07 \\
DHTau & 23.5 $\pm$ 0.7 & 0.1 $\pm$ 0.5 & 0.36 $\pm$ 0.07 & 0.16 & 81.0 $\pm$ 3.0 & -3.7 $\pm$ 2.1 & 0.53 $\pm$ 0.10 & 0.34 & 0.76 & 1.05 \\
DKTau & 25.4 $\pm$ 0.9 & -5.3 $\pm$ 0.6 & 0.21 $\pm$ 0.20 & 0.13 & 177.0 $\pm$ 7.0 & -24.0 $\pm$ 3.0 & 0.36 $\pm$ 0.35 & 0.45 & 0.89 & 0.96 \\
DPTau & 22.0 $\pm$ 0.5 & 0.5 $\pm$ 0.5 & 0.14 $\pm$ 0.02 & 3.61 & 65.0 $\pm$ 20.0 & 3.0 $\pm$ 5.0 & 0.32 $\pm$ 0.25 & 1.77 & 0.90 & 1.16 \\
DRTau & 12.0 $\pm$ 1.0 & -0.3 $\pm$ 5.0 & $<$0.45 & 0.06 & 53.0 $\pm$ 5.0 & -10.0 $\pm$ 5.4 & $<$0.50 & 0.12 & 0.90 & 1.15 \\
FMTau & 25.0 $\pm$ 0.7 & 0.4 $\pm$ 5.0 & 0.28 $\pm$ 0.03 & 0.20 & 86.0 $\pm$ 3.0 & 5.0 $\pm$ 5.1 & 0.67 $\pm$ 0.04 & 0.41 & 0.58 & 1.85 \\
FZTau & 29.0 $\pm$ 1.5 & -2.5 $\pm$ 3.6 & 0.60 $\pm$ 0.12 & 0.15 & 162.0 $\pm$ 10.0 & 6.0 $\pm$ 4.6 & 0.54 $\pm$ 0.12 & 0.40 & 0.90 & 1.00 \\
GHTau & 47.0 $\pm$ 2.0 & -0.6 $\pm$ 0.7 & 0.16 $\pm$ 0.04 & 0.12 & 146.0 $\pm$ 8.0 & -10.0 $\pm$ 2.1 & 0.36 $\pm$ 0.08 & 0.26 & 0.76 & 0.97 \\
GITau & 32.0 $\pm$ 1.5 & -1.9 $\pm$ 0.6 & 0.40 $\pm$ 0.05 & 0.28 & 123.0 $\pm$ 8.0 & 6.0 $\pm$ 6.0 & 0.51 $\pm$ 0.07 & 0.60 & 0.90 & 1.07 \\
GKTau & 37.0 $\pm$ 4.0 & -4.4 $\pm$ 1.4 & 0.20 $\pm$ 0.15 & 0.08 & 137.0 $\pm$ 10.0 & 13.0 $\pm$ 4.0 & 0.35 $\pm$ 0.18 & 0.22 & 0.90 & 0.99 \\
ITTau & 67.0 $\pm$ 4.0 & 5.0 $\pm$ 5.1 & -- & 0.20 & 223.0 $\pm$ 17.0 & 6.0 $\pm$ 6.4 & -- & 0.33 & 0.90 & 1.00 \\
RNO90 & 47.0 $\pm$ 2.0 & -4.0 $\pm$ 1.0 & 0.41 $\pm$ 0.11 & 0.21 & 288.0 $\pm$ 28.0 & -25.0 $\pm$ 8.0 & 0.35 $\pm$ 0.21 & 0.30 & 1.14 & 0.87 \\
RULup & 15.0 $\pm$ 1.0 & -12.0 $\pm$ 5.0 & -- & 0.06 & 44.0 $\pm$ 6.0 & -36.0 $\pm$ 5.6 & -- & 0.07 & 0.88 & 1.14 \\
RXJ1842 & 26.0 $\pm$ 1.0 & 0.5 $\pm$ 0.6 & 0.08 $\pm$ 0.05 & 0.20 & 84.0 $\pm$ 6.0 & -11.0 $\pm$ 2.1 & 0.34 $\pm$ 0.11 & 0.28 & 1.03 & 0.98 \\
V773Tau & 46.0 $\pm$ 2.0 & 3.0 $\pm$ 5.4 & -- & 0.29 & 200.0 $\pm$ 14.0 & 16.0 $\pm$ 7.1 & -- & 0.39 & 0.99 & 1.00 \\
V853Oph & 18.0 $\pm$ 1.0 & -0.9 $\pm$ 1.1 & 0.15 $\pm$ 0.14 & 0.39 & 79.0 $\pm$ 8.0 & -6.0 $\pm$ 4.1 & 0.39 $\pm$ 0.35 & 1.05 & 0.76 & 0.96 \\
VVCrAS & 18.0 $\pm$ 2.0 & -4.6 $\pm$ 5.0 & -- & 0.18 & 36.0 $\pm$ 6.0 & -23.0 $\pm$ 6.4 & -- & 0.27 & 1.07 & 1.13 \\
\enddata

\tablecomments{The centroids, FWHM, and EW$_m$ values are reported for the 6300 \AA\ line. The last two columns report the correction factors for the 55/63$_m$ ratios as shown in Equation \ref{eqn: 55/63_corr} and Figure \ref{fig: EWcorr_plots}.
}
\end{deluxetable*}

\begin{deluxetable*}{l c c c c c c}
\tabletypesize{\small}
\tablewidth{0pt}
\tablecaption{\label{tab: OIsingleJ_param} Properties of [OI] SCJ emission.}
\tablehead{\colhead{Name} & \colhead{FWHM} & \colhead{Centroid} & \colhead{55/63$_{m}$} & \colhead{EW$_{m}$} & \colhead{Cont} & \colhead{Veil}
\\ 
  & (km/s) & (km/s) &  & (\AA) &  corr & corr  }
\tablecolumns{7}
\startdata
AS209 & 15.0 $\pm$ 4.0 & -8.0 $\pm$ 2.1 & $<$0.45 & 0.03 & 0.93 & 0.98 \\
DLTau & 29.0 $\pm$ 2.0 & -9.0 $\pm$ 1.1 & $<$0.33 & 0.16 & 0.93 & 1.00 \\
DOTau & 42.3 $\pm$ 0.6 & -25.5 $\pm$ 5.0 & 0.27 $\pm$ 0.02 & 0.99 & 0.90 & 1.25 \\
EXLup & 17.5 $\pm$ 3.0 & -1.5 $\pm$ 1.1 & -- & 0.03 & 0.90 & 0.94 \\
EXLup08 & 49.0 $\pm$ 2.0 & -12.0 $\pm$ 5.0 & -- & 0.27 & 0.90 & 1.00 \\
FNTau & 26.0 $\pm$ 1.0 & -11.0 $\pm$ 1.3 & 0.17 $\pm$ 0.03 & 0.33 & 0.68 & 1.00 \\
HMLup & 23.0 $\pm$ 1.0 & -3.8 $\pm$ 1.8 & $<$0.33 & 0.23 & 0.68 & 0.93 \\
HNTau & 50.0 $\pm$ 1.0 & -6.0 $\pm$ 3.6 & 0.10 $\pm$ 0.01 & 1.31 & 1.03 & 0.99 \\
SCrAN & 133.0 $\pm$ 18.0 & -19.0 $\pm$ 3.2 & 0.43 $\pm$ 0.05 & 0.68 & 1.03 & 0.97 \\
Sz73 & 53.0 $\pm$ 3.0 & -17.0 $\pm$ 5.1 & $<$0.16 & 1.17 & 0.90 & 0.92 \\
Sz98 & 28.0 $\pm$ 2.0 & -6.6 $\pm$ 1.0 & $<$0.41 & 0.12 & 0.90 & 1.05 \\
WaOph6 & 25.0 $\pm$ 2.0 & 0.4 $\pm$ 10.0 & $<$0.43 & 0.03 & 0.90 & 0.95 \\
\enddata

\tablecomments{Columns as in Table \ref{tab: OIdouble_param}, but for SCJ components.
}
\end{deluxetable*}

\begin{deluxetable*}{l c c c c c c c}
\tabletypesize{\small}
\tablewidth{0pt}
\tablecaption{\label{tab: OIsingle_param} Properties of [OI] SC emission.}
\tablehead{\colhead{Name} & \colhead{FWHM} & \colhead{Centroid} & \colhead{55/63$_{m}$} & \colhead{EW$_{m}$} & \colhead{Cont} & \colhead{Veil} & \colhead{R$_{kepl}$}
\\ 
  & (km/s) & (km/s) &  & (\AA) &  corr & corr & (au) }
\tablecolumns{8}
\startdata
CoKuTau4 & 17.5 $\pm$ 0.7 & -1.7 $\pm$ 1.7 & $<$0.09 & 0.11 & 0.86 & 0.97 & -- \\
CYTau & 47.0 $\pm$ 1.2 & -2.2 $\pm$ 2.5 & 0.35 $\pm$ 0.06 & 0.46 & 0.76 & 0.98 & 0.1 \\
DMTau & 26.1 $\pm$ 0.4 & -0.1 $\pm$ 0.5 & 0.16 $\pm$ 0.03 & 0.54 & 0.68 & 0.92 & 0.6 \\
DoAr44 & 68.0 $\pm$ 2.7 & -0.5 $\pm$ 1.3 & 0.26 $\pm$ 0.10 & 0.16 & 1.06 & 0.97 & 0.7 \\
DSTau & 113.0 $\pm$ 5.0 & 1.0 $\pm$ 2.1 & 0.58 $\pm$ 0.09 & 0.22 & 0.90 & 1.01 & 0.2 \\
FPTau & 70.0 $\pm$ 4.0 & -1.5 $\pm$ 3.4 & 0.30 $\pm$ 0.20 & 0.42 & 0.76 & 0.73 & 0.2 \\
GMAur & 37.0 $\pm$ 1.0 & -1.5 $\pm$ 0.6 & 0.17 $\pm$ 0.05 & 0.35 & 0.90 & 0.91 & 1.6 \\
GOTau & 77.0 $\pm$ 3.0 & -2.7 $\pm$ 1.9 & 0.38 $\pm$ 0.15 & 0.20 & 0.76 & 0.96 & 0.2 \\
GQLup & 90.0 $\pm$ 10.0 & -9.0 $\pm$ 3.0 & 0.25 $\pm$ 0.10 & 0.23 & 0.93 & 0.92 & 0.3 \\
GWLup & 87.0 $\pm$ 3.5 & -6.0 $\pm$ 2.1 & 0.42 $\pm$ 0.17 & 0.32 & 0.76 & 0.92 & 0.1 \\
HQTau & 76.0 $\pm$ 3.0 & -0.5 $\pm$ 9.8 & $<$0.11 & 0.39 & 1.06 & 1.00 & -- \\
LkCa15 & 47.0 $\pm$ 2.0 & 2.3 $\pm$ 1.1 & $<$0.11 & 0.14 & 0.93 & 0.87 & 0.8 \\
LkHa330 & 25.0 $\pm$ 2.0 & -5.4 $\pm$ 0.6 & $<$0.08 & 0.04 & 1.22 & 0.74 & 0.5 \\
RXJ1615 & 55.0 $\pm$ 2.6 & 0.7 $\pm$ 1.2 & $<$0.15 & 0.20 & 0.93 & 0.90 & 0.6 \\
RXJ1852 & 24.0 $\pm$ 3.0 & 0.1 $\pm$ 0.8 & 0.18 $\pm$ 0.13 & 0.10 & 0.99 & 1.00 & 0.4 \\
RYLup & 31.0 $\pm$ 2.0 & 1.0 $\pm$ 2.5 & $<$0.20 & 0.09 & 1.06 & 0.70 & 4.4 \\
Sz76 & 86.0 $\pm$ 1.4 & -7.3 $\pm$ 1.3 & 0.26 $\pm$ 0.03 & 1.54 & 0.68 & 0.83 & -- \\
Sz111 & 32.0 $\pm$ 1.0 & -0.4 $\pm$ 0.6 & $<$0.19 & 0.52 & 0.86 & 1.00 & 1.1 \\
TWA3A & 57.0 $\pm$ 5.0 & 0.6 $\pm$ 1.2 & 0.26 $\pm$ 0.20 & 0.51 & 0.58 & 0.77 & -- \\
TWHya & 13.0 $\pm$ 0.2 & -1.3 $\pm$ 0.5 & 0.20 $\pm$ 0.04 & 0.43 & 0.90 & 1.11 & 0.2 \\
UXTauA & 32.0 $\pm$ 2.0 & -1.0 $\pm$ 1.1 & $<$0.21 & 0.11 & 1.08 & 0.90 & 2.1 \\
V836Tau & 68.0 $\pm$ 1.0 & -4.5 $\pm$ 0.8 & 0.21 $\pm$ 0.15 & 0.63 & 0.90 & 0.95 & 0.3 \\
VYTau & 76.0 $\pm$ 4.0 & -17.0 $\pm$ 5.4 & $<$0.60 & 0.15 & 0.86 & 2.19 & -- \\
\enddata

\tablecomments{Columns as in Table \ref{tab: OIdouble_param}, but for SC components. Keplerian radii estimated from the HWHM (Section \ref{sec: disc_kepl}) are added in the last column.
}
\end{deluxetable*}

\begin{deluxetable*}{l c c c c c c}
\tabletypesize{\small}
\tablewidth{0pt}
\tablecaption{\label{tab: HVC_param} Properties of [OI] HVC blue-shifted emission.}
\tablehead{\colhead{Name} & \colhead{FWHM} & \colhead{Centroid} & \colhead{55/63$_{m}$} & \colhead{EW$_{m}$} & \colhead{EW$_{tot}$} & \colhead{\# comp}
\\ 
  & (km/s) & (km/s) &  & (\AA) & (\AA)  &   }
\tablecolumns{7}
\startdata
AATau & 30.0 $\pm$ 4.0 & -32.0 $\pm$ 2.0 & $<$0.11 & 0.11 & 0.11 & 1 \\
AS205N & 58.0 $\pm$ 3.0 & -223.0 $\pm$ 1.0 & $<$0.14 & 0.11 & 0.47 & 2 \\
AS209 & 70.0 $\pm$ 3.0 & -55.0 $\pm$ 2.0 & $<$0.26 & 0.08 & 0.08 & 1 \\
AS353A & 47.0 $\pm$ 1.0 & -280.0 $\pm$ 1.0 & -- & 0.30 & 0.90 & 2 \\
CWTau & 71.0 $\pm$ 2.0 & -116.0 $\pm$ 1.0 & 0.06 $\pm$ 0.04 & 0.40 & 0.40 & 1 \\
DFTau & 57.0 $\pm$ 1.0 & -114.0 $\pm$ 1.0 & $<$0.08 & 0.40 & 0.40 & 1 \\
DGTau & 80.0 $\pm$ 1.0 & -165.0 $\pm$ 1.0 & 0.06 $\pm$ 0.02 & 2.43 & 6.15 & 2 \\
DKTau & 41.0 $\pm$ 4.0 & -133.0 $\pm$ 1.0 & $<$0.07 & 0.07 & 0.07 & 1 \\
DLTau & 61.0 $\pm$ 5.0 & -147.0 $\pm$ 2.0 & $<$0.40 & 0.24 & 0.77 & 2 \\
DOTau & 59.0 $\pm$ 0.5 & -96.0 $\pm$ 0.2 & 0.07 $\pm$ 0.01 & 2.55 & 3.64 & 2 \\
DPTau & 73.0 $\pm$ 3.0 & -84.0 $\pm$ 2.0 & $<$0.10 & 1.15 & 1.15 & 1 \\
DRTau & 63.0 $\pm$ 14.0 & -262.0 $\pm$ 4.0 & $<$1.00 & 0.04 & 0.19 & 2 \\
EXLup & 153.0 $\pm$ 10.0 & -49.0 $\pm$ 5.0 & -- & 0.23 & 0.23 & 1 \\
EXLup08 & 87.0 $\pm$ 3.0 & -183.0 $\pm$ 2.0 & -- & 0.24 & 0.54 & 2 \\
FNTau & 49.0 $\pm$ 1.0 & -74.0 $\pm$ 1.0 & $<$0.07 & 0.56 & 0.56 & 1 \\
GITau & 69.0 $\pm$ 5.0 & -69.0 $\pm$ 4.0 & $<$0.03 & 0.18 & 0.18 & 1 \\
GKTau & 70.0 $\pm$ 13.0 & -144.0 $\pm$ 5.0 & $<$0.30 & 0.05 & 0.05 & 1 \\
HMLup & 90.0 $\pm$ 3.0 & -118.0 $\pm$ 1.2 & $<$0.08 & 0.60 & 0.60 & 1 \\
HNTau & 141.0 $\pm$ 1.0 & -65.0 $\pm$ 1.0 & 0.06 $\pm$ 0.01 & 3.30 & 3.30 & 1 \\
IPTau & 117.0 $\pm$ 2.0 & -34.0 $\pm$ 1.0 & 0.18 $\pm$ 0.03 & 0.61 & 0.61 & 1 \\
RULup & 80.0 $\pm$ 2.0 & -148.0 $\pm$ 1.0 & -- & 0.31 & 0.97 & 2 \\
RXJ1842 & 68.0 $\pm$ 8.0 & -121.0 $\pm$ 4.0 & $<$0.30 & 0.08 & 0.08 & 1 \\
SCrAN & 31.0 $\pm$ 1.0 & -139.0 $\pm$ 0.3 & $<$0.03 & 0.47 & 1.63 & 3 \\
Sz73 & 54.0 $\pm$ 1.0 & -90.0 $\pm$ 1.0 & $<$0.10 & 2.95 & 2.95 & 1 \\
Sz98 & 160.0 $\pm$ 3.0 & -32.0 $\pm$ 1.1 & 0.23 $\pm$ 0.15 & 1.10 & 1.10 & 1 \\
Sz102 & 80.0 $\pm$ 10.0 & -135.0 $\pm$ 5.0 & 0.08 $\pm$ 0.02 & 4.40 & 4.40 & 1 \\
V409Tau & 86.0 $\pm$ 4.0 & -76.0 $\pm$ 2.0 & -- & 0.17 & 0.17 & 1 \\
V853Oph & 32.0 $\pm$ 2.0 & -30.0 $\pm$ 0.6 & $<$0.02 & 1.28 & 1.28 & 1 \\
VVCrAS & 216.0 $\pm$ 7.0 & -290.0 $\pm$ 4.0 & -- & 1.57 & 4.36 & 3 \\
WaOph6 & 163.0 $\pm$ 8.0 & -56.0 $\pm$ 4.0 & $<$0.20 & 0.20 & 0.20 & 1 \\
\enddata

\tablecomments{This table includes the most blue-shifted HVC component in the 6300 \AA\ line in each object, and the total EW for the approaching (blue-shifted) part of the jet, as explained in the text (Section \ref{sec: OI_params}). The last column indicates if more Gaussian components are found in the blue-shifted side of HVC in each object (i.e. if any red-shifted HVC is detected it is not included here, e.g. in FZTau, BPTau, GITau, DFTau). The value reported for EW$_{tot}$ has been corrected for FeI contamination in three objects, EXLup08, RULup, and VVCrA (see Appendix \ref{App: FeI_contamin}).
}
\end{deluxetable*}

\begin{deluxetable*}{l l l c c}
\tablewidth{0pt}
\tablecaption{\label{tab: fit_params} Linear fit parameters $y = a+bx$.}
\tablehead{\colhead{} & \colhead{y} & \colhead{x} & \colhead{a} & \colhead{b}}
\tablecolumns{5}
\startdata
Fig.\ref{fig: BCNCSCJ_HVC_corr} & centroid(BC) & log EW$_{tot,corr}$(HVC) & -10.5 $\pm$ 2.9 & -14.4 $\pm$ 3.6  \\
   & centroid(SCJ) & log EW$_{tot,corr}$(HVC) & -8.0 $\pm$ 1.7 & -10.8 $\pm$ 2.6 \\
   & log FWHM(NC) & log EW$_{tot,corr}$(HVC) & 1.33 $\pm$ 0.03  & -0.18 $\pm$ 0.03 \\
   & log FWHM(BC) & log EW$_{tot,corr}$(HVC) & 1.89 $\pm$ 0.03 & -0.24 $\pm$ 0.04 \\
   & log FWHM(SCJ) & log EW$_{tot,corr}$(HVC) & 1.46 $\pm$ 0.04 & 0.37 $\pm$ 0.05 \\
   & log EW$_{meas}$(SCJ) & log EW$_{tot,corr}$(HVC) & -0.52 $\pm$ 0.07 & 1.19 $\pm$ 0.14 \\
   & log FWHM(SCJ) & centroid(HVC) & 0.91 $\pm$ 0.15 & -0.007 $\pm$ 0.001 \\
   & log EW$_{corr}$(SCJ) & centroid(HVC) & -2.07 $\pm$ 0.40 & -0.018 $\pm$ 0.004 \\
Fig.\ref{fig: OI_Lacc_corr} & centroid(NC) & log L$_{acc}$ & -6.2 $\pm$ 1.1 & -3.7 $\pm$ 0.8  \\
   & centroid(BC) & log L$_{acc}$ & -26.1 $\pm$ 4.5 & -15.9 $\pm$ 3.2  \\
   & centroid(HVC) & log L$_{acc}$ & -232 $\pm$ 17 & -101 $\pm$ 13  \\
   & log FWHM(NC) & log L$_{acc}$ & 0.98 $\pm$ 0.07 & -0.34 $\pm$ 0.05  \\
   & log FWHM(BC) & log L$_{acc}$ & 1.53 $\pm$ 0.08 & -0.34 $\pm$ 0.06  \\
   & log FWHM(SCJ) & log L$_{acc}$ & 2.14 $\pm$ 0.15 & 0.57 $\pm$ 0.13  \\
Fig.\ref{fig: BCNC_correlations} & centroid(BC) & centroid(NC) & -0.66 $\pm$ 2.32 & 3.58 $\pm$ 0.54  \\
   & log FWHM(BC) & log FWHM(NC) & 0.23 $\pm$ 0.21 & 1.26 $\pm$ 0.15  \\
   & log EW$_{meas}$(BC) & log EW$_{meas}$(NC) & 0.16 $\pm$ 0.08 & 0.78 $\pm$ 0.10  \\
Fig.\ref{fig: OI_incl_corr} & centroid(HVC) & disk incl. & 254 $\pm$ 33 & 3.16 $\pm$ 0.73  \\
   & log FWHM(SC) & disk incl. & 1.04 $\pm$ 0.11 & 0.014 $\pm$ 0.002  \\ 
Fig.\ref{fig: SC_n1331_corr} & log FWHM(SC) & $n_{13-31}$ & 1.84 $\pm$ 0.04 & -0.26 $\pm$ 0.03  \\
   & log FWHM(BC) & $n_{13-31}$ & 1.74 $\pm$ 0.06 & -0.62 $\pm$ 0.10  \\
   & log EW$_{corr}$(SC) & $n_{13-31}$ & -0.30 $\pm$ 0.07 & -0.38 $\pm$ 0.06  \\
   & log 55/63$_{corr}$(NC) & $n_{13-31}$ & -0.87 $\pm$ 0.07 & -0.64 $\pm$ 0.12  \\
   & log 55/63$_{corr}$(SC) & $n_{13-31}$ & -0.57 $\pm$ 0.06 & -0.30 $\pm$ 0.05  \\
Fig.\ref{fig: JETvelocities} & deproj. centroid(HVC) & log L$_{acc}$ & -284 $\pm$ 24 & -115 $\pm$ 19  \\  
Fig.\ref{fig: SC_discussion} & log Kepl. radius(SC) & $n_{13-31}$ & -0.66 $\pm$ 0.10 & 0.45 $\pm$ 0.08  \\  
   & centroid(SC) & log Kepl. radius(SC) & -0.09 $\pm$ 0.71 & 4.6 $\pm$ 1.3  \\  
   & log EW$_{corr}$(SC) & log Kepl. radius(SC) & -0.80 $\pm$ 0.09 & -0.78 $\pm$ 0.16  \\
   & log 55/63$_{corr}$(SC) & log Kepl. radius(SC) & -0.92 $\pm$ 0.05 & -0.51 $\pm$ 0.09  \\
\enddata
\tablecomments{The table includes the best fit parameters of the statistically significant correlations shown in the figures as labeled in the first column.}
\end{deluxetable*}

\end{document}